\documentclass[12pt,a4paper,twoside,openany]{book}
\usepackage[final]{MCthesis}				%

\title{The impact of the spatial resolution of wind data on multi-decadal wind power forecasts in Germany}
\author{Sofia Morelli}
\matnr{4154334}
\date{\today} 

\supervisor{N. Effenberger}{L. Schmidt}{Dr. N. Ludwig}
\reviewer{Dr. N. Ludwig}{Prof. Dr. H. Brandt}{Prof. Dr. A. Kelava}

\begin{document}
\maketitle

\chapter{Introduction}\label{chapter:background}
One of the primary strategies for mitigating climate change is the transition to renewable energy \citep{IPCC2023}.
Wind power is a crucial natural resource, with global cumulative wind power capacity having exceeded \SI{1}{\tera\watt} due to \SI{117}{\giga\watt} of new wind power installations in 2023 \citep{GlobalWindReport2024}.
To integrate this capacity effectively into the power grid, to find new installation sites, and to gain insight into the future energy output potential, using accurate wind power predictions is essential.
For multi-decadal forecasts, the potential impacts of climate change on wind resources must be considered. Several studies have therefore employed global climate models for such multi-decadal wind power predictions \citep[for a review see][]{reviwWindResourceUnderClimateChange}. However, these models can only provide discretized data at coarse resolutions due to computational and storage constraints. This prompts the question of what wind speed data resolution is sufficient to produce reliable multi-decadal wind power forecasts. Therefore, the present work aims to evaluate the impact of spatial resolution on the forecast quality building on the work of \citet{effenberger2023mind} who have previously investigated the influence of temporal resolution.

Before delving into the investigation, the following \Crefrange{sec:wpp}{sec:resolution} provide the necessary background information. \Cref{sec:wpp} on wind power forecasting addresses the following questions: Why are forecasts needed? How can the power output of wind turbines be predicted from the wind speed? And what makes such forecasts difficult? The focus then turns to the difficulty of obtaining precise wind speed data. \Cref{sec:weather_data} outlines how globally complete and temporally consistent weather hindcasts can be obtained through a blend of observations and numerical weather predictions. Such a hindcast was employed as a reference for validating the climate model data. How climate projections differ from weather predictions and which climate data sets are available is discussed in \Cref{sec:climate_data}. Finally, \Cref{sec:resolution} elaborates on the problem of low data resolution by explaining the underlying causes and presenting downscaling methods to increase the resolution found in the literature.

The research question and related works are presented in \Cref{chapter:research_question}. In \Cref{chapter:methods} the investigated models (\Cref{sec:models}), the applied pre-processing (\Cref{sec:preprocessing}), and the methods (\Cref{sec:analysis}) are introduced. The results are presented in \Cref{chapter:results} and discussed in \Cref{chapter:discussion}. The conclusion is provided in \Cref{chapter:conclusion}.

\section{Wind Power Forecasting}
\label{sec:wpp}

This section provides a comprehensive overview of the methods and challenges associated with wind power forecasting. 
\Cref{subsec:wpp_purpose} discusses the necessity for forecasts on different time scales, introducing the multi-decadal scale that is investigated in this thesis. \Cref{subsec:wpp_generation} describes the process of wind power generation with wind turbines, highlighting critical parameters that must be considered by the prediction model. \Cref{subsec:wpp_model} describes the mathematical relationship between wind speed and power in the form of the theoretical power curve used to predict wind turbine power output in this work. Finally, \Cref{subsec:wpp_challenges} discusses the challenges of wind power forecasting, including the difficulty of obtaining reliable wind data, which is the focus of this thesis.

\subsection{From Short-Term to Multi-Decadal Forecasts}
\label{subsec:wpp_purpose}

The purpose of wind power forecasts varies with the length of the forecast interval. The existing literature distinguishes between ultra/very short-term, short-term, medium-term, and long-term forecasts. However, a clear definition of these timescales is lacking. \citet{timescales} have examined studies from 2010 to 2014 and proposed a classification based on these findings, which is outlined in the following.

According to the definition in  \citet{timescales}, very short-term forecasts range from seconds to an hour. They are used for real-time monitoring and grid control to ensure grid stability. Short-term forecasts of up to six hours are used to optimize power generation to meet demand and minimize costs through load prioritization and coordination of exchanges between different power systems. Medium-range forecasts covering six to 72 hours, are crucial for managing system disruptions, scheduling maintenance, and determining additional power needs. Long-term forecasts spanning 72 hours to several years, are used to plan the expansion and integration of wind power generation, including new technologies, infrastructure renewal, and site selection. 

In addition to the established categories, I would like to introduce the term `multi-decadal predictions', as used in \citet{effenberger2023mind} and \citet{reviwWindResourceUnderClimateChange}. This category refers to time spans of several years to decades into the future. At this scale, the long-term effects of climate change significantly influence the forecasts and are critical to the management decisions based on them. This multi-decadal time scale is the focus of this thesis.

\subsection{Generating Wind Power with Wind Turbines}
\label{subsec:wpp_generation}

\begin{figure}[htp]
    \centering
    \includegraphics[width=0.9\textwidth]{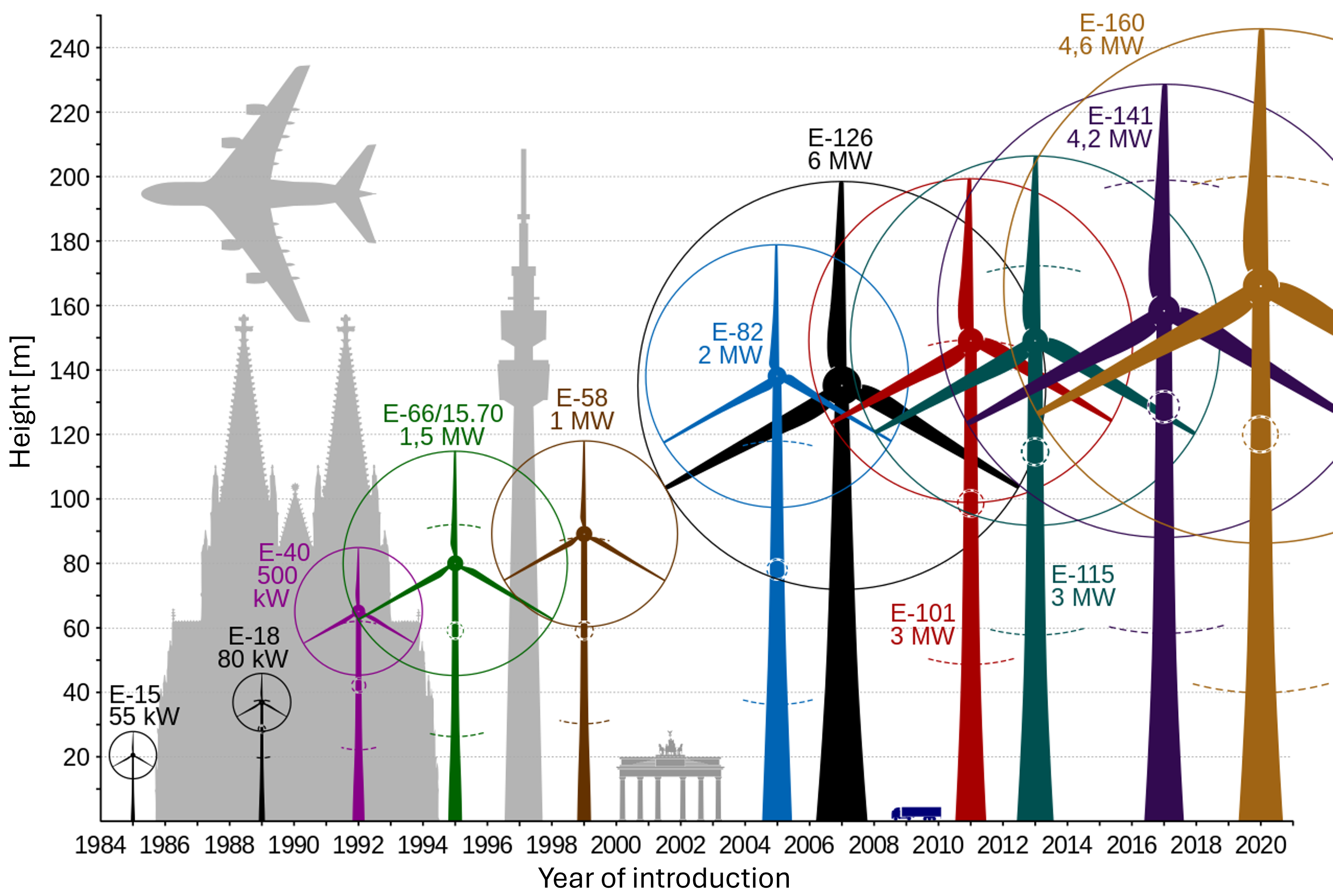}
    \caption{Size evolution of Enercon wind turbines. Turbine types are sorted by year of introduction and drawn at the highest available hub height, with the shortest hub height indicated by dashed lines according to the manufacturer's website. Background elements include an Airbus A380, the Cologne Cathedral, the Berlin Television Tower, and the Brandenburg Gate. Source: \cite{jahobr_enercon_sizes}.}
    \label{fig:enercon_wind_turbines}
\end{figure}

Wind power generation can be defined as the process of converting the wind's kinetic energy into usable electrical energy \citep{wind_power_definiton}. In practice, this is primarily accomplished by horizontal axis wind turbines \citep{hawt_vawt} which are depicted in \Cref{fig:enercon_wind_turbines}.
Their principal components are rotor blades, which capture the movement of air as low-speed kinetic energy; a generator that converts this energy into electricity; and a tower to which the blades are attached. The precise shape and materials depend on the manufacturer and the specific turbine type. Aside from the available wind at the site, the characteristics of these components have the greatest impact on how much power can be generated.

Among the most crucial parameters are the rotor diameter and the hub height, which is the distance from the ground at which the axis of the rotor blades is mounted. The average rotor diameter for onshore wind turbines with a capacity of \SI{1}{\mega\watt} or larger connected to the electricity grid in Europe was \SI{118.61}{\meter} in 2019, while the average hub height was \SI{134.0}{\meter} in 2016, according to the most recent data from \citet{JRC_onshore_turbines}. In the European offshore wind energy sector, the average rotor diameter was \SI{123.61}{\meter}, and the average hub height was \SI{88.60}{\meter} in 2020. In the same year, new offshore turbines had an average rotor diameter of \SI{162.36}{\meter}, and an average hub height of \SI{104.03}{\meter}, indicating a trend towards larger turbines \citep{offshore_turbines_europe}. This development is demonstrated in \Cref{fig:enercon_wind_turbines}.

\subsection{Predicting Wind Power with the Power Curve}
\label{subsec:wpp_model}

The basic model for predicting wind power is the turbine-specific power curve. This non-linear transformation takes the wind speed as input and outputs the wind power a turbine is expected to produce at steady-state performance under ideal aerodynamic conditions. As described in \citet{wind_power_curve}, the curve can be divided into four regions separated by the cut-in wind speed $ws_I$, the rated wind speed $ws_R$ and the cut-off wind speed $ws_O$. At very low wind speeds, the predicted wind power is zero because the rotor blades are not yet turning. Once the cut-in wind speed $ws_I$ is exceeded, the wind power increases cubically with the wind speed. This relation between wind speed $ws$ at hub height and wind power $wp$ is given by
\begin{equation}
    wp=\frac{1}{2} \pi r^2 \rho c_p \cdot  ws^3  \text{\,,}
\end{equation}
where $r$ is the rotor radius of the wind turbine and $\rho$ is the air density at the site. The power coefficient $c_p$ is the fraction of power that is extracted from the free stream wind by the rotor \citep{book_windEnergyExplained}. Its ideal limit is $2/3$ \citep{Betz1926}. The precise value of $ws_I$ depends on the turbine design, with \citet{cutInWS} identifying it to be $\SI{3.5}{\meter\per\second}$ for the majority of turbine types. At the rated wind speed $ws_R$, when the maximum power that can be produced by the turbine is reached, the power curve saturates at the maximum value. Above the cut-off wind speed $ws_O$, the turbine is shut down due to the risk of damage, so the predicted power is zero again at very high wind speeds. The specific power curve used for this thesis is shown in \Cref{fig:power_curve} in \Cref{subsec:wp_analysis}.

\subsection{Model-Inherent and Data-Related Prediction Challenges}
\label{subsec:wpp_challenges}

A theoretical model is always a simplification of reality. In the case of the power curve \citep{book_windEnergyExplained}, the turbine is assumed to be a so-called actuator disk with uniform thrust and an infinite number of blades. The airflow is assumed to be constant, homogeneous, and incompressible, without frictional drag and rotating wake. This neglects many influencing factors such as airfoil design, wind direction, and topography of the surroundings.  \citet{performanceHAWT} found that optimal aerodynamic conditions, required for HAWTs to reach the ideal limit, are only achieved at certain wind speeds. In addition, operational failures or maintenance can cause disturbances that are not reflected in the prediction. These simplifications and neglected factors add uncertainty to the predictions with the theoretical power curve. 
Furthermore, \citet{uncertainty_power_curve} found that due to the nonlinearity of the wind power conversion, the relative error compared to the wind speed forecast increases by a factor of between $1.8$ and $2.6$.

Collecting turbine data is another significant challenge. Identifying the wind power potential of a region requires access to specific turbine locations and characteristics, but data from manufacturers is limited. Obtaining accurate wind data as input is equally challenging. Wind speeds vary depending on time, site, and height above the ground. Observations are inhomogeneously distributed (see \Cref{subsec:wd_observations}) and must be interpolated both horizontally and vertically to the turbine locations at hub height. Moreover, historical data is inadequate for future forecasting due to the wind's high temporal variability, necessitating simulated data. Numerical weather prediction models (described in \Cref{subsec:wd_nwp}) are employed for the standard time intervals, while climate models (\Cref{sec:climate_data}) are best suited for multi-decadal predictions because they account for climate change. 
Despite their complexity, these models entail further simplifications that affect the quality of the output. One limitation is the need for discretization (see \Cref{sec:resolution}) leading to coarse resolutions, which is the problem investigated in this thesis.

\section{Weather Hindcasting}
\label{sec:weather_data}

This section explains how historical weather data records used as a benchmark for past climate model output are generated. 
Weather is the state of the atmosphere at a specific point in time, captured through meteorological parameters such as wind speed and direction, temperature, precipitation, and humidity at a particular location. These quantities can either be measured directly, as described in \Cref{subsec:wd_observations}, or predicted with atmospheric models, as outlined in \Cref{subsec:wd_nwp}.
A combination of both, measured observations and numerical predictions, is used to produce high-resolution hindcasts of past weather conditions through data assimilation techniques, presented in \Cref{subsec:wd_data_assimilation}. 

\subsection{Observational Weather Data}
\label{subsec:wd_observations}

Various measurement systems contribute to the collection of observational weather data. Measurements at the Earth's surface are collected by ground stations \citep{waetherStations} on land and by buoys and drifters \citep{PacificOceanWeatherSensorNetwork}, or, less frequently, ships \citep{shipBasesWeatherObservations} in the ocean. Data at higher altitudes are provided by drones \citep{LidarandDroneObservations, weatherDrones}, aircraft \citep{weatherAricrafts}, and periodically launched weather balloons carrying radiosondes \citep{weather_balloons}. Each of these devices is equipped with different instruments that measure the meteorological conditions at its location. In addition, wind speeds can be derived from satellite images and radar by tracking the movement of clouds and water vapor \citep{EUMETSATGlobalAVHRRWindProduct, weather_satellites}. 

As early as the beginning of the 19th century, several nations recognized the need for a standardized measurement procedure using uniform scales. In 1873, the International Meteorological Organization was founded. In 1951, it was replaced by the World Meteorological Organization (WMO), an intergovernmental organization within the United Nations system that continues to play an important role in meteorology and especially climate science (see \Cref{subsec:cmip6}). Nevertheless, the observation data is not comprehensive enough to create complete global weather records. Additional model data is required, which can be simulated using numerical weather prediction described in the next Section.

\subsection{Numerical Weather Prediction}
\label{subsec:wd_nwp}

Future weather conditions are forecasted with numerical weather prediction (NWP) models. These models are based on partial differential equations (PDEs) describing the atmosphere as a fluid, as detailed in \Cref{sec:fluid_dynamics_nwp}. 
As further discussed in \Cref{subsec:gcm_modeling}, NWP models are similar to the atmospheric component of climate models since the fundamental equations in the dynamic core are the same. Both employ numerical methods for integrating the system of PDEs, which necessitates discretization (see \Cref{sec:resolution}). 
An important difference, however, lies in the specification of the additional conditions needed to determine a unique solution. While climate models use a combination of boundary and initial conditions, NWP is solely based on initial conditions. These consist of the values of all model variables at the beginning of the forecast interval and are usually generated with data assimilation techniques - the same method used to construct weather hindcasts, discussed in the next \Cref{subsec:wd_data_assimilation}. 

\begin{figure}[htp]
    \centering
    \includegraphics[width=0.6\textwidth]{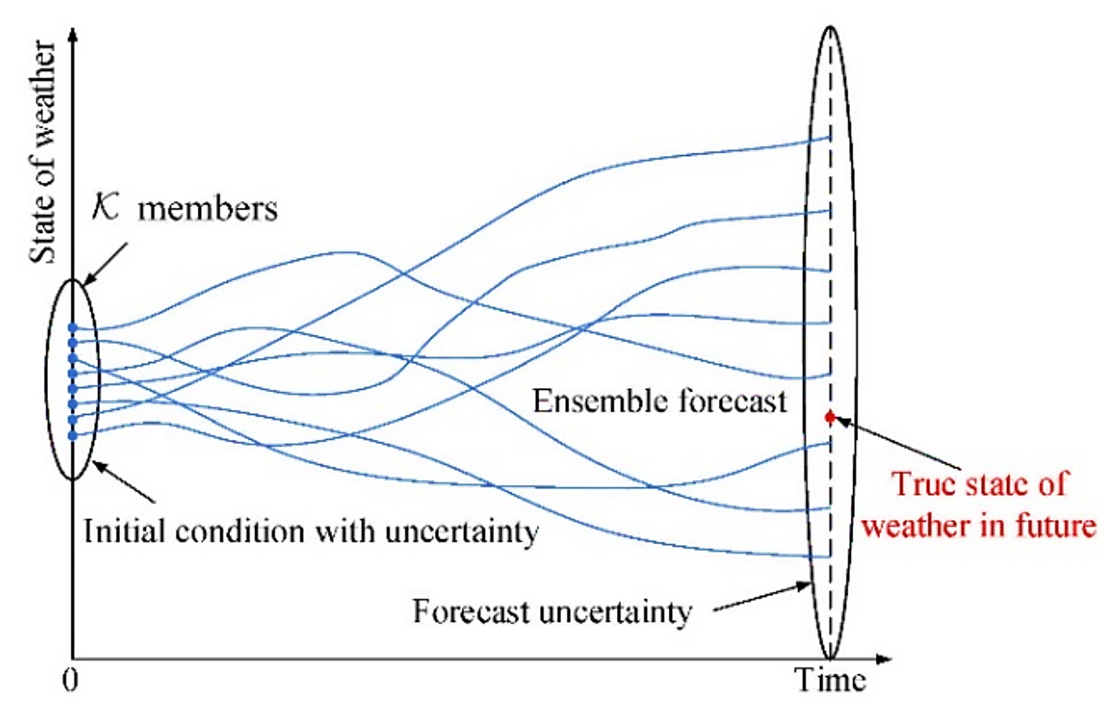}
    \caption{Multi-Member Ensemble Weather Forecasting. Ensemble members starting from different initial conditions are drawn as blue lines. The trajectories show how the predictions diverge over time, leading to an increase in uncertainty. Each ensemble forecast represents a possible future weather state; the spread indicates the forecast uncertainty. The true state of the weather in the future, marked in red, should lie within the uncertainty window. Source: \cite{image_ensemble_forcasts}.}
    \label{fig:ensemble_forecasts}
\end{figure}

The correct choice of initial conditions is essential, as even small deviations can lead to large changes in the forecast \citep{instableInitialConditions}.
To account for this sensitivity and for the fact that initial conditions and model settings are never an exact representation of reality, a technique known as multi-member ensemble forecasting is employed. This approach involves running multiple simulations with slightly different initial conditions or model parameters to assess the spectrum of potential outcomes and to estimate the associated uncertainties. This procedure is demonstrated in \Cref{fig:ensemble_forecasts}.

\subsection{Blending Observations and Predictions for Weather Hindcasts}
\label{subsec:wd_data_assimilation}

In principle, weather conditions that have already occurred can be most accurately described by observations. However, measurements are subject to noise and irregular distribution. For instance, met mast wind data are only available locally, yet observations from a single site rarely provide sufficient insight. Therefore, meteorological services employ data assimilation techniques to generate globally complete and temporally consistent maps of weather data. If the data are used as initial conditions when running a NWP model to produce future predictions, they are referred to as analysis. If data assimilation is employed to generate a comprehensive data set of past weather conditions, the resulting product is referred to as a reanalysis.

\begin{figure}[htp]
    \centering
    \includegraphics[width=0.7\textwidth]{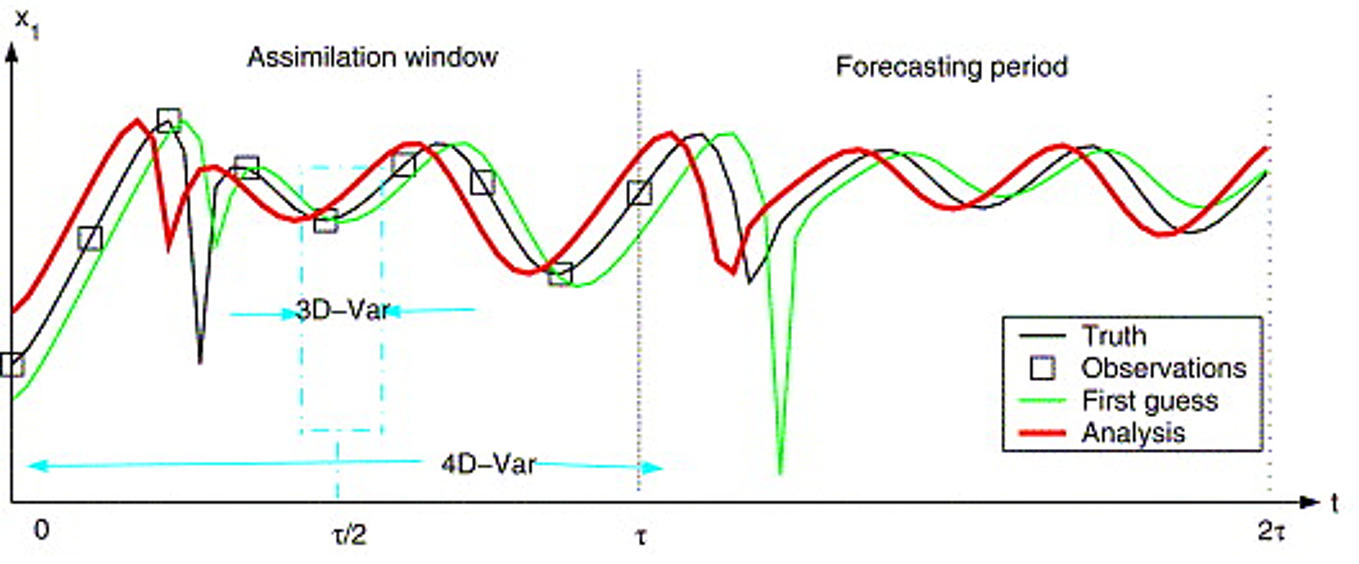}
    \caption{Data assimilation process integrating the first guess from the NWP model (green line) with recent observations (black squares) to produce an improved forecast (red line). The 3D-Var method updates predictions using individual data points, while the 4D-Var method simultaneously considers all observations within the assimilation interval. Source: \cite{dataAssimilation}.}
    \label{fig:data_assimilation}
\end{figure}

The reanalysis data set used in this thesis called ERA5 \citep{era5} was generated through variational data assimilation. This approach, which can be either three-dimensional (3D-Var) or four-dimensional (4D-Var), is briefly described in the following and sketched in \Cref{fig:data_assimilation} (for a detailed discussion, please refer to \citet{dataAssimilation}). To illustrate, consider a set of new observations in a specific time period called assimilation interval. First, previous data is input into a NWP model to create a preliminary estimate for the assimilation interval. The subsequent step is to combine this so-called background estimate with the most recent observations. This objective is represented by a cost function consisting of the logarithmic probability densities of the background and the observation. 3D-Var takes the cost function at one specific time point and minimizes it with respect to the (re)analysis vector using an iterative algorithm such as conjugate gradient or quasi-Newton methods. 4D-Var extends this approach by including all observations within the assimilation interval into the cost function and minimizing it with respect to the initial condition, which must satisfy the model equations as a constraint.

\section{Climate Projections}
\label{sec:climate_data}
After two weeks or less, the exact knowledge of the initial state of the atmosphere becomes irrelevant for future weather conditions \citep{longRangeWeatherForecasts}. This phenomenon is known as the `butterfly effect'. As a result, multi-decadal weather time series are inherently unpredictable. However, one can study the average weather conditions, i.e.\ the climate.  \Cref{subsec:gcm_modeling} explains how the modeling procedure changes in this case, while \Cref{subsec:cmip6} outlines the international coordination of climate modeling.

\subsection{Differences in the Modeling of Weather and Climate}
\label{subsec:gcm_modeling}

While using the same governing equations to describe the atmosphere as in NWP, global climate models (GCMs) also incorporate other spheres of the climate system described in \Cref{subsec:climate_system}. Most notably, they account for ocean currents and their interactions with the atmosphere, such as the surface heat exchange \citep{NWP_climate}. In addition, they include components for simulating land surfaces, sea ice, vegetation, and biogeochemical cycles.

In addition to initial conditions, GCMs employ boundary conditions \citep{NWP_climate}. These are fixed constraints at the edges of the model's domain, including land surface properties like vegetation and soil types, land-sea distribution, ice sheet and sea ice extent, ocean temperatures, and atmospheric composition. In addition, they incorporate external forcings like greenhouse gas concentrations, solar radiation, and volcanic activity \citep{forcing_IPSL} to model changes in the climate system like anthropogenic climate change (see \Cref{subsec:climate_system} for an explanation).

\subsection{Multi-Ensemble Modeling through CMIP6}
\label{subsec:cmip6}

Climate modeling has been coordinated on an international level early on.
In 1980, the WMO established the World Climate Research Program (WCRP) in collaboration with the International Council for Science (ICSU), which launched the Coupled Model Intercomparison Project (CMIP) in 1995. CMIP facilitates global climate research by collecting model data from various institutions in standardized formats and making them publicly available. This standardization allows for the creation of multi-model ensemble predictions, which aim to mitigate errors originating from model-specific biases. Additionally, CMIP supports multi-run predictions,  which involve variations in initial conditions and model parameters similar to the multi-member ensembles in NWP. Together multi-run and multi-ensemble climate projections provide probabilistic forecasts that encompass the full range of potential future climate states and offer robust uncertainty estimates.

Data collection has occurred in several phases, with CMIP6 being the most recent. It is organized as follows \citep{CMIP6}: Each participating model must be run with configurations specified by four baseline experiments called DECK (Diagnostic, Evaluation, and Characterization of Climate), including a pre-industrial control simulation. In addition, a historical simulation covering the period from 1850 to 2015 must be submitted. Models that meet these requirements can participate in different CMIP-Endorsed Model Intercomparison Projects (MIPs), which address specific scientific questions. These historical simulations are considered in this thesis.

The experiment protocols are closely linked to the Climate Assessment Reports of the Intergovernmental Panel on Climate Change (IPCC), which was established in 1988 by the WMO and the United Nations Environment Programme (UNEP) to summarize the current state of research and inform policymakers. For instance, the future scenario experiments of ScenarioMIP define configurations that are consistent with the Shared Socioeconomic Pathways \citep{SSPs} used as narratives in the 6th Assessment Report (AR6) by applying forcing levels of according Representative Concentration Pathways \citep{IPCC2023}.

\section{The Problem of Low Spatial Resolution}
\label{sec:resolution}

The preceding section emphasized the necessity of using simulated wind speed data for wind power prediction. However, this approach is constrained by the limited number of data points available at discrete times and locations. In \Cref{subsec:discretiziation} the limiting factors to the resolution are discussed, and in \Cref{subsec:downscaling}, some methods for downscaling are presented.

\subsection{Limiting Factors to the Resolution}
\label{subsec:discretiziation}

Increasing the resolution of climate models presents significant computational challenges, affecting both run time and data storage.
The computation time increases quickly since numerical operations must be performed at each step in space and time. Doubling the spatial resolution necessitates twice as many time points, and a three-dimensional grid requires eight additional data points \citep{book_parametrization}.

Moreover, the storage of the complete time series is a significant challenge for both the institutions providing the data and the researchers analyzing it. To illustrate, a file of the low-resolution climate model JAP (see \Cref{subsec:gcm}) containing the values of one variable for one year with 6-hourly time steps has a size of 145 MB. This number increases rapidly when considering multiple years, additional variables, and especially finer resolutions.

In addition, spatial and temporal resolutions are inherently linked through the transport speed. The Courant-Friedrichs-Lewy (CFL) condition governs this relationship in models using Eulerian integrators\footnote{Semi-Lagrangian schemes allow stable and accurate integrations with larger time steps \citep{SemiLagrangianIntegrationSchemes}. However, the other limits to discretization still apply.} (for an explanation of the integration schemes see \Cref{sec:fluid_dynamics_nwp}). It states that the ratio between the spatial step and the time step must be smaller than the transport speed. Otherwise, information would be lost as it travels further than one grid cell in a single time step. This condition, while necessary for integration stability, leads to the problem that small time steps must be chosen for high spatial resolutions. For example, assuming maximum wind speeds of \SI{100}{\m\per\s}, the CFL condition mandates a temporal resolution of \SI{2}{\min} to achieve a spatial resolution of \SI{12}{\kilo\m}.

These computational and storage constraints explain why global climate models often employ particularly coarse spatial resolutions.

\subsection{Increasing the Spatial Resolution via Downscaling}
\label{subsec:downscaling}

The downscaling of global climate data serves as a valuable tool in bridging the gap between large-scale climate models and the need for more detailed local predictions, which often differ from global projections due to regional factors such as topography and local climate patterns. This process aims to translate information from coarse-resolution global climate models to smaller spatial scales
\citep{downscalingMethodsInClimateChangeResearch}. Mathematically, this can be expressed as the relation 
\begin{equation}
    y(r)=f(X, g(r))+n(r)
\end{equation}
between the meterological variable of interest $y$ at a specific location $r$ and the large-scale climate conditions $X$, where $g$ is the local geography and $n$ is the local noise \citep{book_DownscalingClimateInformation}. 

\begin{figure}[htp]
    \centering
    \includegraphics[width=0.6\textwidth]{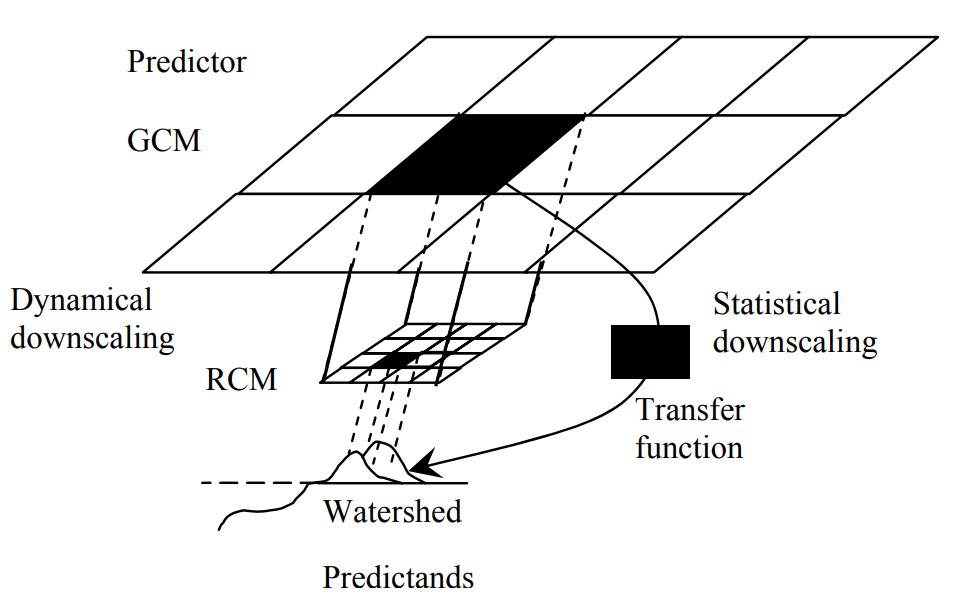}
    \caption{Schematic of downscaling methods, illustrating the process of dynamical downscaling from Global Climate Models (GCM) to Regional Climate Models (RCM), and the application of statistical downscaling using transfer functions to model the variable of interest called predictand (e.g. local watershed conditions). Source: \cite{downscaling_sketch}.}
    \label{fig:downscaling}
\end{figure}

Two different approaches to reducing the resolution are distinguished: statistical downscaling and dynamical downscaling. Both are illustrated in \Cref{fig:downscaling} and are briefly explained below.

Statistical downscaling approaches focus on modeling the transfer
function $f(X,\cdot)$, which is applied post hoc to global or regional climate model data. 
The large-scale climate conditions $X$ are interpreted as predictors for the variable of interest $y(r)$ called predictand. Perfect prognosis approaches \citep{firstPerfectProgPaper} establish an empirical relationship between them, which is assumed to be an instantaneous, physically motivated relationship derived from historical observations and/or reanalysis data. This comprises the analog method, which identifies analogous past weather patterns using the Euclidean distance, weather typing with K-means clustering, and multiple linear regression \citep{statisticalDownscalingPerfectProgComparison}. Model output statistics (MOS) approaches, in contrast, make use of statistical relationships between the local and the global scale with methods such as statistical regression \citep{downscalingMOSHubHeigt} and decision trees \citep{DTdownscalingM5}. Most recently, deep learning approaches have also been employed, including convolutional neural networks \citep{DLdownscalingCNNWindChina} or bidirectional gated recurrent unit \citep{DLdownscalingBiGRUChina}.

Dynamical downscaling approaches simultaneously model the large scales $f(X,g(r))$ and local conditions $n(r)$ by employing nested grid models \citep{book_DownscalingClimateInformation}. These Dynamical Regional Climate Models (RCMs) are very similar to global climate models (GCMs) run on a confined model domain, such as a single continent, using GCM output as boundary conditions. Similar to CMIP, the Coordinated Regional Climate Downscaling Experiment (CORDEX) \citep{CORDEX} provides a standardized framework for RCMs from different institutions. The most recent protocol established for AR6 \citep{IPCC2023} includes an evaluation experiment for the period 1980-2010 with the European Centre for Medium-Range Weather Forecasts (ECMWF) ERA-Interim reanalysis as boundary conditions. Furthermore, it encompasses a historical experiment spanning 1950-2005 and scenario experiments for future climate conditions from 2006-2100, both with boundary conditions provided by GCMs.

\chapter{Related Works and Research Question}\label{chapter:research_question}
Only a handful of studies have examined the influence of spatial resolutions on the output of climate models in terms of wind speed and the quality of wind power prediction. 

\citet{SpatialResoltionRCMs} evaluated the effect of different resolutions by using the Rossby Center Version 3 RCM at four resolutions between \SI{6}{\kilo\meter} and \SI{50}{\kilo\meter}. They extracted wind speed values in a region centered around Denmark from 1987 to 2008. Increasing the resolution led to an increase in mean wind speed and a significant increase in extremely high wind speeds. However, it did not consistently improve the agreement between simulated and observed wind speed power spectra at five locations. In particular, the variance at scales between $6$ to \SI{12}{\hour} was not necessarily in better agreement with observations for the finer resolution runs. 

\citet{AddedValueCordex} compared the normalized wind probability density function (PDF) of daily mean wind speed from historical EURO-CORDEX high-resolution simulations at \SI{0.11}{\degree} to their respective coarser-gridded \SI{0.44}{\degree} counterparts. The higher-resolution models exhibited superior performance in capturing the measured wind speed probability density function, particularly in the upper tail. Moreover, the RCMs demonstrated added value compared to their corresponding boundary GCM, with the extent of this improvement varying depending on the model, region, and season. The same was shown for convection-permitting regional climate models of the CORDEX Flagship Pilot Study (FPS)-Convection compared to their global driving reanalysis ERA-Interim and respective coarser-resolution RCM counterparts in \citet{AddedValueFPS}.

When comparing the wind power predictions with observations from a met-mast in Karlsruhe, \citet{Frisius2024} found that the model ensemble of CORDEX FPS-Convection simulations demonstrated a slight advantage over the coarser-resolution driving RCMs. However, they identified the need to extrapolate the near-surface wind speeds to the hub height as a significant limitation of the high-resolution data. Another convection-permitting model with wind speeds available at higher altitudes performed worse than the EURO-CORDEX simulations. They concluded that an intermediate RCM resolution of about \SI{12}{\kilo\meter} is sufficient for assessing wind energy climate in non-complex terrain.

\citet{ResolutionDependenceofExtremeWindSpeedProjections} compared future projections of extreme near-surface wind speeds from a standard uniform coarse-resolution simulation and a variable high-resolution simulation with the global NCAR Community ESM over the Great Lakes in the US. They found that the fine-resolution run projected stronger high wind speeds and attributed this difference to local processes that the coarse-resolution model cannot resolve. 
\citet{CMIP6vsWRF-RCM} found that CMPI6 GCMs overestimated the observed near-surface wind speed magnitude for the period between 1985 and 2014 over the Iberian Peninsula. In contrast, the RCM WRF-CESM2 underestimated it. As the RCM did not outperform the GCMs, the authors advise exercising caution when using regionalized products. 

\newpage
In summary, these previous studies on spatial resolution have primarily focused on the performance of dynamically downscaled wind speed data compared to lower-resolution model outputs, yielding mixed results. Some studies report improved accuracy with higher resolution, while others do not, with findings often specific to particular regions, resolution ranges, metrics, and models. Moreover, many of the models these studies investigated are either niche products or outdated versions, such as the CORDEX RCMs that rely on CMIP5 GCMs for global boundary conditions. Given that CMIP6 has been shown to significantly outperform CMIP5 in modeling near-surface wind speeds \citep{miao2023evaluation}, it is clear that further investigation is needed to determine the appropriate spatial resolution for accurate multi-decadal wind power predictions using the latest data.

This study seeks to fill that gap by exploring the spatial resolution of the latest generation of GCMs provided by CMIP6 for the German region.

The research is inspired by the work of \citet{effenberger2023mind}, which evaluated the impact of temporal wind speed resolution on wind power forecasting. Their findings suggest that using three-hourly or six-hourly instantaneous wind speed values closely aligns with the distribution of highly resolved data, whereas temporal averaging distorts this distribution significantly. Building on this approach, this study aims to offer analogous recommendations regarding the spatial resolution of GCMs.

However, several challenges complicate this investigation. At the temporal scale, the GCM output is available at high resolutions of up to ten minutes. This allows for a straightforward benchmark comparison by downsampling the low-resolution data from the model outcome and comparing these low-resolution samples to the original high-resolution values of the same model run. In addition, enough data points are available for statistical analyses at single locations. In contrast, at the spatial scale, GCM output is available only at coarse resolutions. This limitation presents two primary issues: obtaining comparable data at varying resolutions and determining a suitable ground truth.

One way to generate data at different spatial resolutions would be through artificial downsampling, using methods such as averaging or omitting data points, as I did in a previous research project. However, these methods do not reflect the effects of varying step sizes in the numerical integration of the complex partial differential equation system used in climate models. This is also true for more sophisticated statistical downscaling techniques. Furthermore, the field of statistical downscaling is characterized by competing approaches based on different output characteristics. Before assessing the validity of these methods, it is essential to evaluate the quality of the original data. Therefore, this thesis focuses on comparing the original GCM model outcomes without any alterations. 

Some institutions provide CMIP6 data sets of the same models run at two different spatial resolutions, which are particularly interesting for this investigation. However, to obtain more data points and answer the question of how much influence spatial resolution has compared to the choice of model, I also compare data originating from different models.

\newpage
The second challenge lies in selecting an appropriate reference data set. Climate data should ideally reflect historical weather patterns, and reanalysis data products are widely recognized as the most accurate representations of past atmospheric conditions (see \Cref{subsec:wd_data_assimilation}). However, the output of different reanalysis products varies, and the quality of the results differs depending on the region considered. 

Based on several studies assessing the quality of reanalysis data with respect to wind speed output for Germany, I selected ERA5 as a validation data set. From a comparison with the HadISD wind observation data set collected from 245 weather stations across Europe, \citet{ERA5Europe} concluded that the ERA5 reanalysis represents the data well enough "to be used, for example, to validate climate models simulating present climate conditions, with dynamical spatial consistency". However, the HadISD data set used in \citet{ERA5Europe} contains only very few locations in Germany. 

\citet{Era5DWDdata} compared ERA5 with wind measurements recorded at 128 German weather stations from January 1995 to August 2019 and found that 6-hourly near-surface wind speeds derived from ERA5 had an average Pearson correlation with the data of over $80\%$. This correlation was significantly better than that observed for its own predecessor reanalysis, ERA-Interim, and slightly worse than that of COSMO-REA6, a regional reanalysis system for continental Europe using ERA-Interim as a boundary model. 
\citet{ERA5Merra2} show that ERA5 performs better than the competing product from NASA, the global reanalysis MERRA-2 when modeling the wind power output of individual wind turbines. However, \citet{ERA5BiasCorrection} demonstrate that a bias correction should be applied to compensate for the overestimation of wind power in flat terrain and the underestimation in topographically complex regions. Yet, I decided not to employ bias correction because of the lack of a standard procedure and thus the risk of altering the results in a nongeneralizable way. Overall, these results let us conclude that ERA5 is the most adequate available choice of reanalysis data when comparing wind speed distributions in Germany. 

In conclusion, I examined historical wind speed simulations from CMIP6 GCMs and compared them with ERA5 reanalysis data, using 6-hourly instantaneous values which are sufficient on the time scale according to \citet{effenberger2023mind}. The primary objective was to address the fundamental research question: What impact does the spatial resolution have on the characteristics of wind speed distributions and wind power predictions? Furthermore, the study assesses the performance of the selected data sets for the German region.

\chapter{Methods}\label{chapter:methods}
This chapter describes the data (\Cref{sec:models}) and the steps taken to pre-process (\Cref{sec:preprocessing}) and analyze (\Cref{sec:analysis}) them. 
The data consist of wind speed samples from historical runs of ten CMIP6 GCMs (\Cref{subsec:gcm}) and the ERA5 reanalysis data set serving as a reference (\Cref{subsec:era5}).
When pre-processing and analyzing them, I proceeded in two steps: First, I gathered the models' wind speed outcomes from 2005 to 2015 in Germany extrapolated to the hub height at their original resolutions (\Cref{subsec:nsws_original_res}) and evaluated the wind speed distributions with several parametric and non-parametric metrics (\Cref{subsec:ws_comparison}). Then, I interpolated the wind speeds to the turbine locations (\Cref{subsec:interpolation}) and compared the cumulative wind power (\Cref{subsec:wp_analysis}).

\section{Data}
\label{sec:models}
This section provides an overview of the analyzed wind speed data sets and the original spatial resolutions at which they are provided. The global CMIP6 models under consideration are presented in \Cref{subsec:gcm}, while the reference data set ERA5 is described in \Cref{subsec:era5}.

\subsection{Global Climate Model Wind Speed Output}
\label{subsec:gcm}
I examined ten CMIP6 GCM data sets, which I selected based on the availability of historical model runs at a 6-hourly temporal resolution to ensure comparability.
The spatial resolutions are listed in \Cref{tab:global_models} (see \Cref{sec:numerical_grids} for an explanation of the different grid types). 
The data are provided by seven research institutions whose abbreviated names are used as model labels (the full names and locations of the institutions can be found in Appendix \Cref{tab:institutions}). Three of the models were available in two different spatial resolutions. In this case, the data sets are distinguished by the additional labels LR and HR for low and high resolution, respectively.

\begin{table}[htp]
    \centering
    \caption{CMIP6 GCM data set labels as used in this thesis (first column), names of atmospheric model components (second column), original spatial resolutions in terms of grid specification (third column), number of longitudes and latitudes (fourth column), and native resolution (fifth column) as indicated in the references given in the last column.}
    \label{tab:global_models}
    \begin{tabu} to \textwidth { l | X  X[1.2]  l  l  X[1.6]}
    \toprule
        Label & Model  & Exact  & Lon/Lat & Native & Reference \\
         & & Resolution & & Resol. &  \\
    \midrule
        CMCC & CM2-SR5 & 0.9° × 1.25° (regular) & 288 x 192 &  100 km & \citet{CMCC} \\
        EC-EARTH & EC-Earth3.3 & T255 (spectral)
        & 512 x 256 & 100 km & \citet{EC-Earth} \\
        IPSL & CM6A-LR & 2.5° × 1.3° (regular)
        & 144 x 143 & 250 km & \citet{IPSL} \\
        JAP & MIROC6 &  T85 (spectral) & 256 x 128 & 250 km & \citet{JAP}\\
        MOHC-LR & HadGEM3-GC31-LL & N96 (regular)
        & 192 x 144 & 250 km & \citet{MOHC-LR} \\
        MOHC-HR & HadGEM3-GC31-ML & N216 (regular)
        & 432 x 324 & 100 km & \citet{MOHC-MR} \\
        MPI-LR & ESM1.2-LR & T63 (spectral) & 192 x 96 & 250 km & \citet{MPI-LR} \\
        MPI-HR & ESM1.2-HR & T127 (spectral) & 384 x 192 & 100 km & \citet{MPI-HR} \\
        NCC-LR  & NorESM2-LM &  2.5° × 1.875° (regular) & 144 x 96 & 250 km & \citet{NCC-LR} \\
        NCC-HR  & NorESM2-MM & 1.25° × 0.9375° (regular) & 288 x 192 & 100km & \citet{NCC-MR} \\
    \bottomrule
    \end{tabu}
\end{table}

\subsection{ERA5 Reanalysis serving as Reference}
\label{subsec:era5}

I used ERA5 reanalysis weather data set \citep{era5} produced by the European Centre for Medium-Range Weather Forecasts (ECMWF) and distributed by Copernicus, the Earth Observation component of the European Union's space program, for the validation of the GCM output because it was identified as the most accurate for Germany in the literature (see \Cref{chapter:research_question}).

ERA5 is based on 4D-Var data assimilation (see \Cref{subsec:wd_data_assimilation}) using the Integrated Forecasting System (IFS) NWP model, version Cycle 41r2, which has been operational at ECMWF since 2016  \citep{era5_paper}. Originally, the model is discretized as T639 spectral coefficients corresponding to a reduced N320 Gaussian grid (see \Cref{sec:numerical_grids}). However, the data available for download have been re-gridded to a regular \SI{0.25}{\degree} latitude-longitude grid by bilinear interpolation. A degree of latitude has the length \SI{111.319}{\kilo\meter}, while the length of a degree of longitude is given by $111.319\cdot \cos(\text{latitude}) \si{km}$.
Thus, in the region of Germany, the ERA5 data have a resolution of \SI{27.83}{\kilo\meter} in the north-south direction and between \SI{16.06}{\kilo\meter} and \SI{18.98}{\kilo\meter} in the east-west direction.  

\section{Pre-Processing the Wind Speed Data}
\label{sec:preprocessing}
The relevant quantity for wind power predictions is the wind speed $ws$. It is defined as the absolute value of the two orthogonal wind velocity components $v$ and $u$ as follows
\begin{equation}
    ws = \sqrt{v^2+u^2} \text{\,.}
\end{equation}
In light of the findings presented in \citet{effenberger2023mind} regarding the temporal resolution, I extracted 6-hourly instantaneous values.
To compare the climate model results with the reanalysis data, I used the same six-hourly resolved output of the historical model runs over a ten-year range from 2005 to 2015.

\subsection{Wind Speed Samples at their Original Resolution}
\label{subsec:nsws_original_res}
The GCMs typically provide wind speeds only at \SI{10}{\meter} above the ground. However, for wind power predictions, wind speeds at turbine hub height are necessary. Therefore, I extrapolated the wind speeds $ws_{10}$ from their original height $h_{10}$ to a typical hub height of $h_{100}=\SI{100}{\meter}$ according to the theoretical wind profile power law
\begin{equation}
    ws_{100}=\left(\frac{h_{100}}{h_{10}}\right)^{\alpha}ws_{10}  \text{\,,} 
\end{equation}
where the coefficient $\alpha$ is empirically derived to be about $1/7\approx0.143$ for neutral stability conditions \citet{exponent_power_law}. This corresponds to a linear transformation with a factor of $10^{1/7}\approx1.4$. 

To cover the region of Germany, I chose the area encompassed by latitudes between $47.3$ and $55.1$ degrees and longitudes between $5.9$ and $15.0$ degrees. 
All models were considered at their original resolution, except for the MOHC models, which employ an Arakawa C grid \citep{ARAKAWA1977173}. This grid defines the temperature points at the center of the grid cells, the $u$ components at the sides, and the $v$ components at the top and bottom. Consequently, I re-gridded the $u$ and $v$ components to align with the temperature coordinates (as proposed by Malcolm Roberts from MOHC) using bilinear interpolation. 

\begin{table}[ht]
\centering
\caption{Number of spatial data points in the region of Germany for GCM and ERA5.}
\label{tab:wind_speed_values_germany_rectangle_2005}
\begin{tabularx}{\textwidth}{X|p{0.8cm}|X|p{0.6cm}|p{1.2cm}|p{1.2cm}|X|X|p{1.2cm}|p{1.2cm}|X} 
\toprule
NCC-LR & IPSL & MPI-LR & JAP & MOHC-LR & CMCC & NCC-HR & MPI-HR & EC-EARTH & MOHC-HR & ERA5 \\
\midrule
 16 & 24 & 25 & 30 & 30 & 64 & 64 & 80 & 143 & 154 & 1147 \\
\bottomrule
\end{tabularx}
\end{table}

\begin{figure}[htp]
    \centering
    \begin{subfigure}[b]{0.24\textwidth}
        \centering
        \includegraphics[width=4cm]{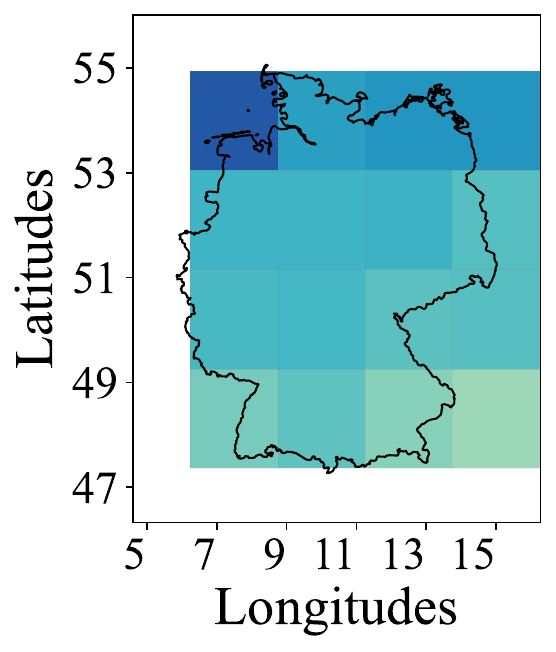}
        \caption{NCC-LR}
    \end{subfigure}
    \hfill
    \begin{subfigure}[b]{0.24\textwidth}
        \centering
        \includegraphics[width=4cm]{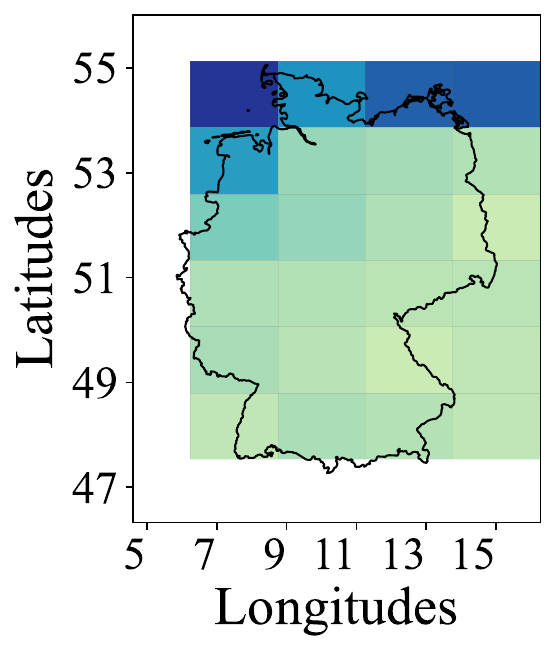}
        \caption{IPSL}
    \end{subfigure}
    \hfill
    \begin{subfigure}[b]{0.24\textwidth}
        \centering
        \includegraphics[width=4cm]{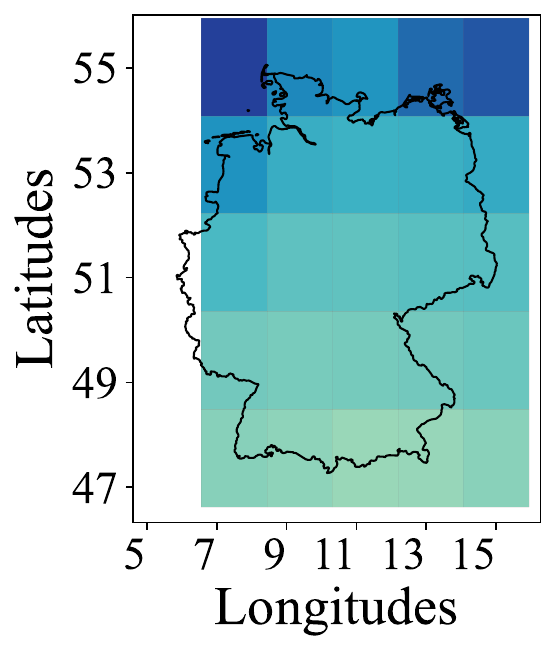}
        \caption{MPI-LR}
    \end{subfigure}
    \hfill
    \begin{subfigure}[b]{0.24\textwidth}
        \centering
        \includegraphics[width=4cm]{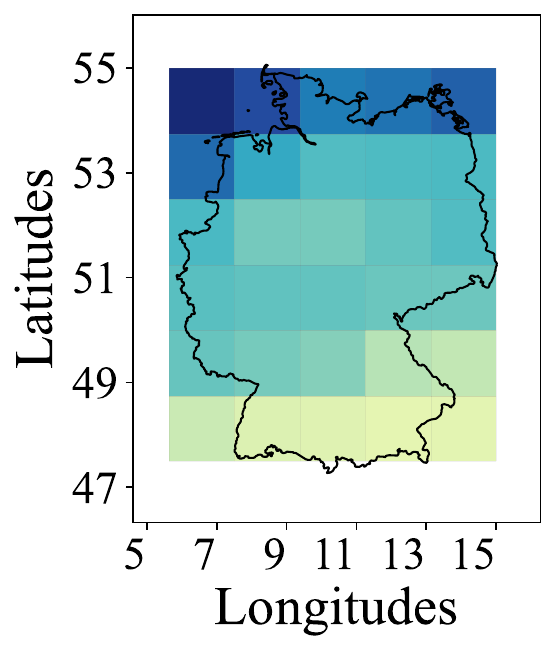}
        \caption{MOHC-LR}
    \end{subfigure}
    \par\bigskip
    \begin{subfigure}[b]{0.24\textwidth}
        \centering
        \includegraphics[width=4cm]{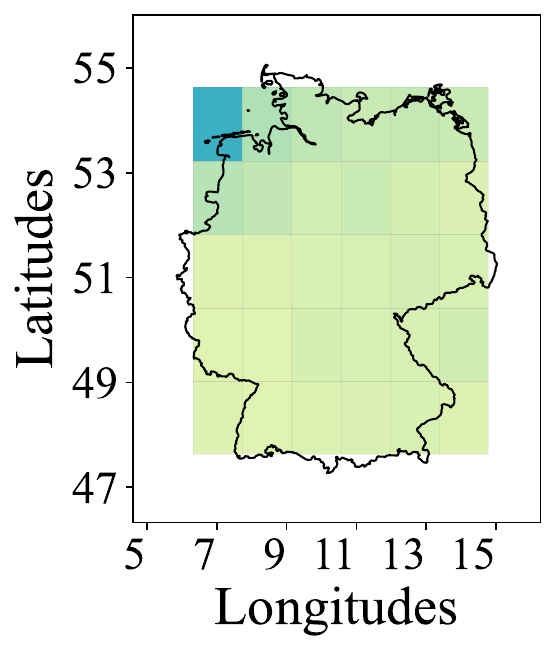}
        \caption{JAP}
    \end{subfigure}
    \hfill
    \begin{subfigure}[b]{0.24\textwidth}
        \centering
        \includegraphics[width=4cm]{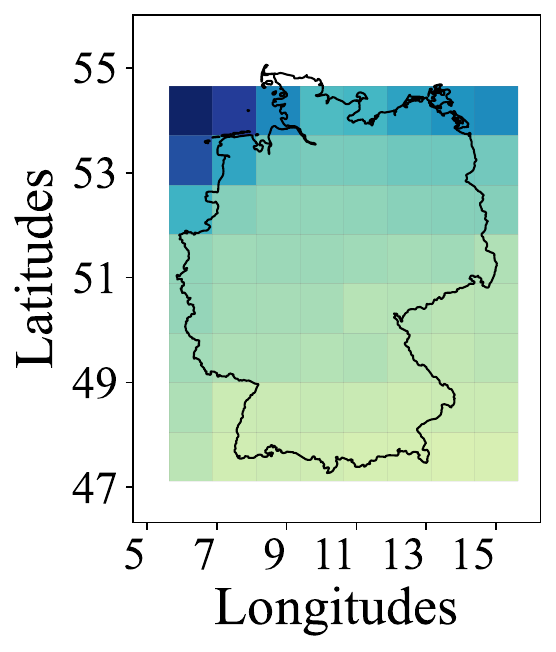}
        \caption{CMCC}
    \end{subfigure}
    \hfill
    \begin{subfigure}[b]{0.24\textwidth}
        \centering
        \includegraphics[width=4cm]{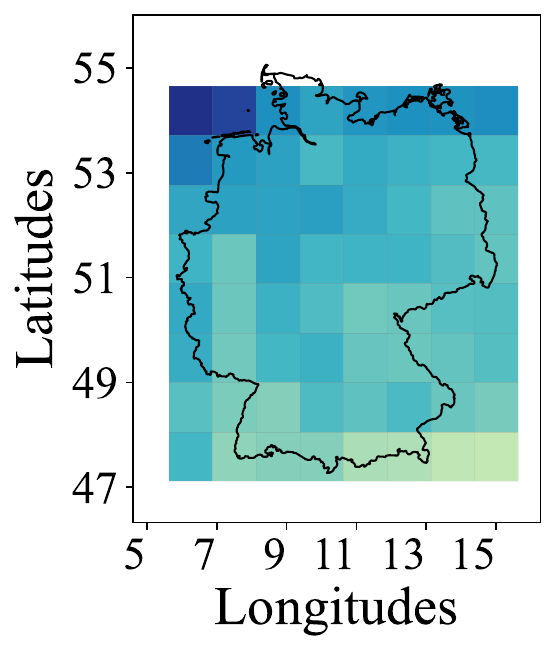}
        \caption{NCC-HR}
    \end{subfigure}
    \hfill
    \begin{subfigure}[b]{0.24\textwidth}
        \centering
        \includegraphics[width=4cm]{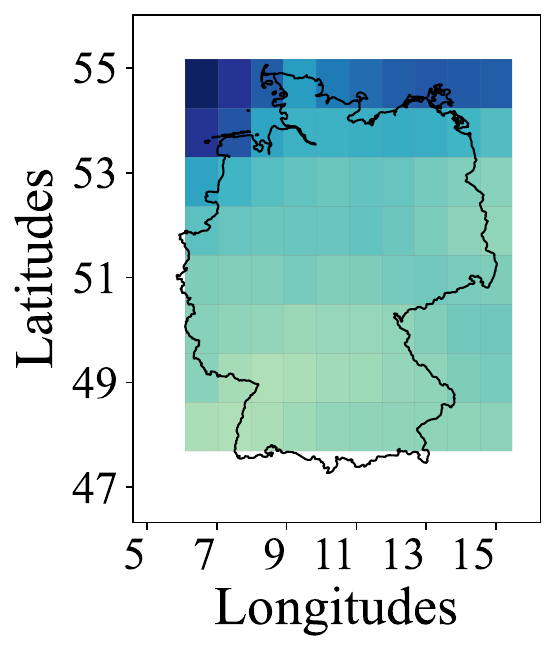}
        \caption{MPI-HR}
    \end{subfigure}
    \par\bigskip
    \begin{subfigure}[b]{0.24\textwidth}
        \centering
        \includegraphics[width=4cm]{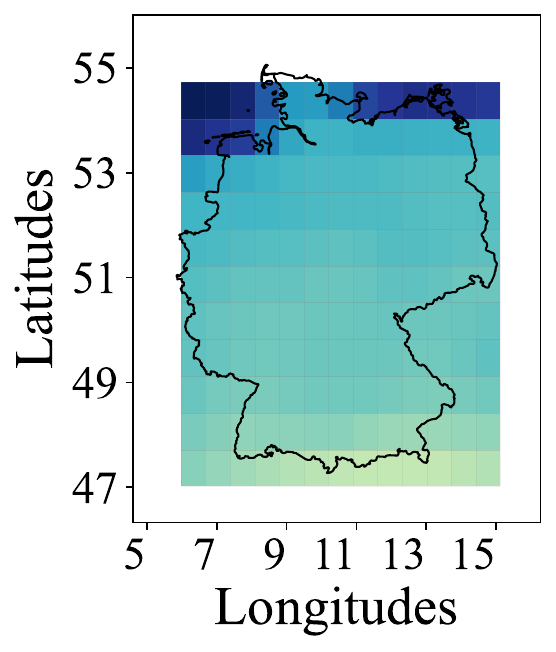}
        \caption{EC-EARTH}
    \end{subfigure}
    \begin{subfigure}[b]{0.24\textwidth}
        \centering
        \includegraphics[width=4cm]{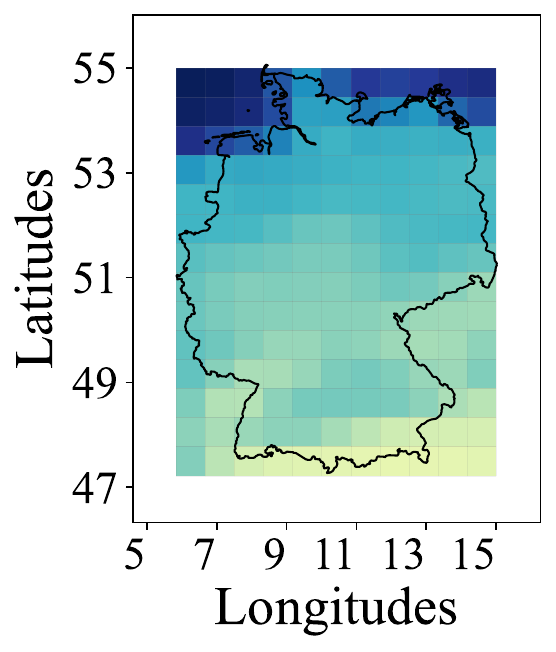}
        \caption{MOHC-HR}
    \end{subfigure}
    \begin{subfigure}[b]{0.24\textwidth}
        \centering
        \includegraphics[width=4cm]{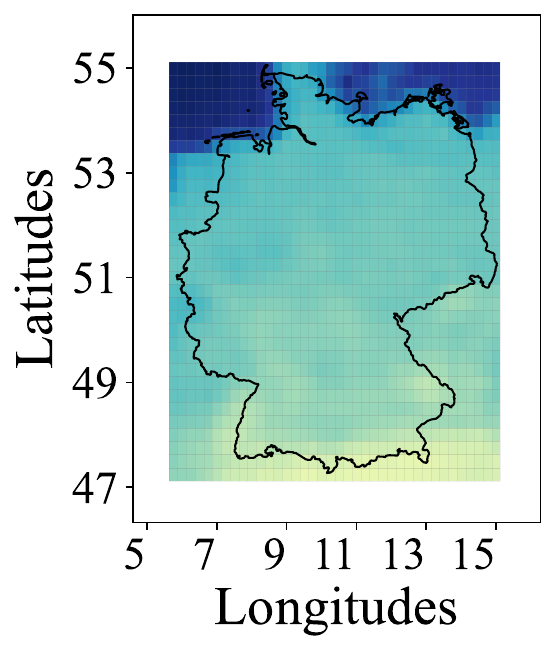}
        \caption{ERA5}
    \end{subfigure}
    \begin{subfigure}[b]{\textwidth}
        \centering
        \includegraphics[height=2.2cm]{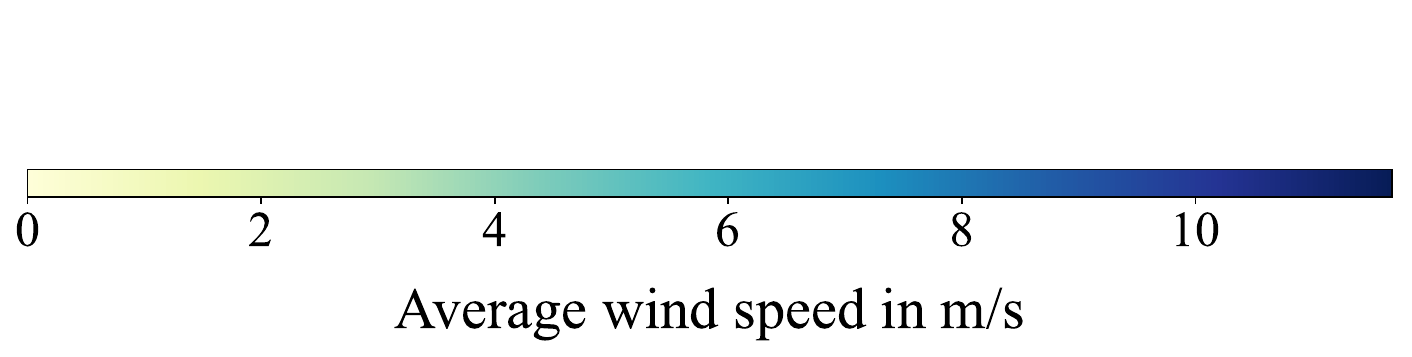}
    \end{subfigure}
    \caption{Maps of all compared models at their original resolution. The colored dots represent the annual average wind speed values in each cell.}
    \label{fig:grid_maps}
\end{figure}

The maps in \Cref{fig:grid_maps} illustrate the discrepancies in the original spatial resolution between the models. The first row displays low-resolution global climate models with cell lengths between $150$ and \SI{200}{\kilo\meter}, encompassing the area of Germany with between $16$ and $30$ data points (see the exact values in \Cref{tab:wind_speed_values_germany_rectangle_2005}). The higher-resolution global climate models in the second row have cell lengths between $60$ and \SI{100}{\kilo\meter}, which provides between $64$ and $154$ data points. The last column contains the map of ERA5 with a resolution of \SI{0.25}{\degree} in latitude and longitude, corresponding to $1147$ data points in Germany.

\subsection{Interpolation to Turbine Locations}
\label{subsec:interpolation}
\begin{figure}[htp]
    \centering
    \begin{subfigure}[b]{0.24\textwidth}
        \centering
        \includegraphics[width=4cm]{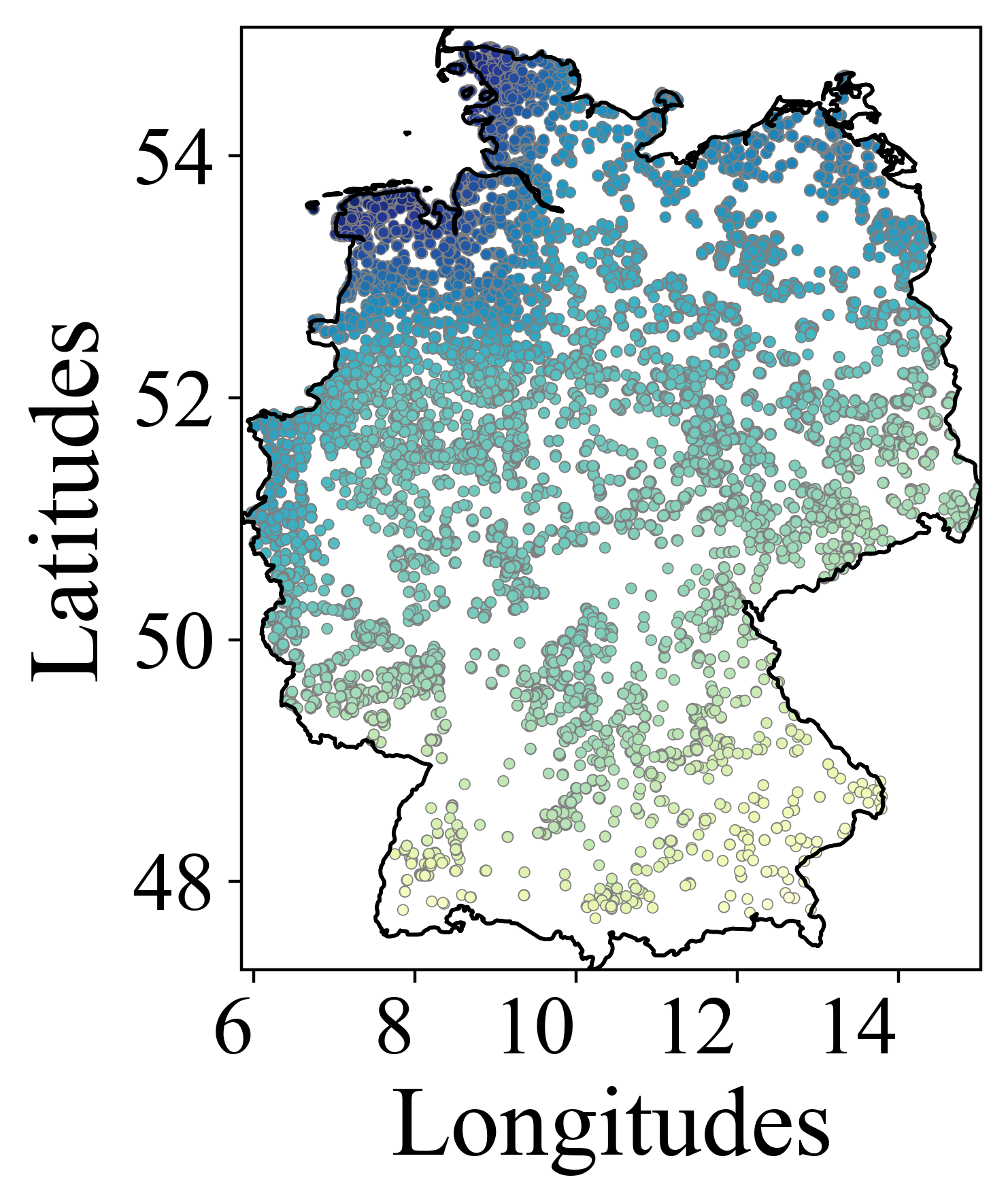}
        \caption{NCC-LR}
    \end{subfigure}
    \hfill
    \begin{subfigure}[b]{0.24\textwidth}
        \centering
        \includegraphics[width=4cm]{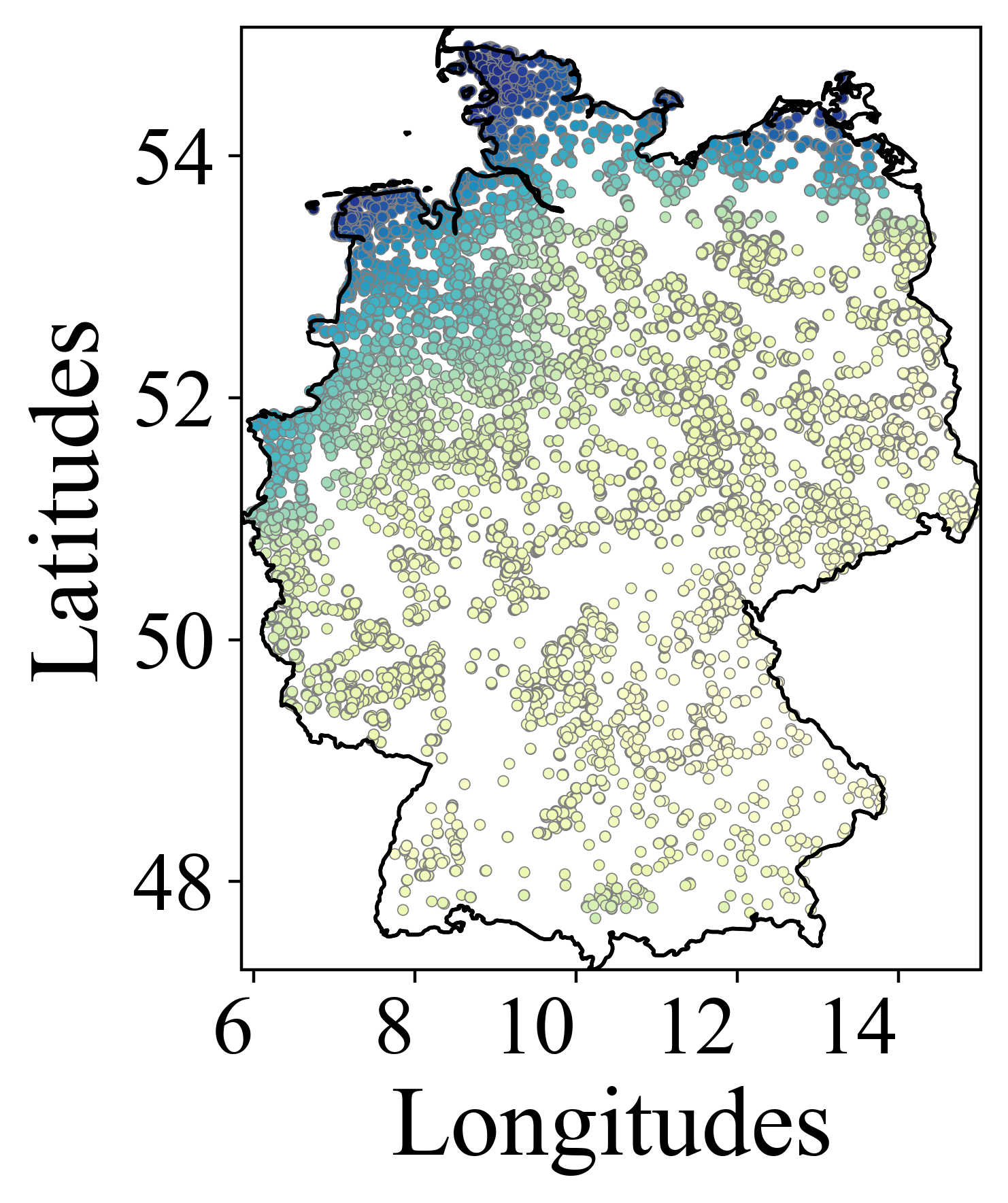}
        \caption{IPSL}
    \end{subfigure}
    \hfill
    \begin{subfigure}[b]{0.24\textwidth}
        \centering
        \includegraphics[width=4cm]{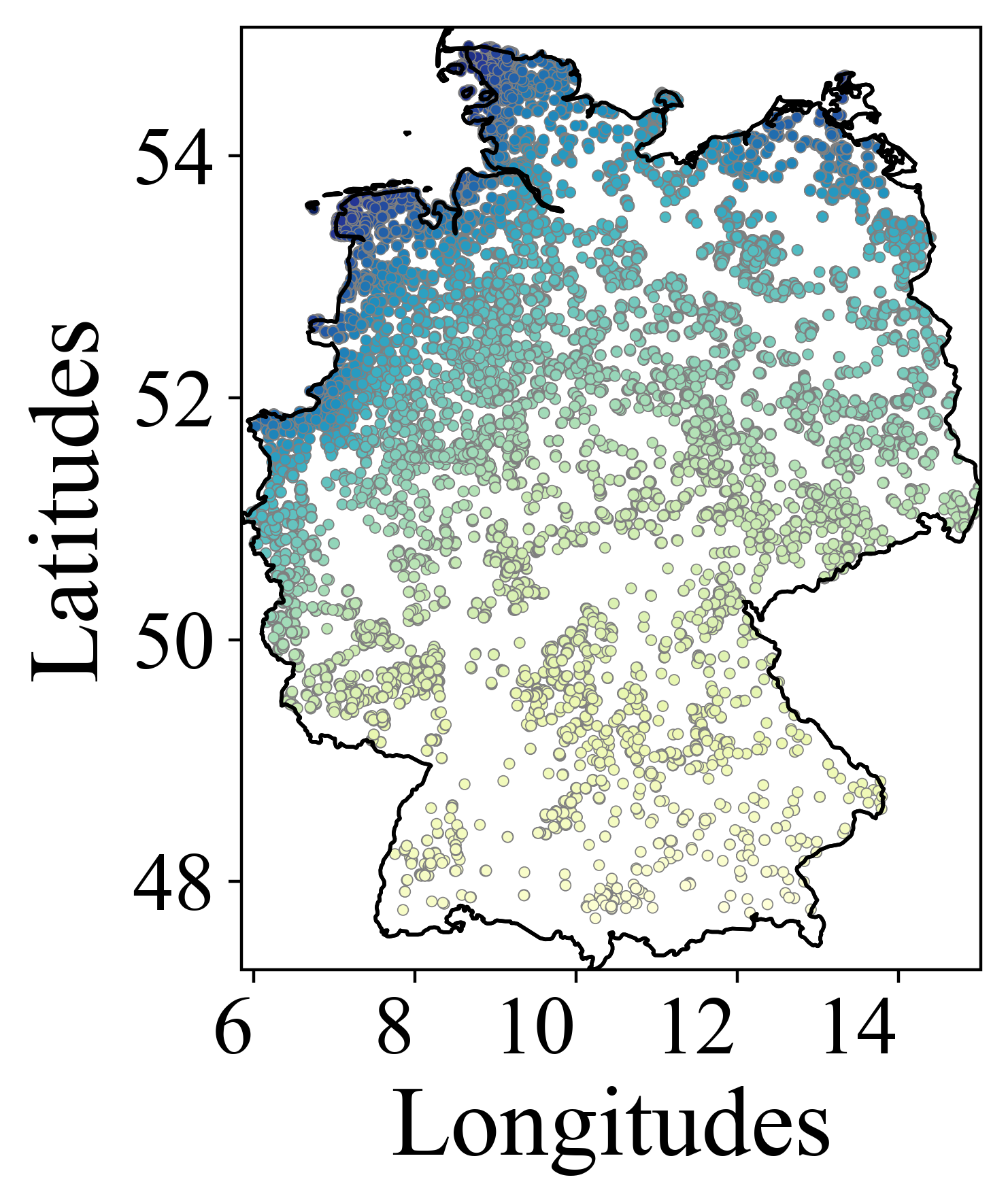}
        \caption{MPI-LR}
    \end{subfigure}
    \hfill
    \begin{subfigure}[b]{0.24\textwidth}
        \centering
        \includegraphics[width=4cm]{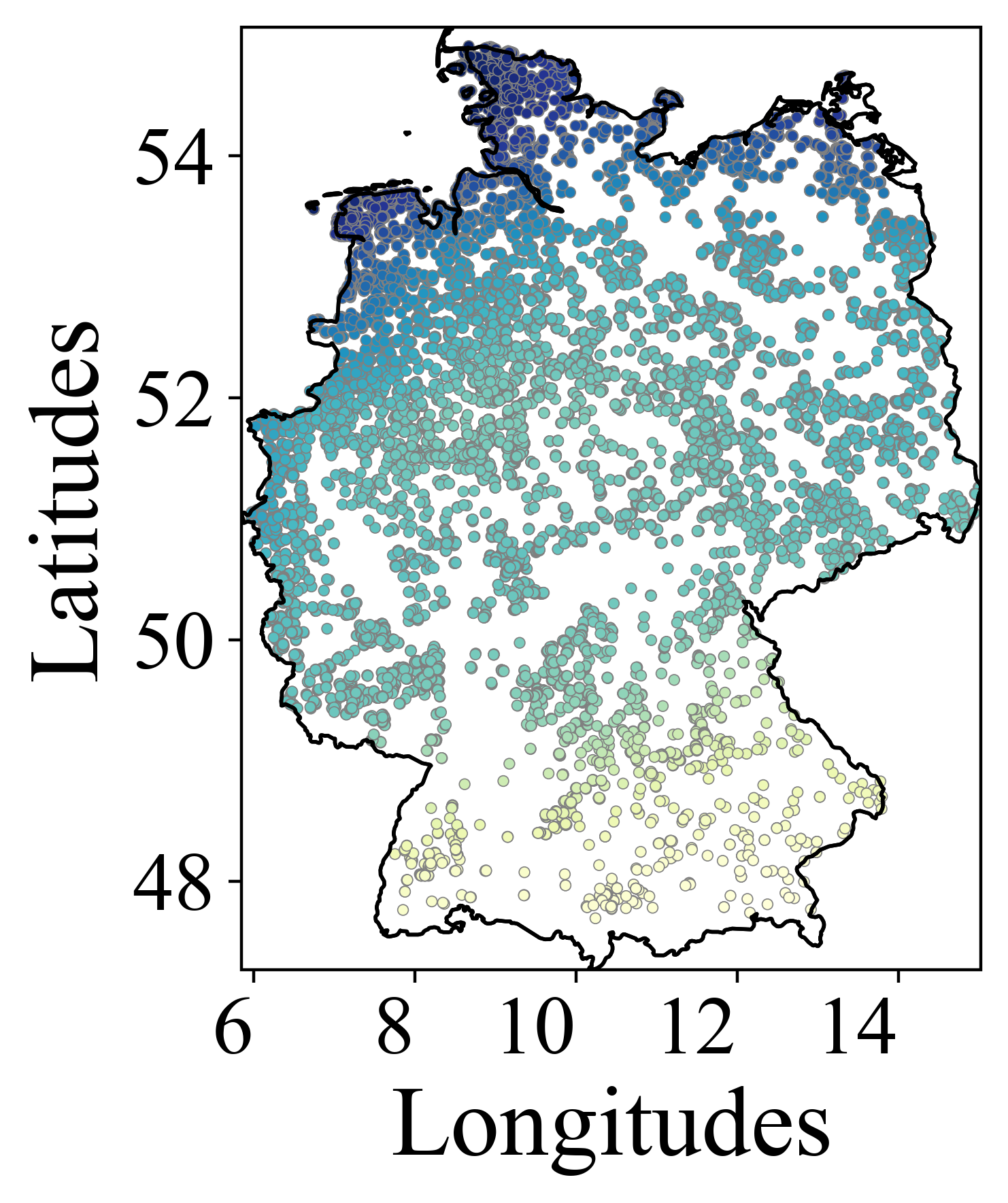}
        \caption{MOHC-LR}
    \end{subfigure}
    \par\bigskip
    \begin{subfigure}[b]{0.24\textwidth}
        \centering
        \includegraphics[width=4cm]{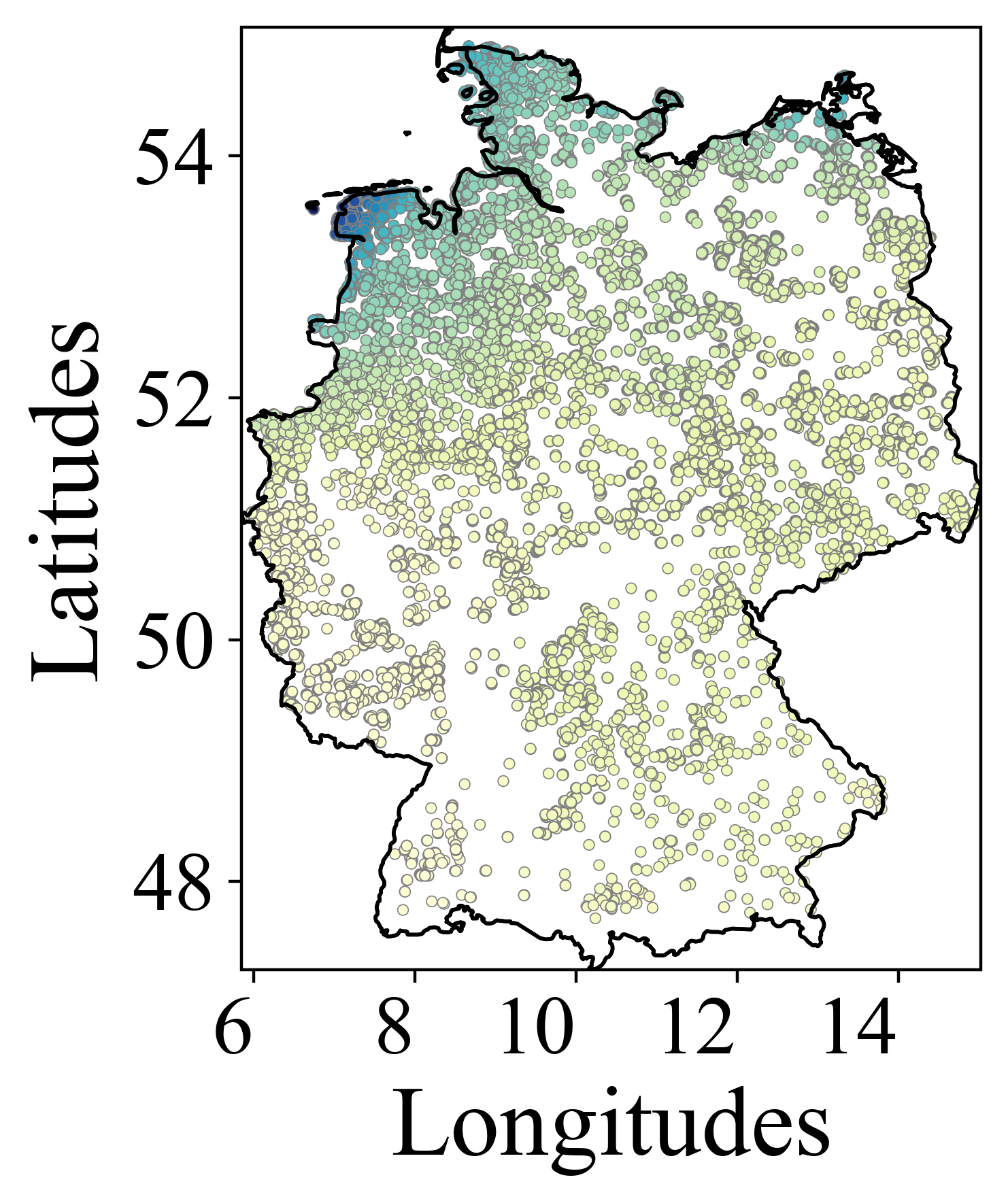}
        \caption{JAP}
    \end{subfigure}
    \hfill
    \begin{subfigure}[b]{0.24\textwidth}
        \centering
        \includegraphics[width=4cm]{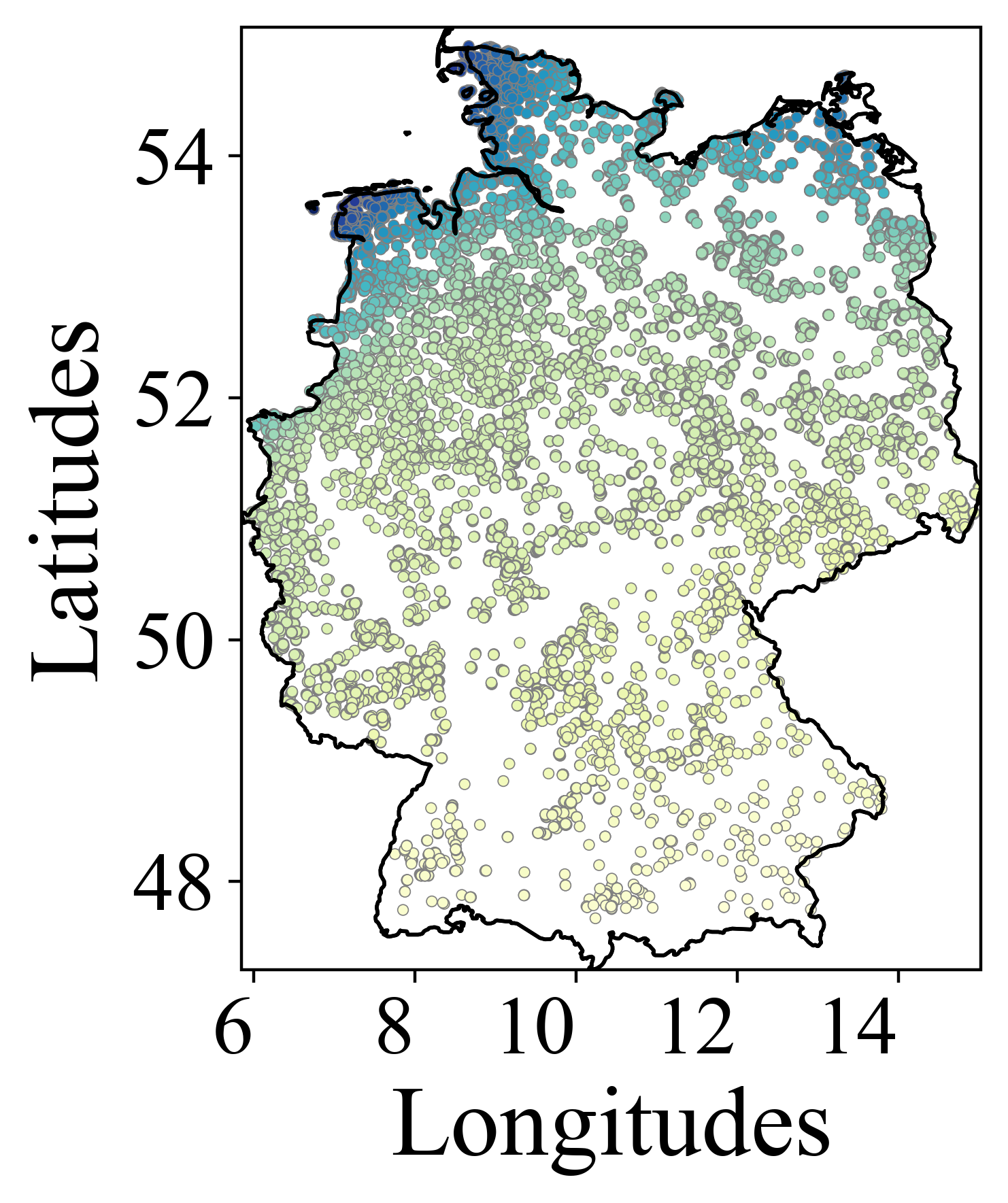}
        \caption{CMCC}
    \end{subfigure}
    \hfill
    \begin{subfigure}[b]{0.24\textwidth}
        \centering
        \includegraphics[width=4cm]{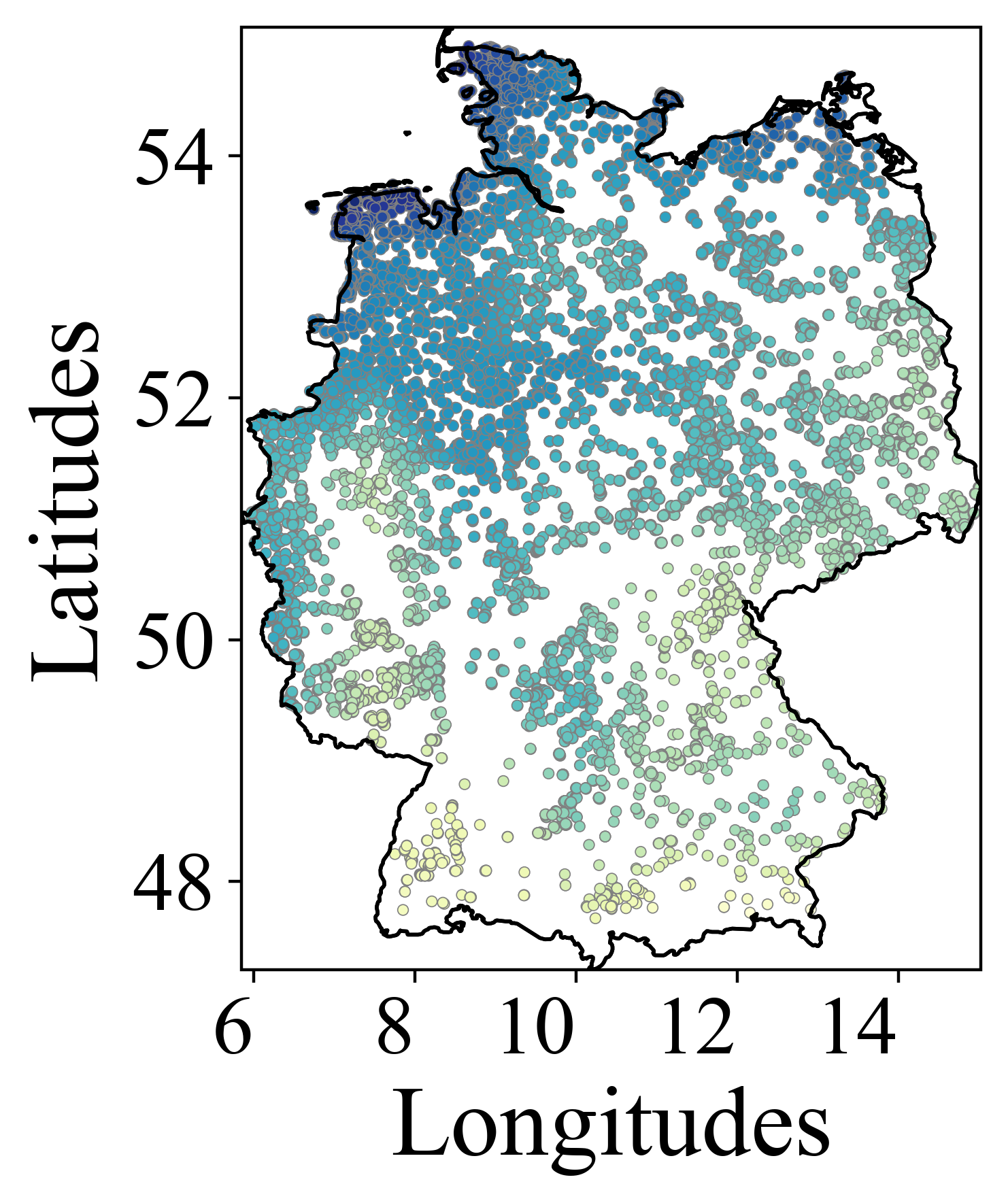}
        \caption{NCC-HR}
    \end{subfigure}
    \hfill
    \begin{subfigure}[b]{0.24\textwidth}
        \centering
        \includegraphics[width=4cm]{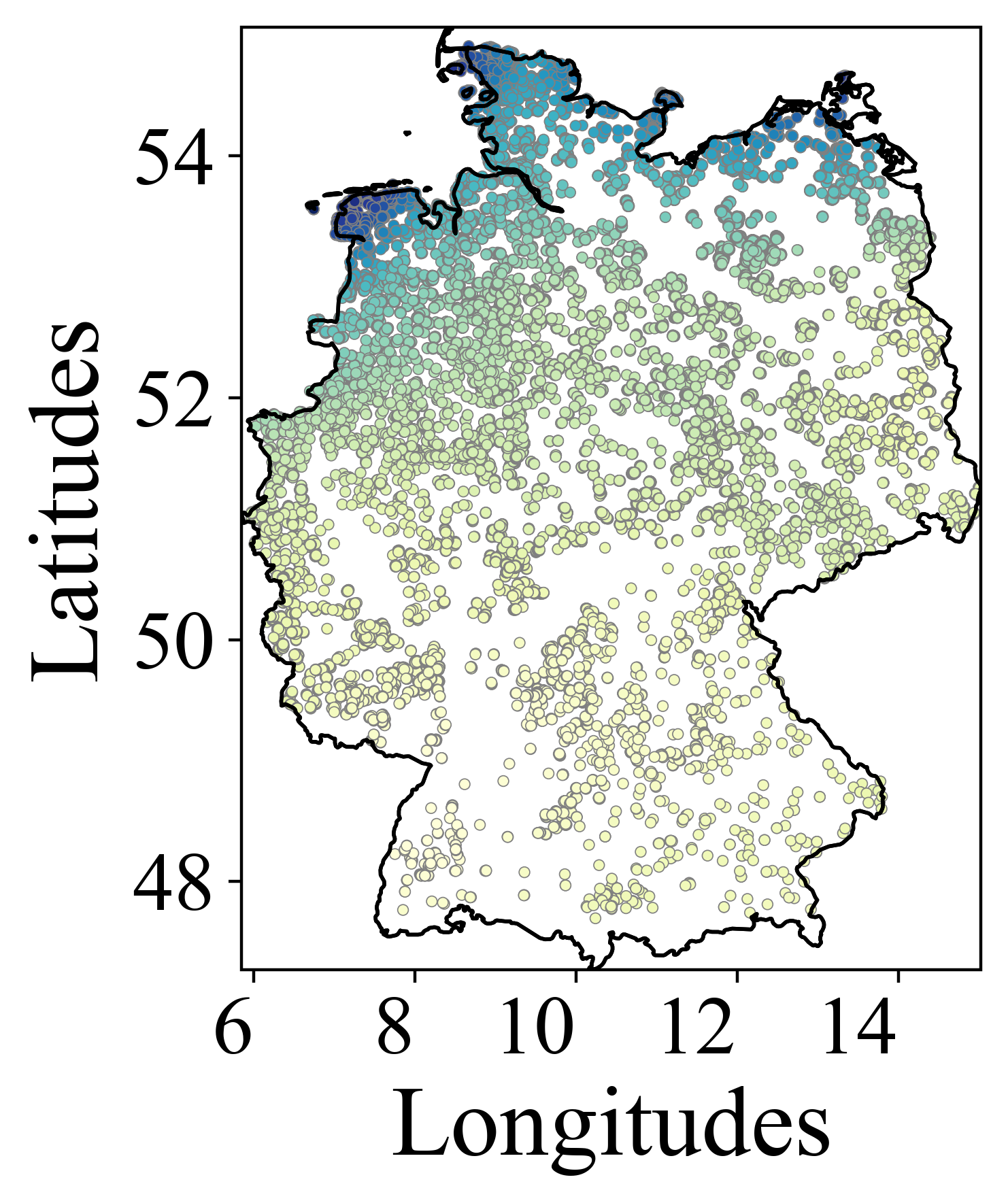}
        \caption{MPI-HR}
    \end{subfigure}
    \par\bigskip
    \begin{subfigure}[b]{0.24\textwidth}
        \centering
        \includegraphics[width=4cm]{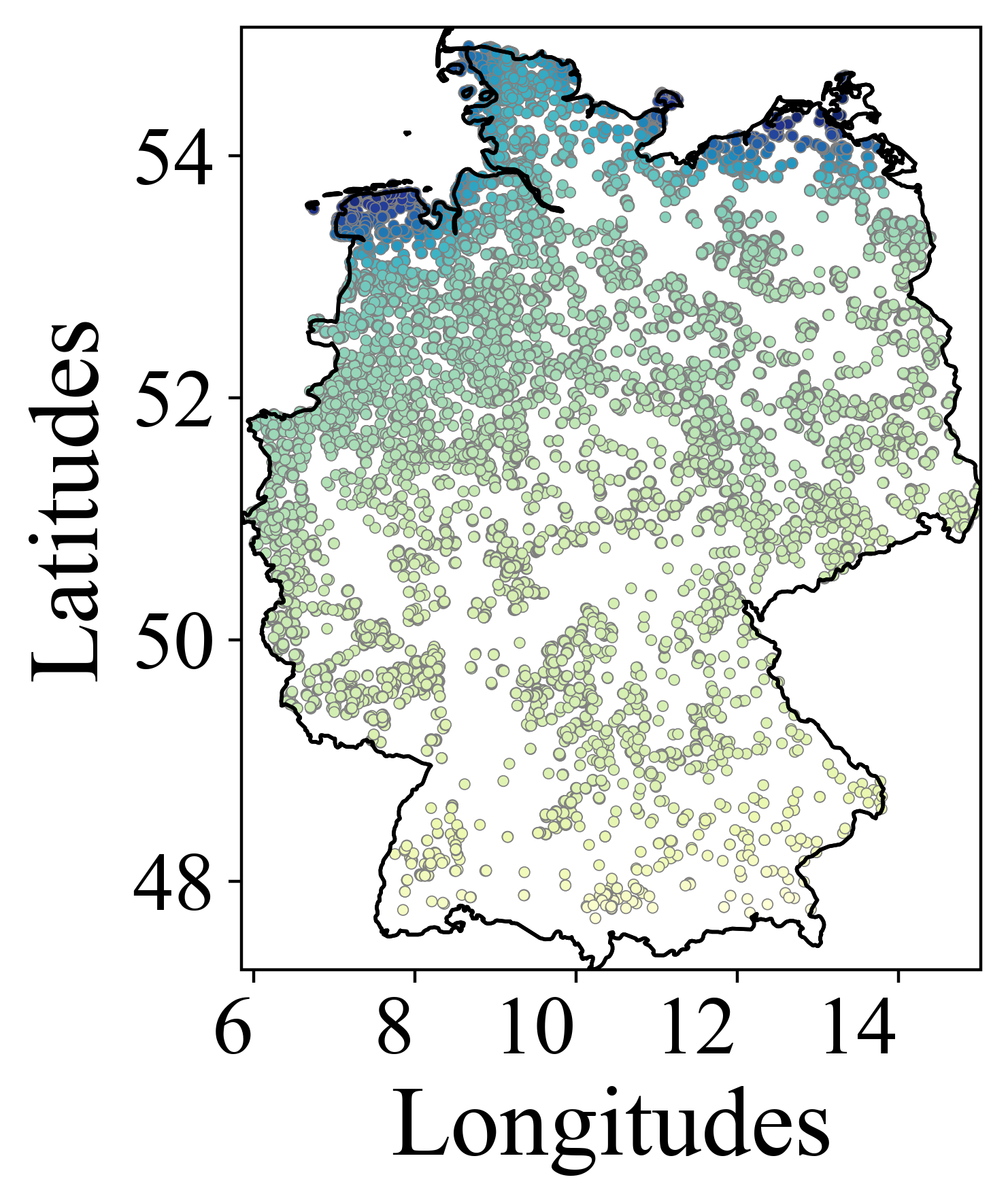}
        \caption{EC-EARTH}
    \end{subfigure}
    \begin{subfigure}[b]{0.24\textwidth}
        \centering
        \includegraphics[width=4cm]{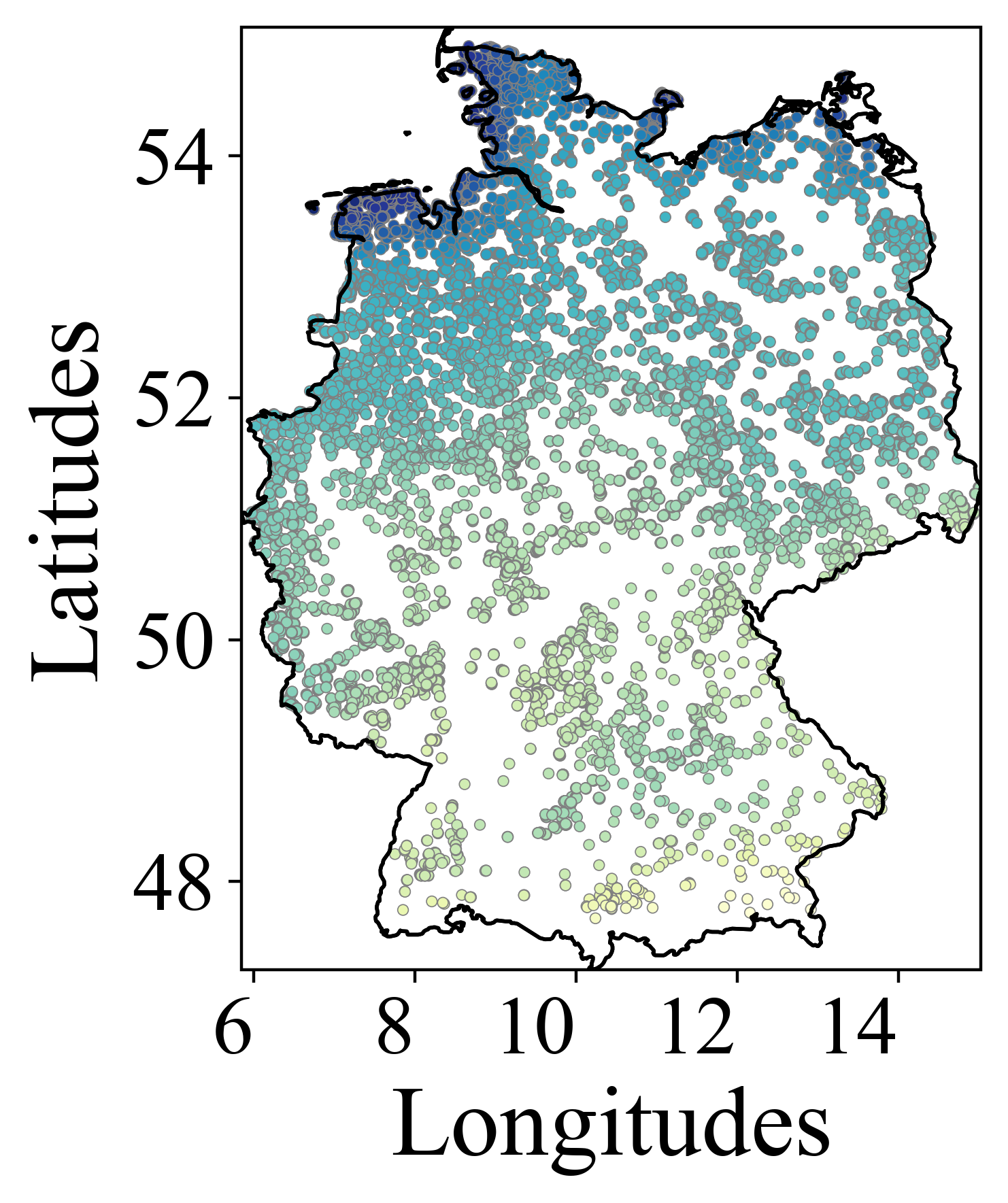}
        \caption{MOHC-HR}
    \end{subfigure}
    \begin{subfigure}[b]{0.24\textwidth}
        \centering
        \includegraphics[width=4cm]{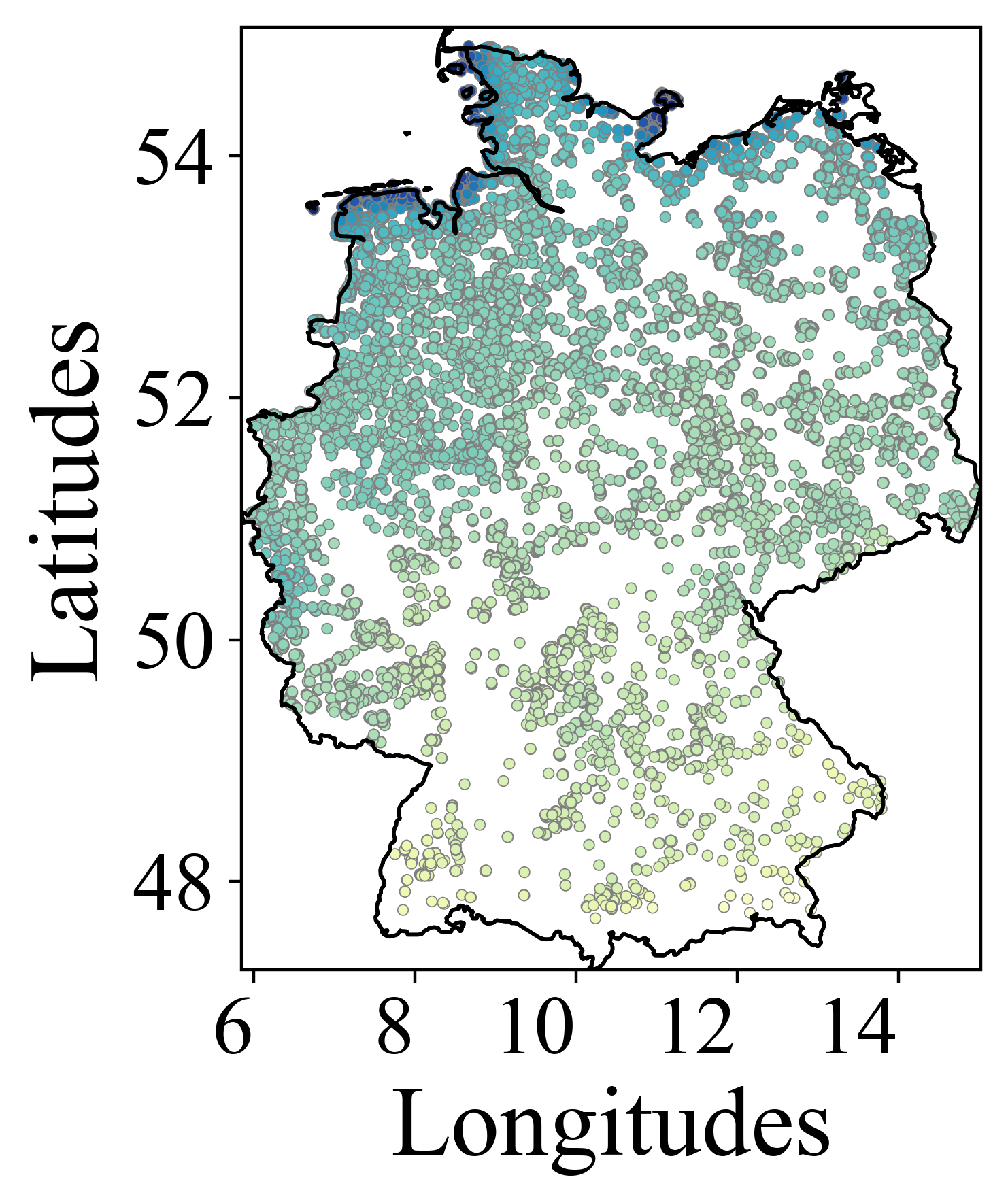}
        \caption{ERA5}
    \end{subfigure}
    \begin{subfigure}[b]{\textwidth}
        \centering
        \includegraphics[height=2.2cm]{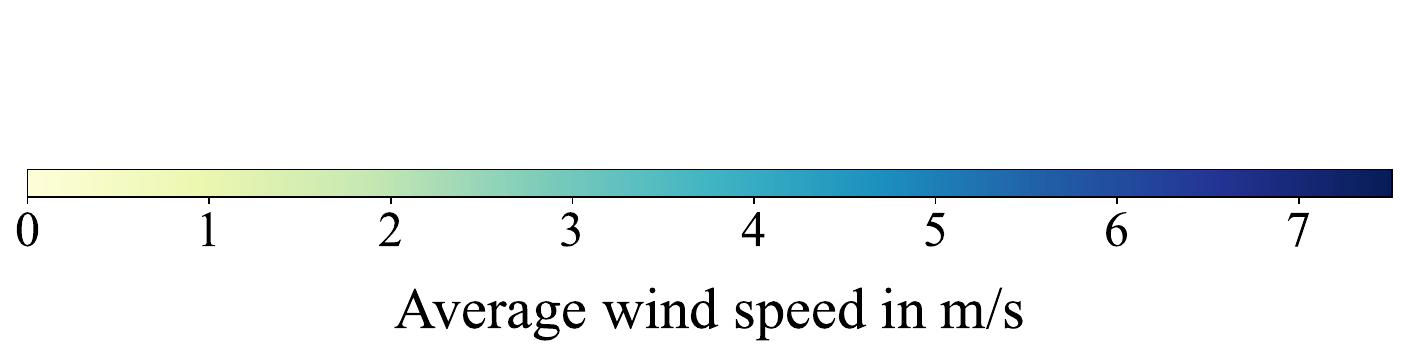}
    \end{subfigure}
    \caption{Maps of all compared models when the wind speed data are interpolated to the turbine locations at \SI{100}{\meter} hub height.}
    \label{fig:grid_maps_turbines}
\end{figure}

For the wind power prediction, I interpolated the original NSWS data to the turbine locations in Germany installed before 2016, as provided by \citet{turbineLocations}, using a piecewise linear interpolator. \Cref{fig:grid_maps_turbines} shows the resulting time averages of the interpolated wind speeds at the turbine locations.

\section{Analysis}
\label{sec:analysis}
The analysis is divided into two parts: First, the characteristics of the NSWS distributions were investigated with various parametric and nonparametric measures presented in  \Cref{subsec:ws_comparison}. Then, the wind power was predicted and compared as discussed in \Cref{subsec:wp_analysis}.

\subsection{Wind Speed Distribution Comparison}
\label{subsec:ws_comparison}
Conventional time series scoring rules, such as those proposed by \citet{gneiting2007strictly}, are inadequate for analyzing climate model output. This is because climate models aim to model distributions over time rather than producing precise point forecasts. Therefore, I compare the distributions of the wind speed samples in  several other ways presented below. They are intended to serve three main purposes: to show whether the distributions change systematically with spatial resolution, to evaluate the agreement with the reference data, and to highlight differences in wind speed ranges that are particularly important for wind power prediction.

\subsubsection{Kolmogorov-Smirnov Tests}
The similarity of model outcomes was tested with Kolmogorov-Smirnov tests \citep{kolmogorov}. The null hypothesis states that the samples from both models are drawn from the same distribution. The test statistic $D_n$ is calculated as the supremum of the absolute differences between the cumulative distributions $\widehat{F}^1$ and $\widehat{F}^0$
\begin{equation}
    D_n = \sup_{ws} |\widehat{F}^1(ws)-\widehat{F}^0(ws)| \text{\,,}
\end{equation}
where the empirical cumulative distribution functions of $n$ wind speed samples $ws_i$ is estimated by the cumulative sum
\begin{equation}
\label{eq:cumulative}
    \widehat{F}_n(ws) = \frac{1}{n} \sum_{i=1}^n \mathbf{1}_{(-\infty,ws]}(ws_i)  \text{\,.}
\end{equation}

\subsubsection{Kernel Density Estimates}
For visualizing the wind speed distributions, the probability density function was approximated by the kernel density estimate (KDE)
\begin{equation}
    \text{KDE}(ws) = \frac{1}{nh} \sum_{i=1}^n K\Big(\frac{ws-ws_i}{h}\Big)  \text{\,,}
\end{equation}
using Scott's rule \citep{scottRule} for estimating the bandwidth $h$ and the Gaussian kernel
\begin{equation}
K(x) = \frac{1}{\sqrt{2\pi}}\exp \left(-\frac{1}{2}x^2\right)  \text{\,.}
\end{equation}

\subsubsection{Summary Statistics}
The considered summary statistics are the sample mean
\begin{equation}
    \bar{ws} = \frac{1}{n}\sum_{i=0}^n ws_i  \text{\,,}
\end{equation}
the biased sample variance
\begin{equation}
    \text{var}  = \frac{1}{n}\sum_{i=0}^n(ws_i-\bar{ws})^2  \text{\,,}
\end{equation}
the third standardized empirical moment
\begin{equation}
    \text{skewness} =\frac{1}{n} \sum_{i=1}^n\left(\frac{ws_i-\bar{ws}}{\sqrt{\hat{\text{var}}}}\right)^3  \text{\,,}
\end{equation}
the fourth standardized empirical moment
\begin{equation}
    \text{kurtosis} = \frac{1}{n} \sum_{i=1}^n\left(\frac{ws_i-\bar{ws}}{\sqrt{\hat{\text{var}}}}\right)^4  \text{\,,}
\end{equation}
and the maximum value of the entire distribution
\begin{equation}
    \text{maxval} = \max_{i<n}\left(ws_i\right)  \text{\,.}
\end{equation}

\subsubsection{Exponentiated Weibull Distribution}
An analytical distribution function to fit the wind speed data was determined according to the following argument: The asymmetric distribution of temporal measurements at a given location is known to be well captured by gamma or Weibull distributions \citep{weibull2017, weibull1984}. When merging data from different locations, several such temporal distributions with individual parameterizations overlap. To account for this overlap, the exponentiated Weibull distribution was found to be appropriate \citep{expontiatedWeibull}. Its probability density function
$$
f(ws;c,a) = a c \left(1- e^{-ws^c} \right)^{a-1} ws^{c-1} e^{-ws^c}  \text{\,.}
$$ 
is determined by the `Weibull parameter' $c$ and the `exponential parameter' $a$, where $a=1$ gives the standard Weibull distribution. 
The complete fitting function additionally allowed the scaling and shifting of the input $s$ by applying the substitution
$$
    \Tilde{ws} \leftarrow \frac{ws-loc}{scale}
$$
with the scaling parameter $scale$ and the location parameter $loc$. For the fitting procedure, I applied maximum likelihood expecting the iid assumption to hold since the data points were taken at six-hour intervals. The likelihood of all $n$ samples and parameters $c,a,loc,scale \in \theta$ is given by 
$$
    L(\theta) = \prod_{i=0}^n f(\Tilde{ws}_i;c,a)  \text{\,.}
$$
The maximum likelihood estimate
$
    \hat\theta_{\text{ML}} = \underset{\theta \in \Theta}{\arg\max} L(\theta)
$
was numerically calculated with the Trust Region Reflective algorithm \citep{trustRegionAlgorithm} suitable for finding a local minimum of non-convex, once continuously differentiable functions. Notably, the maximum likelihood fitting procedure yields different parameters in each run, as multiple combinations lead to local minima. 

\subsubsection{Linear Regression}
To identify systematic trends with spatial resolution, I used linear least-squares regression. Since the spatial resolution was defined as the number of data points on a logarithmic scale, the independent variable required transformation resulting in the model
\begin{equation}
    Y_i = \alpha + \beta \log X_i + \varepsilon_i  \text{\,,}
\end{equation}
where $Y_i$ is the dependent variable, $X_i$ is the independent variable (the number of spatial data points), $\alpha$ is the intercept, $\beta$ is the slope and $\varepsilon$ is the error. The regression parameters $\alpha$ and $\beta$ were estimated through ordinary least squares, assuming the residuals $\varepsilon_i$ to be homoscedastic and normally distributed. The standard error of the slope was used to indicate the uncertainty of the estimation. The coefficient of determination $R^2$ served as a measure of the goodness of fit. It measures the variance explained by the model. Furthermore, the Overall-$F$-test was considered, which tests for the null hypothesis that the slope of the regression line is zero.

\subsubsection{Jensen-Shannon Distance}
Since there is no unique metric to quantify the differences between distributions, I considered two common metrics to estimate the statistical distance between the GCM wind speed distribution $P$ and reference distribution $Q$. One of them is the Jensen-Shannon (JS) distance which can be seen as a symmetric version of the Kullback–Leibler (KL) divergence \citep{KL} defined as
\begin{equation}
    D_{\mathrm{KL}}(P \| Q)=\sum_{x \in \mathcal{X}} P(x) \log \left(\frac{P(x)}{Q(x)}\right)  \text{\,.}
\end{equation}
Symmetrization is necessary to account for the entire wind speed range, as $D_{\mathrm{KL}}$ becomes infinite if either $P(x)$ or $Q(x)$ is zero. 
The JS distance is given by
\begin{equation}
    D_{\mathrm{JS}}(P \| Q) = \sqrt{\frac{1}{2} D_{\mathrm{KL}}(P \| M)+\frac{1}{2} D_{\mathrm{KL}}(Q \| M)}  \text{\,,}
\end{equation}
where $M$ is the mixture distribution
\begin{equation}
    M =\frac{1}{2}(P+Q)  \text{\,.}
\end{equation}
Since both distributions must be defined on the same set, it is necessary to bin and normalize the wind speed samples before calculating the distance. This is achieved by constructing histograms. I used $35$ bins in the wind speed range from $0$ to \SI{30}{\m\per\s}.

\subsubsection{Wasserstein Distance}
The second metric I used to measure the distance between distributions is the Wasserstein-1 distance \citep{Wasserstein1969}, also known as the Earth mover's distance or the optimal transport distance. This metric quantifies the cost required to transform one distribution into another by optimally moving probability mass. For one-dimensional data, it can be calculated by solving the integral
\begin{equation}
    D_{\mathrm{WS_1}}=\int_{-\infty}^{\infty} |\hat{F}_n^1(ws)-\hat{F}_n^0(ws)|
\end{equation}
over the absolute value of the difference between the empirical cumulative distribution functions $\hat{F}_n^i(ws)$ as defined in \Cref{eq:cumulative}.
\subsubsection{Quantile-Quantile Plots}
To visualize the differences between wind speed distributions, I calculated the quantiles $Q$ of the cumulative distribution function $F$, defined as
\begin{equation}
    Q(p)=F^{-1}(p)=\inf \{s: F(ws) \geq p\}, \quad 0<p<1  \text{\,,}
\end{equation}
at integer percentage points ($p \in \{0.01,0.02,0.03,\cdots,0.98,0.99\}$) and compared them in Q-Q plots. This was done by sorting the $n$ samples in ascending order and taking the value at index $p(n-1)+1$ \citep{quantiles}. 

\subsubsection{KDE Difference in Critical Wind Speed Range}
For the selected turbine type, the cut-in wind speed $ws_I$ is \SI{1}{\meter\per\second} and the cut-off wind speed is about $ws_O$ is \SI{25}{\meter\per\second} (see \Cref{fig:power_curve}). Therefore, I considered values from \SI{1}{\meter\per\second} to \SI{25}{\meter\per\second} as critical wind speed range.
Deviations in this range were displayed as kernel density estimate differences 
\begin{equation}
    \text{KDE}_{\text{GCM}}-\text{KDE}_{\text{ERA5}} \text{\,.}
\end{equation}

\subsubsection{Visualization of the Distribution Tail}
The values exceeding the cut-off wind speed $ws_O$ of \SI{25}{\meter\per\second} were considered to represent extreme wind speeds in the upper tail of the distribution. This upper tail region was visualized by the log ratio between the kernel density estimates 
\begin{equation}
    \text{KDE}_{\text{GCM}}/\text{KDE}_{\text{ERA5}}  \text{\,,}
\end{equation}
as well as by the log of the survival function 
\begin{equation}
    1 - \widehat{F}_n(ws)  \text{\,,}
\end{equation}
where $\widehat{F}_n(ws)$ is defined in \Cref{eq:cumulative}.

\subsection{Wind Power Prediction}
\label{subsec:wp_analysis}

\begin{figure}[htp]
    \centering
    \includegraphics[width=0.9\textwidth]{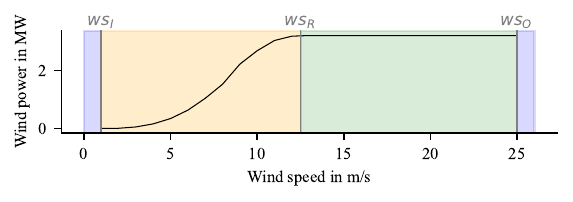}
    \caption{Wind power curve of the Enercon E-115 turbine at a hub height of \SI{100}{\meter}: Zero power at low wind speeds (blue) until the cut-in wind speed $ws_I$, increasing power (orange) until saturation at rated wind speed $ws_R$ (green), and shut down after cut-off wind speed $ws_O$ (blue).}
    \label{fig:power_curve}
\end{figure}

The power curve (see \Cref{subsec:wpp_model}) which was used for the wind power prediction is shown in Figure \ref{fig:power_curve}. The turbine type Enercon E-115/3200 was chosen because it is common in Germany\footnote{Enercon is the producer that installed the most turbines in Germany since 1990 which sum up to a total of 12.606 \citep{WindenergieHersteller}. Until March 2019 the company implemented 1016 turbines of the type E-115. \citep{wiki:ListederWindkraftanlagentypen}
}.

To estimate the long-term effect of the change in spatial resolution on the predicted wind power, the cumulative wind power $CWP$ was calculated. For time point $t_I$ it is given by the sum over the value at $t_I$ and all previous time points
\begin{equation}
    CWP(t_I) = \sum_{i=0}^I \text{windpower}(t_i) \text{\,.}
\end{equation}

\chapter{Results}\label{chapter:results}
The analysis consists of two parts: First, the wind speed data sets are compared in \Cref{sec:ws_sample_comparison}. Then, the corresponding predicted wind power biases are presented in \Cref{sec:results_wpp}. 

\section{Wind Speed Distribution Comparison}
\label{sec:ws_sample_comparison}
The wind speed samples derived from the first runs of the CMIP6 GCMs (\Cref{subsec:gcm}) are compared with each other and against the ERA5 reanalysis data (\Cref{subsec:era5}). The comparison was conducted using the original resolution of each data set, considering all 6-hourly time points from 2005 to 2015 in the region of Germany extrapolated to a hub height of \SI{100}{\m}. First, the overall similarity of the wind speed samples is evaluated in \Cref{subsec:distribution_similarities}. Then, several metrics are examined for their development with the spatial resolution \Cref{subsec:results_summary_statistics}. Next, the wind speed samples are individually compared to ERA5 in detail in \Cref{subsec:comparison_to_era5}. Finally, model performance rankings according to different metrics in \Cref{subsec:model_performance}.

\subsection{Similarity between the Entire Distributions}
\label{subsec:distribution_similarities}
The Kolmogorov-Smirnov tests (\Cref{tab:kolmogorov_smirnov_tests_germany_2005-2015}) indicate that the considered wind speed samples originate from different distributions. In all comparisons, the null hypothesis is rejected at any common level of statistical significance. The majority of the $p$-values are below the smallest positive normalized value of $2.23\cdot 10^{-308}$ for IEEE 754 double-precision floating-point numbers.

\begin{figure}[htp]
    \centering
    \begin{subfigure}[b]{\textwidth}
        \centering
        \includegraphics[width=0.8\textwidth]{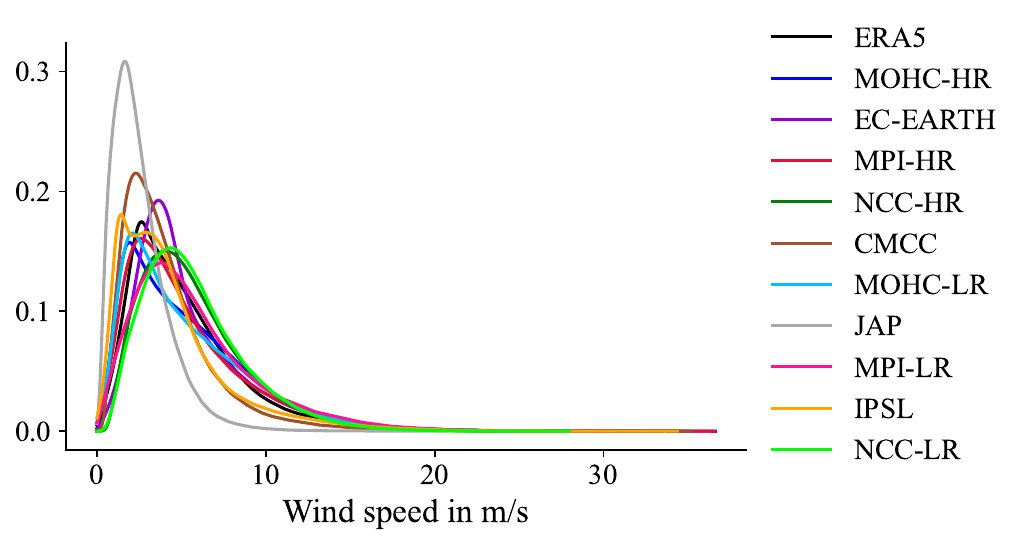}
        \caption{KDEs of all data sets}
        \label{subfig:all kdes}
    \end{subfigure}
    \par\bigskip
    \begin{subfigure}[b]{0.48\textwidth}
        \centering
        \includegraphics[width=\textwidth]{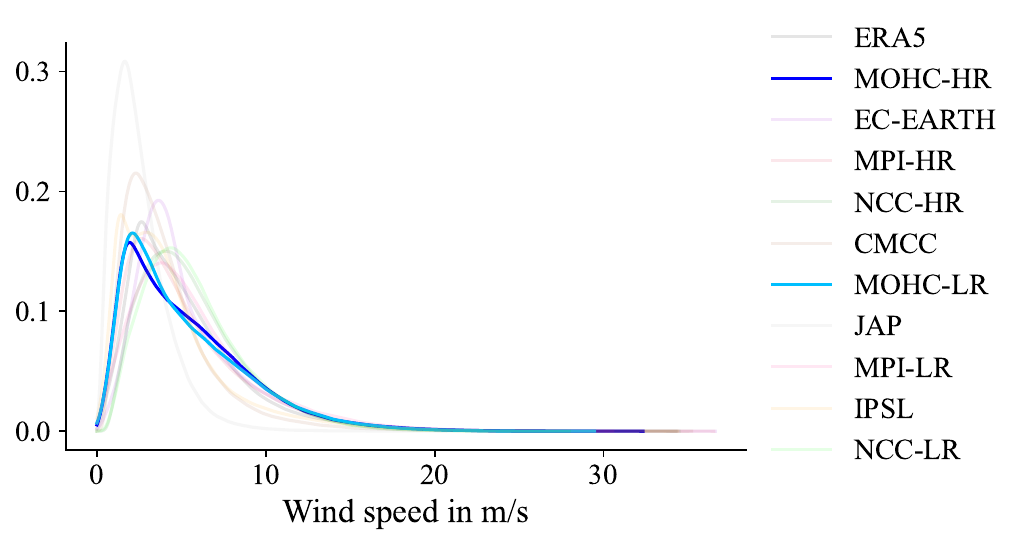}
        \caption{MOHC data sets highlighted}
        \label{subfig:mohc_kdes}
    \end{subfigure}
    \hfill
    \begin{subfigure}[b]{0.48\textwidth}
        \centering
        \includegraphics[width=\textwidth]{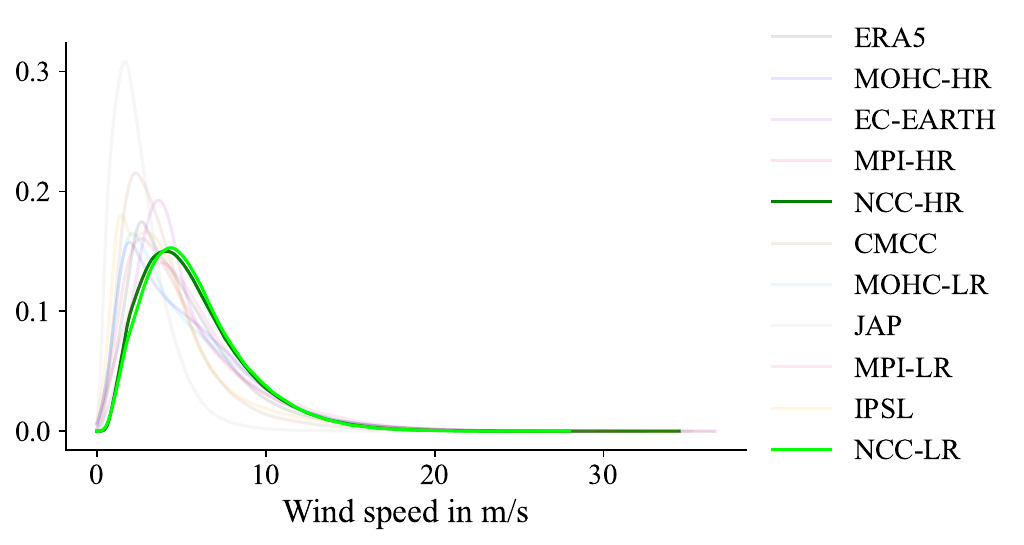}
        \caption{NCC data sets highlighted}
        \label{subfig:ncc_kdes}
    \end{subfigure}
    \caption{Kernel density estimates of wind speed samples derived from different CMIP6 GCMs and ERA5, labels ordered by the models' spatial resolution from top to bottom.}
    \label{fig:all_kdes}
\end{figure}

The kernel density estimates are visualized in \Cref{fig:all_kdes}. This representation demonstrates large differences in the distributions. However, it is important to note that this visualization may not fully accurately represent the shape of the distributions. Despite this limitation, several notable features can be observed. The KDE of the JAP model is particularly distinct, exhibiting a pronounced peak in the low wind speed range. The discrepancies between NCC-LR and NCC-HR and between MOHC-LR and MOHC-HR appear to be the least pronounced (see \Crefrange{subfig:mohc_kdes}{subfig:ncc_kdes}). Revisiting the KS test results, the $p$-values are indeed largest for these two comparisons  (although still well below the threshold for statistical significance).

\begin{landscape}
\begin{table}[H]
\caption{$p$-values of Kolmogorov-Smirnov tests when comparing wind speed samples, where $0.0$ signifies a value below machine precision. The tests reveal that the pairwise wind speed distributions of all models investigated differ significantly.}
\label{tab:kolmogorov_smirnov_tests_germany_2005-2015}
\centering
\begin{tabu} {l|XXXXXXXXXX}
\toprule
& NCC-LR & IPSL & MPI-LR & JAP & MOHC- & CMCC & NCC-HR & MPI-HR & EC- & MOHC- \\
& & & & & LR & &  &  & EARTH & HR \\
\midrule
ERA5       & $0.0$ & $0.0$ & $0.0$ & $0.0$ & $0.0$ & $0.0$ & $0.0$ & $0.0$ & $0.0$ & $0.0$ \\
MOHC-HR    & $0.0$ & $0.0$ & $0.0$ & $0.0$ & $4.6 \cdot 10^{-165}$ & $0.0$ & $0.0$ & $0.0$ & $0.0$ & - \\
EC-EARTH   & $0.0$ & $0.0$ & $0.0$ & $0.0$ & $0.0$ & $0.0$ & $0.0$ & $0.0$ & - & - \\
MPI-HR     & $0.0$ & $0.0$ & $0.0$ & $0.0$ & $4.0 \cdot 10^{-260}$ & $0.0$ & $0.0$ & - & - & - \\
NCC-HR     & $4.1 \cdot 10^{-100}$ & $0.0$ & $0.0$ & $0.0$ & $0.0$ & $0.0$ & - & - & - & - \\
CMCC       & $0.0$ & $0.0$ & $0.0$ & $0.0$ & $0.0$ & - & - & - & - & - \\
MOHC-LR    & $0.0$ & $0.0$ & $0.0$ & $0.0$ & - & - & - & - & - & - \\
JAP       & $0.0$ & $0.0$ & $0.0$ & - & - & - & - & - & - & - \\
MPI-LR     & $0.0$ & $0.0$ & - & - & - & - & - & - & - & - \\
IPSL       & $0.0$ & - & - & - & - & - & - & - & - & - \\
\bottomrule
\end{tabu}
\end{table}

\begin{table}[H]
\caption{Characteristics of the wind speed samples for all considered models, starting with the summary statistics mean, variance, maximum wind speed value, skewness, and kurtosis, followed by the fit parameters of the exponentiated Weibull distribution $a$, $c$, $loc$ and $scale$, and ending with the Jensen-Shannon (JS) and the Wasserstein (WS) distance to ERA5.}
\label{tab:summary_statistics_germany_2005-2015}
\centering
\begin{tabu}{ l|XXXXXXXXXXX }
\toprule
& Mean & Variance & Max. & Skewness & Kurtosis & Parm. & Param. & Param. &  Parm. & JS & WS \\
&  &  & value &  &  & $a$ & $c$ & $loc$ &  $scale$ & Div.& Div. \\
\midrule
ERA5 & 5.19 & 11.34 & 34.30 & 1.49 & 3.08 & 2.43 & 1.10 & 3.15e-05 & 3.29 & - & - \\
MOHC-HR & 5.34 & 13.55 & 32.35 & 1.24 & 1.95 & 2.51 & 0.96 & 1.07e-03 & 3.10 & 0.079 & 0.358 \\
EC-EARTH & 5.49 & 11.95 & 36.62 & 1.65 & 3.84 & 4.90 & 0.86 & 7.55e-04 & 2.08 & 0.074 & 0.296 \\
MPI-HR & 5.23 & 13.12 & 36.53 & 1.47 & 2.86 & 3.14 & 0.90 & 3.88e-04 & 2.55 & 0.042 & 0.209 \\
NCC-HR & 5.71 & 9.84 & 34.48 & 1.18 & 2.01 & 3.29 & 1.04 & 0.23 & 2.92 & 0.107 & 0.592 \\
CMCC & 4.18 & 8.14 & 35.21 & 2.03 & 6.46 & 6.94 & 0.71 & 0.10 & 1.00 & 0.128 & 1.014 \\
MOHC-LR & 5.25 & 13.33 & 29.47 & 1.23 & 1.66 & 2.53 & 0.95 & 2.04e-03 & 2.99 & 0.077 & 0.333 \\
JAP & 2.58 & 2.90 & 20.65 & 1.66 & 5.06 & 2.59 & 1.03 & 7.34e-05 & 1.54 & 0.355 & 2.612 \\
MPI-LR & 5.75 & 13.39 & 30.55 & 1.26 & 1.98 & 1.79 & 1.26 & 3.83e-03 & 4.57 & 0.068 & 0.570 \\
IPSL & 4.41 & 11.30 & 34.36 & 1.83 & 4.58 & 3.12 & 0.86 & 2.39e-03 & 2.03 & 0.126 & 0.782 \\
NCC-LR & 5.80 & 9.36 & 27.98 & 1.10 & 1.66 & 2.52 & 1.23 & 0.27 & 3.78 & 0.124 & 0.702 \\
\bottomrule
\end{tabu}
\end{table}

\begin{table}[H]
\caption{Regression details for summary statistics, fit parameters of the exponentiated Weibull distribution as well as Jensen-Shannon (JS) and Wasserstein (WS) distance including the regression parameters $\alpha$ and $\beta$, the standard error of the slope, the explained variance ($100\cdot R^2$) in \% and the $p$-value of the Overall-$F$-Test.}
\label{tab:regression_details}
\centering
\begin{tabu}{l|XXXXXXXXXXXXX}
\toprule
& Mean & Variance & Max. & Skewness & Kurtosis & Param. & Param. & Param. & Param. & JS & WS \\
& & & value & & & $a$ & $c$ & $loc$ & $scale$ & Div. & Div. \\
\midrule
Intercept $\alpha$ & 4.52 & 8.38 & 24.59 & 1.36 & 3.01 & 2.75 & 1.07 & 0.16 & 2.82 & 0.26 & 1.96 \\
Slope $\beta$ & 0.26 & 1.31 & 4.12 & 0.06 & 0.10 & 0.28 & -0.05 & -0.06 & -0.06 & -0.09 & -0.72 \\
Std. Deviation & 0.59 & 1.93 & 2.61 & 0.19 & 1.00 & 0.91 & 0.10 & 0.06 & 0.64 & 0.09 & 0.68 \\
Explained Var. & 2.15 & 4.87 & 21.66 & 1.16 & 0.12 & 1.03 & 2.22 & 9.68 & 0.10 & 10.85 & 12.21 \\
$p$-value & 0.67 & 0.52 & 0.15 & 0.75 & 0.92 & 0.77 & 0.66 & 0.35 & 0.93 & 0.35 & 0.32 \\
\bottomrule
\end{tabu}
\end{table}

\end{landscape}

\subsection{Trends with Spatial Resolution}
\label{subsec:results_summary_statistics}
\begin{figure}[htp]
    \centering
    \begin{subfigure}[b]{0.48\textwidth}
        \centering
        \includegraphics[height=5cm]{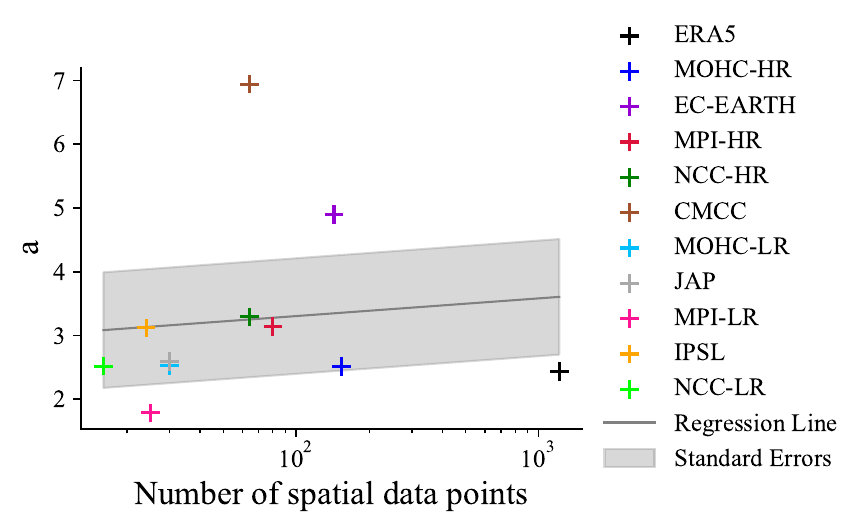}
        \caption{Exponential parameter $a$}
        \label{subfig:weibull_a}
    \end{subfigure}
    \hfill
    \begin{subfigure}[b]{0.48\textwidth}
        \centering
        \includegraphics[height=5cm]{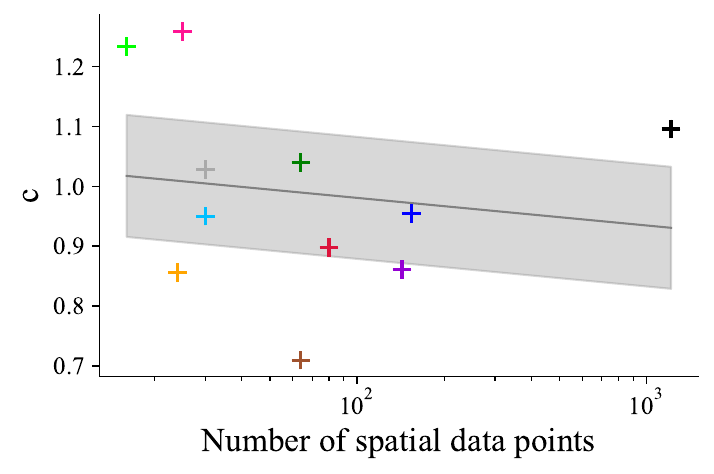}
        \caption{Weibull parameter $c$}
        \label{subfig:weibull_c}
    \end{subfigure}
    \par\bigskip
    \begin{subfigure}[b]{0.48\textwidth}
        \centering
        \includegraphics[height=5cm]{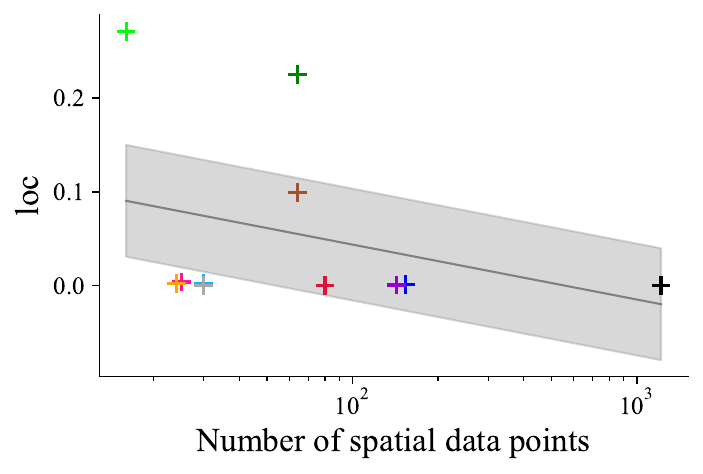}
        \caption{Location parameter $loc$}
        \label{subfig:weibull_loc}
    \end{subfigure}
    \hfill
    \begin{subfigure}[b]{0.48\textwidth}
        \centering
        \includegraphics[height=5cm]{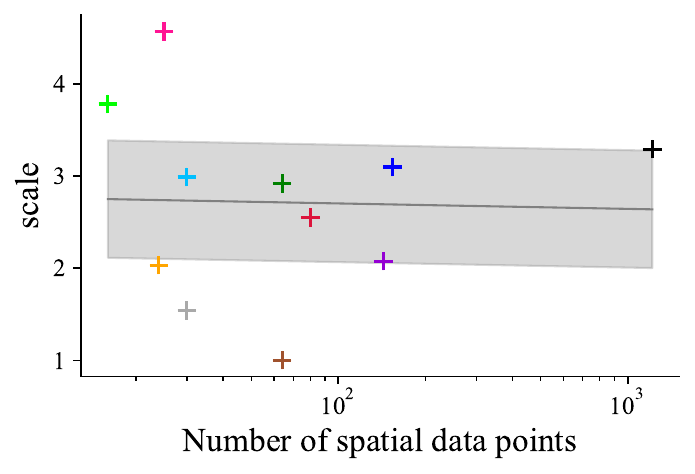}
        \caption{Shape parameter $scale$}
        \label{subfig:weibull_scale}
    \end{subfigure}    
    \caption{Weibull fit parameters for wind speed samples plotted against the spatial resolution with linear regression line with its standard error of the slope in grey.}
    \label{fig:weibull_parameter_trends}
\end{figure}

\begin{figure}[htp]
    \centering
    \centering
    \begin{subfigure}[b]{0.48\textwidth}
        \centering
        \includegraphics[height=5cm]{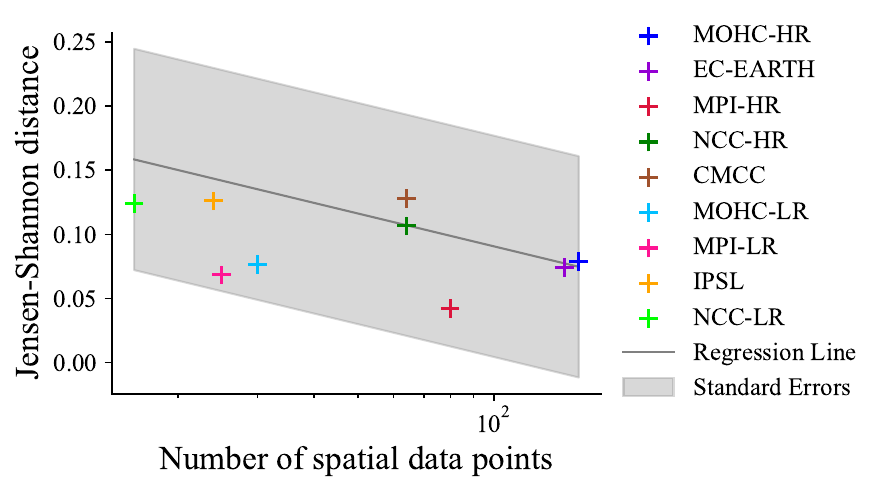}
        \caption{Jensen-Shannon distance}
        \label{subfig:js_div}
    \end{subfigure}
    \hfill
    \begin{subfigure}[b]{0.48\textwidth}
        \centering
        \includegraphics[height=5cm]{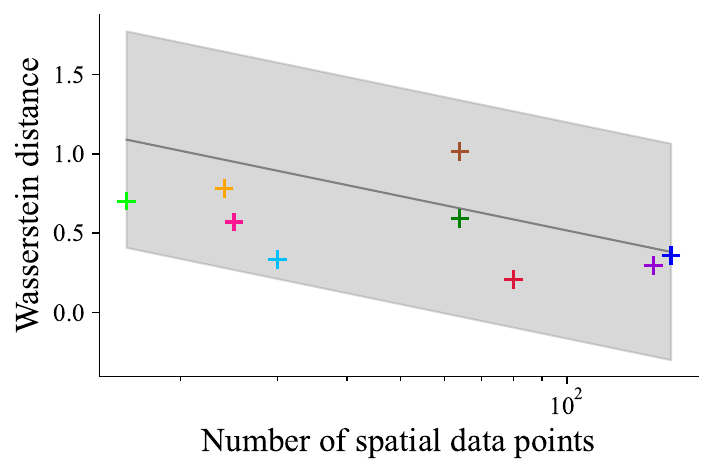}
        \caption{Wasserstein distance}
        \label{subfig:distance_wasserstein}
    \end{subfigure}
    \caption{Jensen-Shannon and Wasserstein distances between wind speed samples from CMIP6 GCMs and ERA5 (value for JAP above the axis limit) plotted against the spatial resolution. Linear regression line with standard error of the slope shown in grey.}
    \label{fig:f_distances_spatial_resolution}
\end{figure}

Several metrics including different summary statistics, fit parameters of the exponentiated Weibull distribution, as well as the Jensen-Shannon and the Wasserstein distance to ERA5 were derived from the wind speed samples (\Cref{tab:summary_statistics_germany_2005-2015}) and plotted against the spatial resolution, measured by the number of data points within the region of consideration (\Crefrange{fig:summary_statistics_ws_Germany}{fig:f_distances_spatial_resolution}). Linear regression was used to identify trends (for details see \Cref{tab:regression_details}).

\subsubsection{Summary Statistics}

The linear regression analysis revealed no statistically significant trends for any of the summary statistics when plotted against spatial resolution. The Overall-$F$-Tests yielded $p$-values ranging from $0.15$ to $0.92$. Even at a high significance level of $10\%$ they are all insignificant, indicating that the null hypothesis of the slope being zero cannot be rejected. Furthermore, the coefficients of determination were low, spanning from $0.0012$ to $0.2166$. These small $R^2$ values suggest that the linear model explains only a minimal portion of the variance in the summary statistics across different spatial resolutions and thus provides a poor fit to the data.

\begin{figure}[htp]
    \centering
    \begin{subfigure}[b]{0.65\textwidth}
        \centering
        \includegraphics[height=5.6cm]{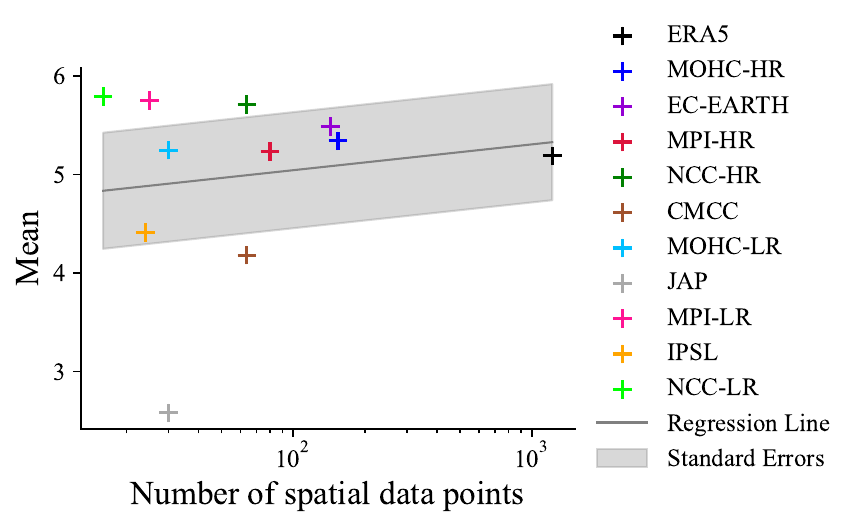}
        \caption{Mean wind speed}
        \label{subfig:mean_ws}
    \end{subfigure}
    \par\bigskip
    \begin{subfigure}[b]{0.48\textwidth}
        \centering
        \includegraphics[height=5cm]{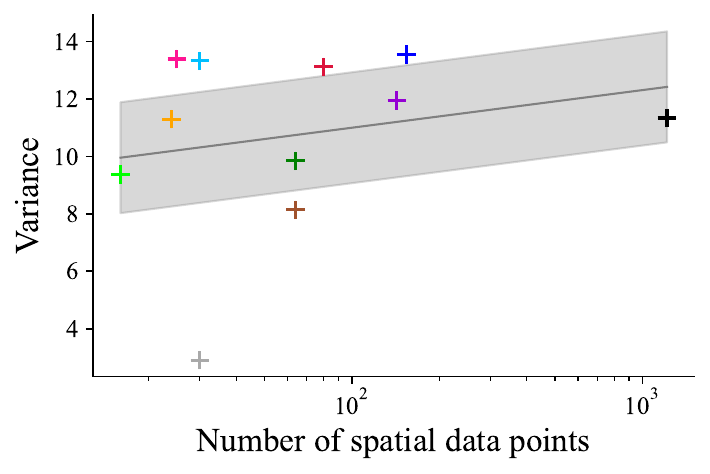}
        \caption{Variance}
        \label{subfig:var_ws}
    \end{subfigure}
    \hfill
    \begin{subfigure}[b]{0.48\textwidth}
        \centering
        \includegraphics[height=5cm]{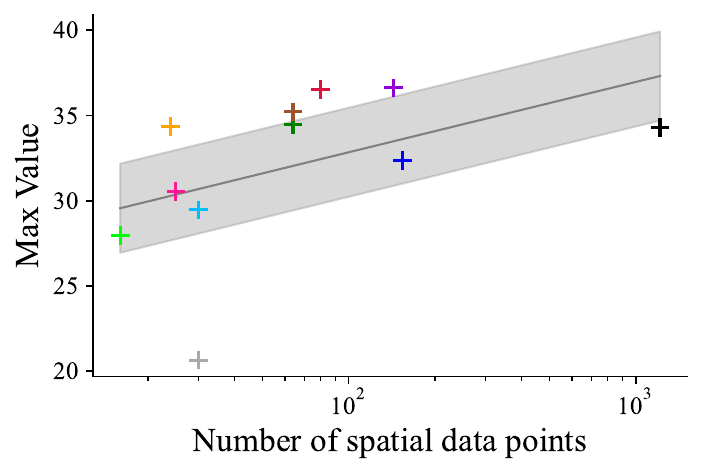}
        \caption{Maximum value}
        \label{subfig:max_ws}
    \end{subfigure}
    \par\bigskip
    \begin{subfigure}[b]{0.48\textwidth}
        \centering
        \includegraphics[height=5cm]{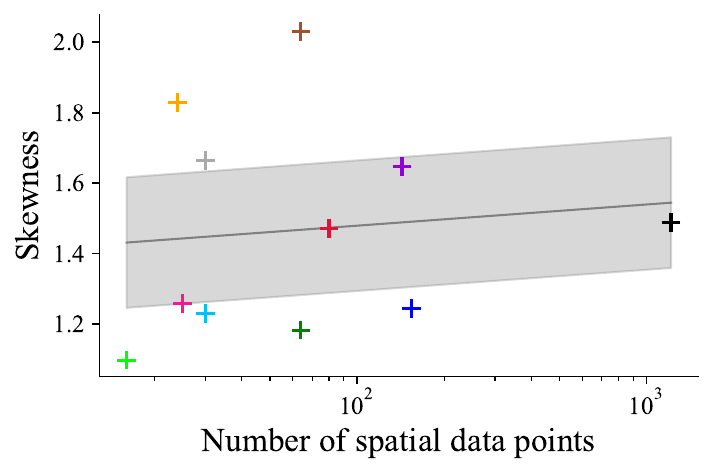}
        \caption{Skewness}
        \label{subfig:skewness_ws}
    \end{subfigure}
    \hfill
    \begin{subfigure}[b]{0.48\textwidth}
        \centering
        \includegraphics[height=5cm]{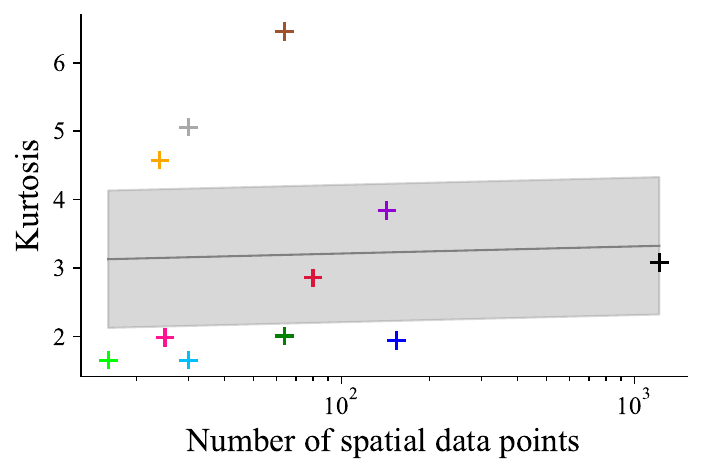}
        \caption{Kurtosis}
        \label{subfig:kurtosis_ws}
    \end{subfigure}    
    \caption{Summary statistics of wind speed samples compared by spatial resolution. The legend in \Cref{subfig:mean_ws} applies to all subplots. The data set labels are sorted by decreasing resolution. The gray line shows the linear regression fit to the data points, with the standard deviation being indicated as a shaded area.}
    \label{fig:summary_statistics_ws_Germany}
\end{figure}

Among the variables analyzed, the maximum wind speed value (\Cref{subfig:max_ws}) showed the most notable, although statistically insignificant, relationship with spatial resolution. It exhibited the smallest $p$-value ($0.15$) and the highest explained variance ($21.66\%$) of all variables considered. The trend suggests an increase in maximum wind speed with the logarithmic number of data points, with an estimated slope of approximately $2.6$. This tendency is further supported by comparisons between data sets from the same GCM run at different spatial resolutions, where the maximum wind speed value consistently appears higher in the high-resolution versions.

\subsubsection{Fit Parameters of Exponentiated Weibull Distribution}

The fits with the exponentiated Weibull distribution capture the individual wind speed samples well, apart from some underestimation at the upper quantiles (see \Cref{sec:weibull_fits}). However, the fit parameters do not reveal any trends with the spatial resolution. The linear regression yields minimal slopes and an explained variance of a maximum $9.68\%$ for the $loc$ parameter. The $p$-values of the Overall-$F$-tests are highly insignificant, ranging from $0.35$ to $0.93$. Other maximum likelihood estimations of the parameters provide similarly unclear results. Fixing one of the parameters decreases the accuracy of the fit, while not providing any additional insights.

\subsubsection{Jensen-Shannon and Wasserstein Distances}
Both the Jensen-Shannon and the Wasserstein distances to ERA5 exhibit a slight decreasing trend with increasing resolution. However, the regression analysis explains only $10.85\%$ and $12.21\%$ of the variance, respectively, with non-significant $p$-values of $0.35$ and $0.32$. The distances of the MPI and NCC high-resolution models are notably smaller than the ones of their respective low-resolution counterparts. In contrast, the MOHC models demonstrate comparable distances, with the low-resolution model exhibiting slightly smaller values. 

\subsection{Detailed Comparison of GCM Outputs to ERA5}
\label{subsec:comparison_to_era5}

The following analysis examines in which way the GCM outputs differ from ERA5. First, QQ plots are constructed, comparing all 1\% quantiles of the GCM models against ERA5. Then, KDE difference plots highlight differences in the critical wind speed range. Finally, the upper tails of the distributions are analyzed using KDE ratio and survival function plots.

\subsubsection{Quantile Comparison across the Full Wind Speed Range}

\begin{figure}[htp]
    \centering
    \begin{subfigure}[b]{0.19\textwidth}
        \centering
        \includegraphics[width=3cm]{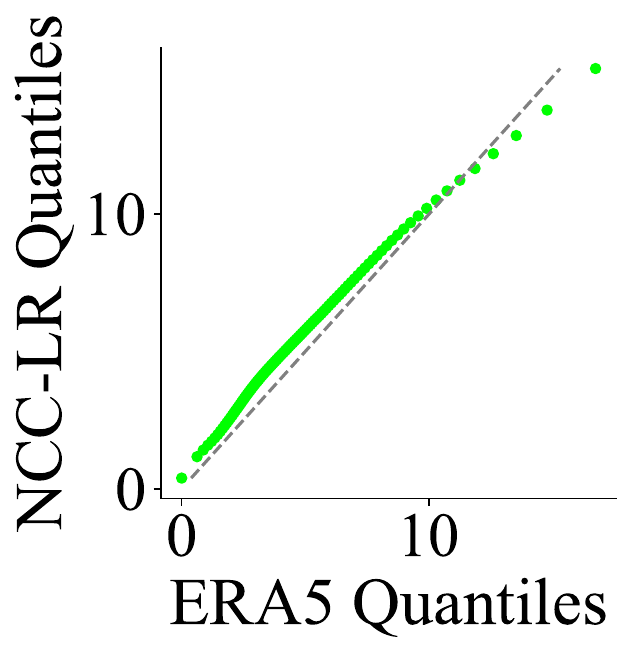}
        \caption{NCC-LR}
    \end{subfigure}
    \hfill
    \begin{subfigure}[b]{0.19\textwidth}
        \centering
        \includegraphics[width=3cm]{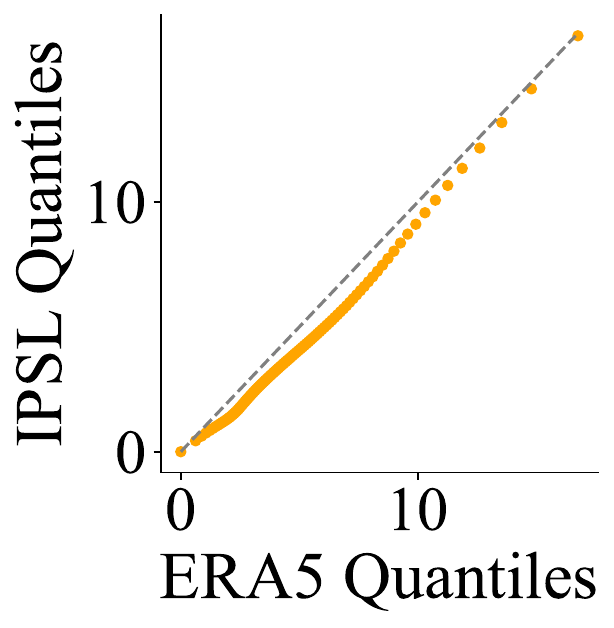}
        \caption{IPSL}
    \end{subfigure}
    \hfill
    \begin{subfigure}[b]{0.19\textwidth}
        \centering
        \includegraphics[width=3cm]{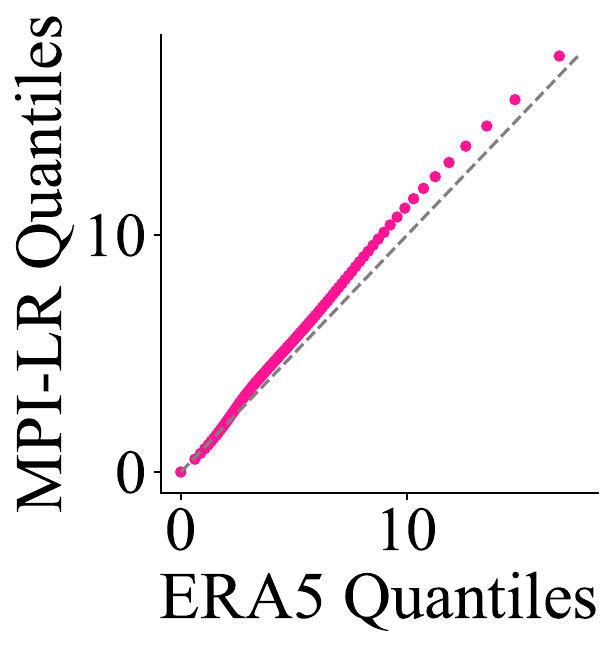}
        \caption{MPI-LR}
    \end{subfigure}
    \hfill
    \begin{subfigure}[b]{0.19\textwidth}
        \centering
        \includegraphics[width=3cm]{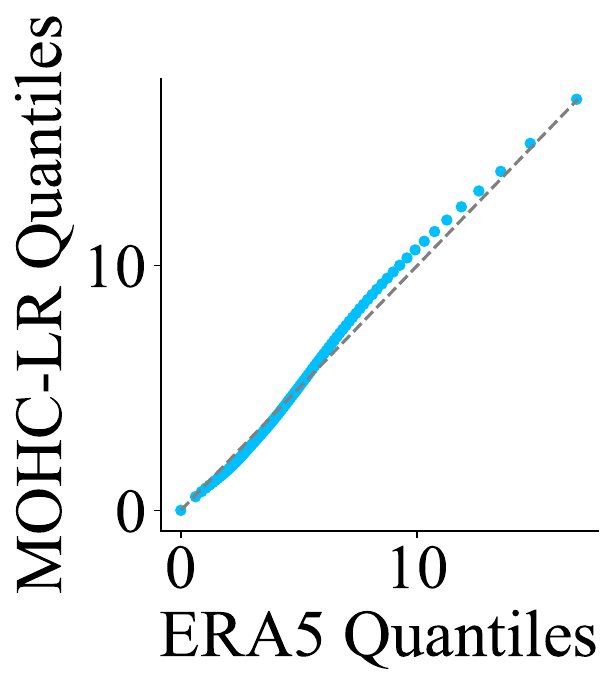}
        \caption{MOHC-LR}
    \end{subfigure}
    \hfill
    \begin{subfigure}[b]{0.19\textwidth}
        \centering
        \includegraphics[width=3cm]{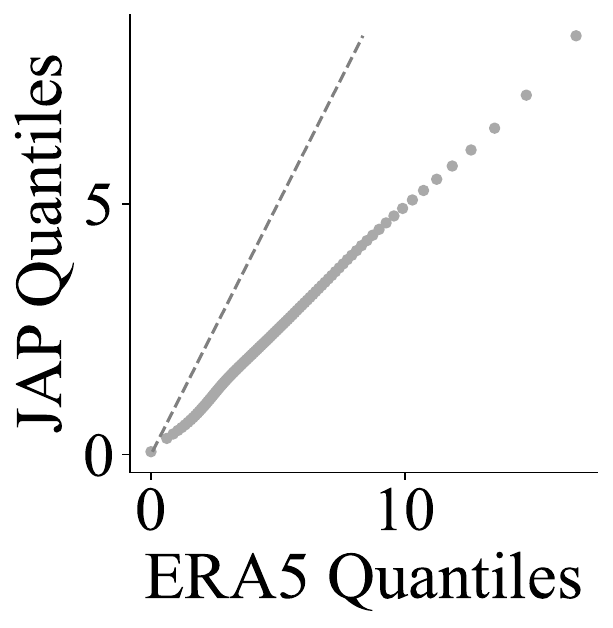}
        \caption{JAP}
    \end{subfigure}
    \par\bigskip
    \begin{subfigure}[b]{0.19\textwidth}
        \centering
        \includegraphics[width=3cm]{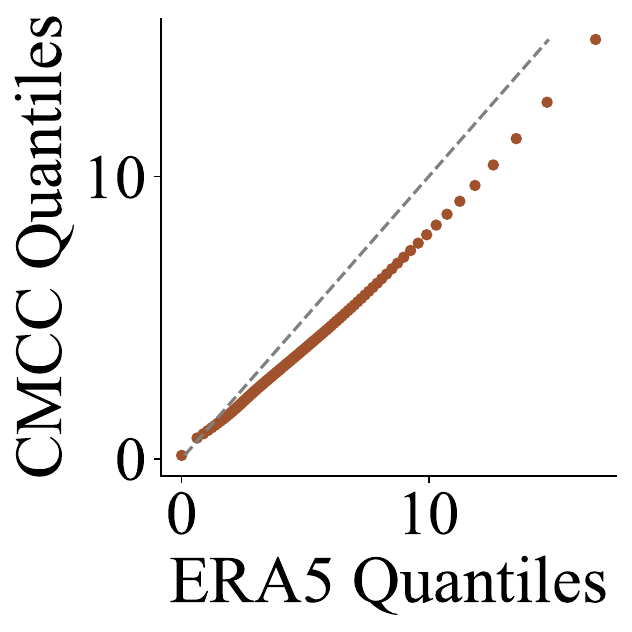}
        \caption{CMCC}
    \end{subfigure}
    \hfill
    \begin{subfigure}[b]{0.19\textwidth}
        \centering
        \includegraphics[width=3cm]{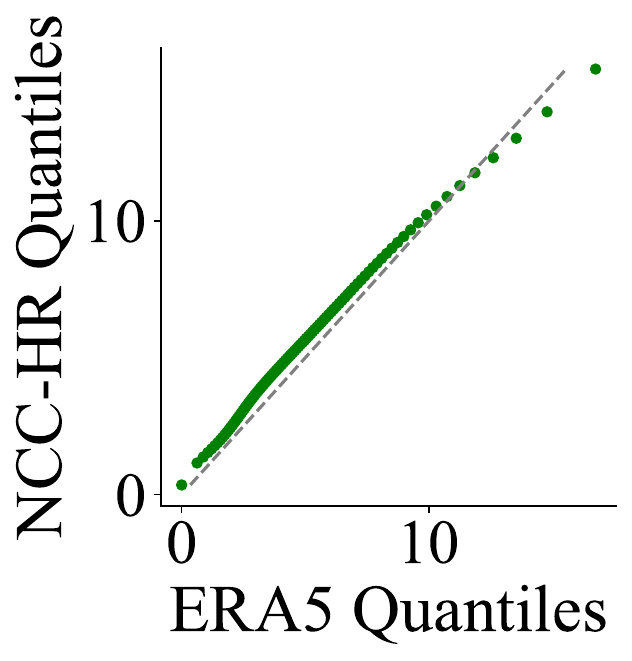}
        \caption{NCC-HR}
    \end{subfigure}
    \hfill
    \begin{subfigure}[b]{0.19\textwidth}
        \centering
        \includegraphics[width=3cm]{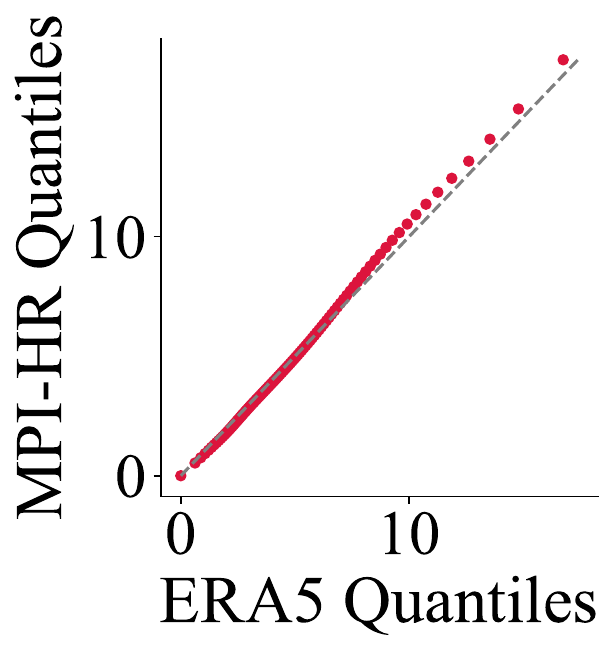}
        \caption{MPI-HR}
    \end{subfigure}
    \hfill
    \begin{subfigure}[b]{0.19\textwidth}
        \centering
        \includegraphics[width=3cm]{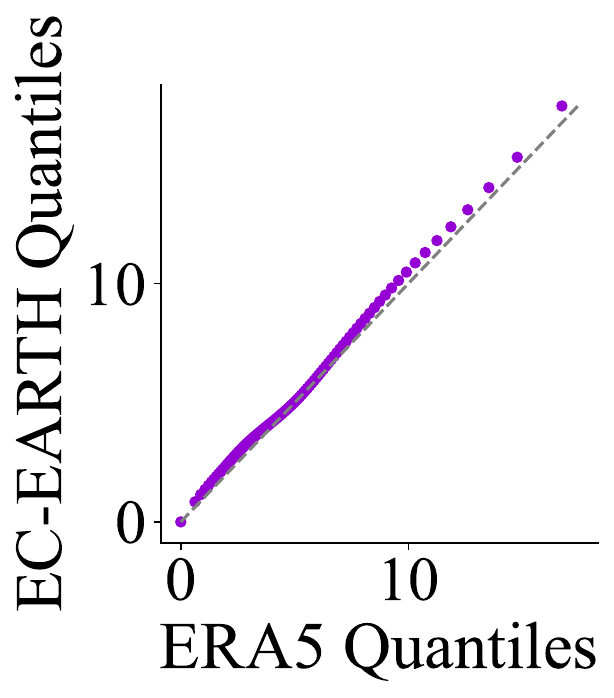}
        \caption{EC-EARTH}
    \end{subfigure}
    \hfill
    \begin{subfigure}[b]{0.19\textwidth}
        \centering
        \includegraphics[width=3cm]{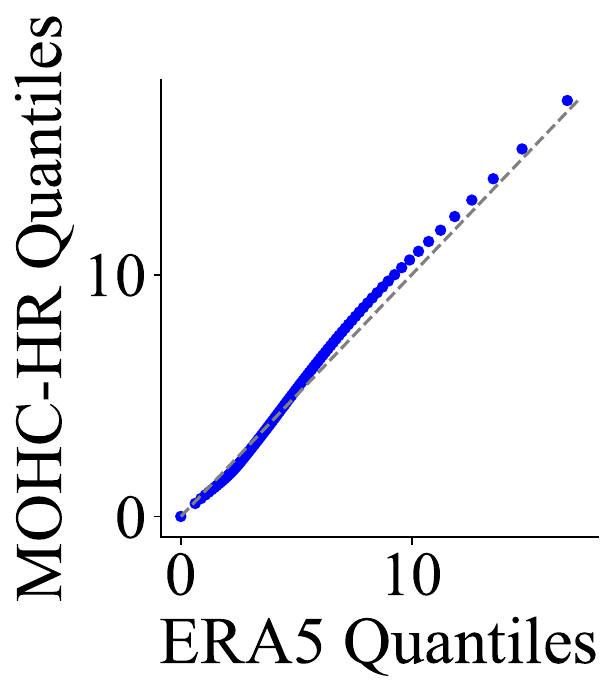}
        \caption{MOHC-HR}
    \end{subfigure}
    \caption{QQ plot comparison of wind speed samples with ERA5, CMIP6 GCMs sorted from lowest to highest spatial resolution from left to right.}
    \label{fig:qq_plot}
\end{figure}

The QQ plots (Figure \ref{fig:qq_plot}) show in which parts of the wind speed distribution the differences to ERA5 are most pronounced. The whole distribution of CMCC and especially JAP is shifted towards lower wind speeds. IPLS also underestimates most parts but is closer to ERA5 in the higher quantiles. EC-EARTH shows a slight overestimation in most quantiles, especially in the higher ones. Both MOHC models slightly underestimate low quantiles and overestimate middle and high quantiles. Both NCC models overestimate low and middle quantiles and underestimate high quantiles.  Both MPI models overestimate medium and high quantiles, with the overestimation being greater for the low-resolution model.

\subsubsection{KDE Difference in the critical Wind Speed Range}

\begin{figure}[htp]
    \centering
    \begin{subfigure}[b]{0.19\textwidth}
        \centering
        \includegraphics[width=3cm]{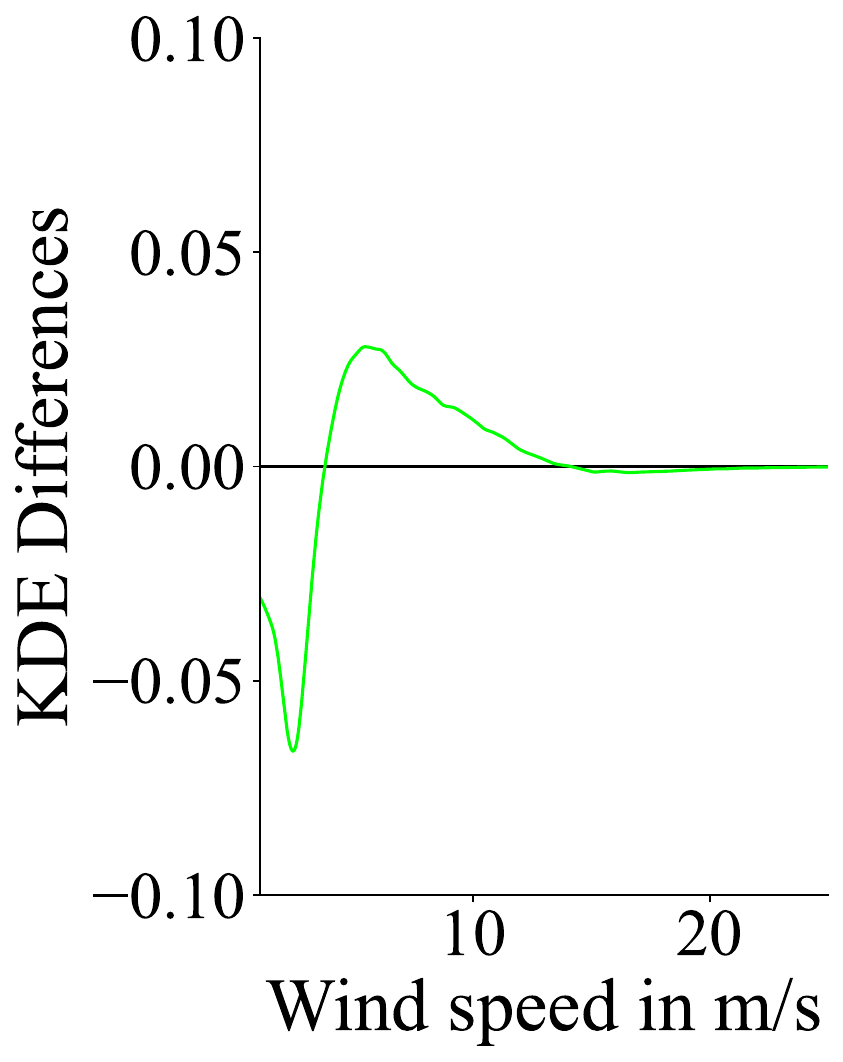}
        \caption{NCC-LR}
    \end{subfigure}
    \hfill
    \begin{subfigure}[b]{0.19\textwidth}
        \centering
        \includegraphics[width=3cm]{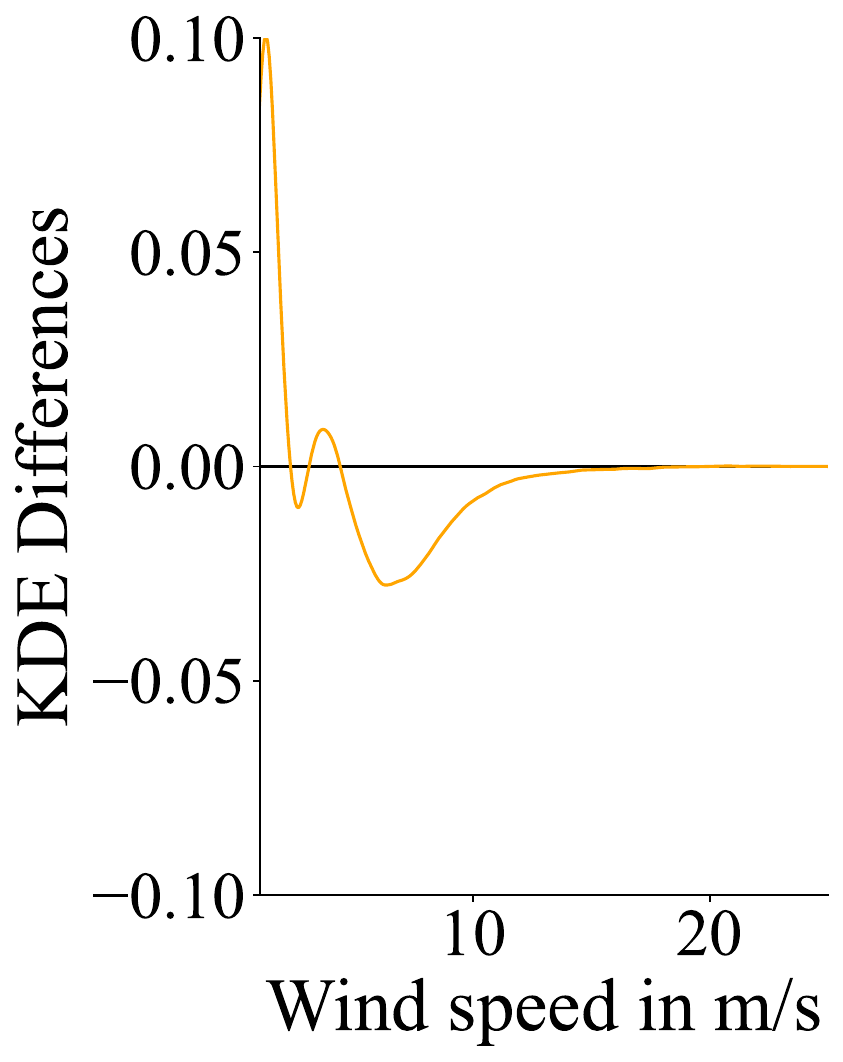}
        \caption{IPSL}
    \end{subfigure}
    \hfill
    \begin{subfigure}[b]{0.19\textwidth}
        \centering
        \includegraphics[width=3cm]{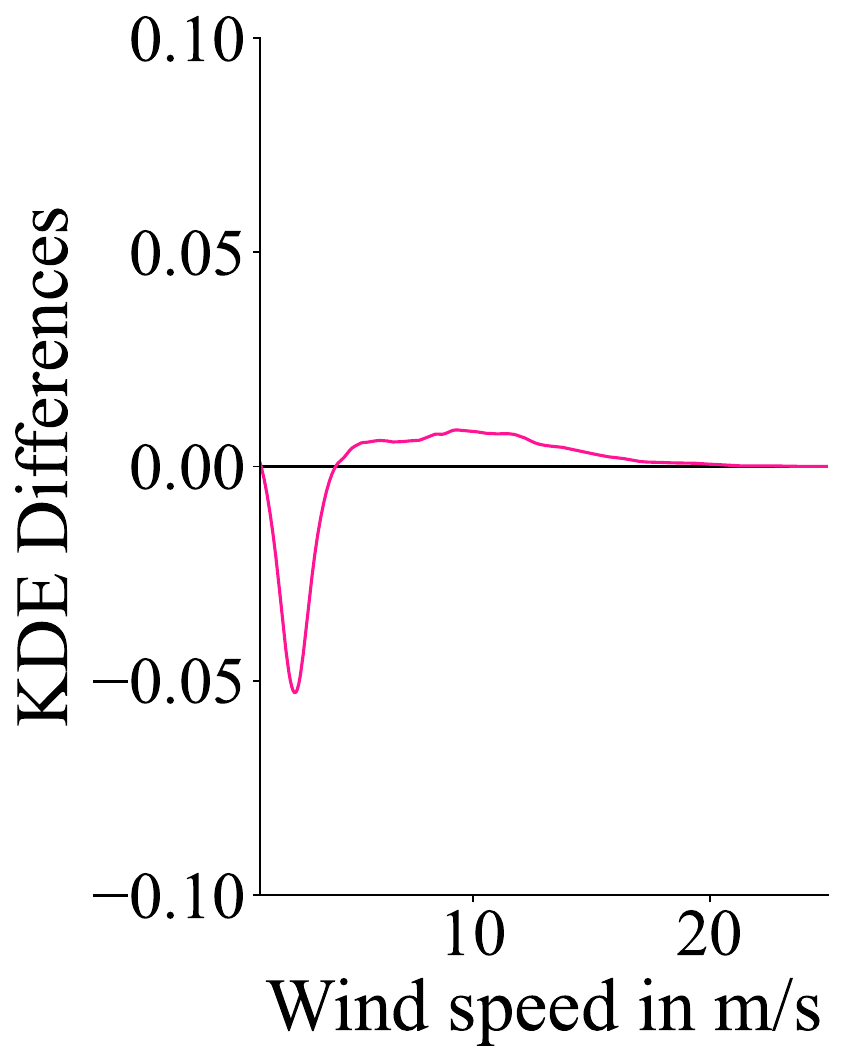}
        \caption{MPI-LR}
    \end{subfigure}
    \hfill
    \begin{subfigure}[b]{0.19\textwidth}
        \centering
        \includegraphics[width=3cm]{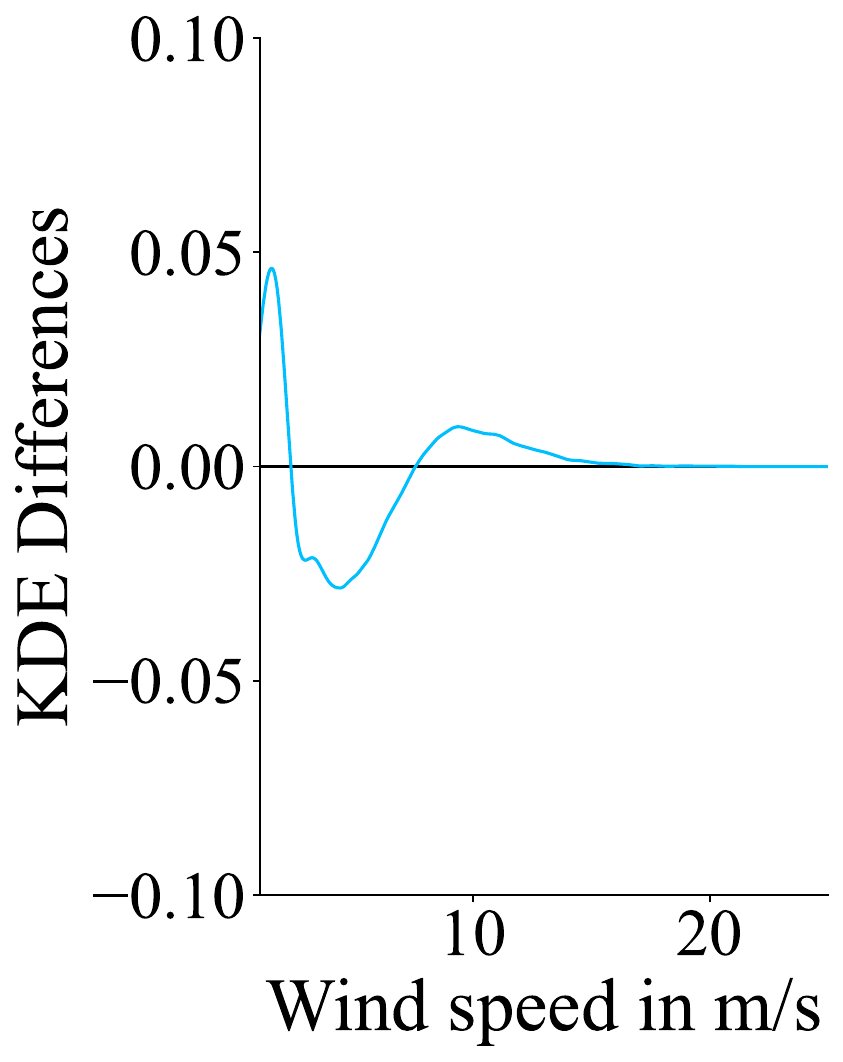}
        \caption{MOHC-LR}
    \end{subfigure}
    \hfill
    \begin{subfigure}[b]{0.19\textwidth}
        \centering
        \includegraphics[width=3cm]{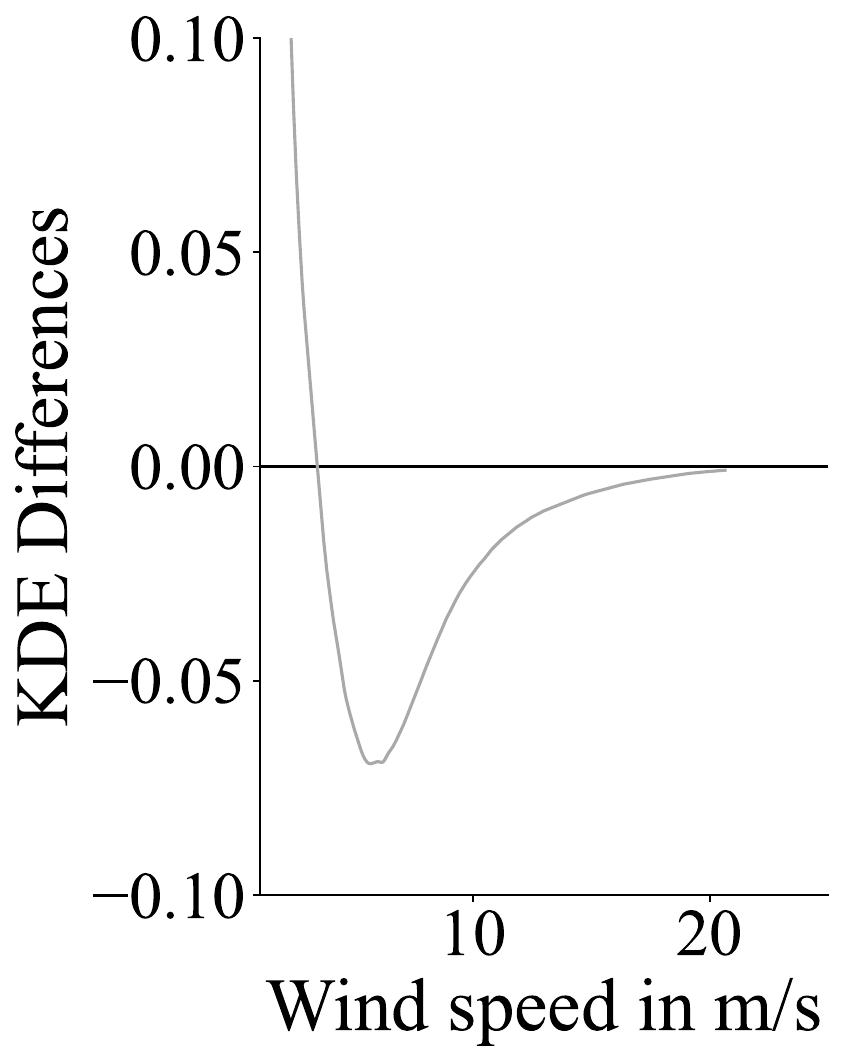}
        \caption{JAP}
    \end{subfigure}
    \par\bigskip
    \begin{subfigure}[b]{0.19\textwidth}
        \centering
        \includegraphics[width=3cm]{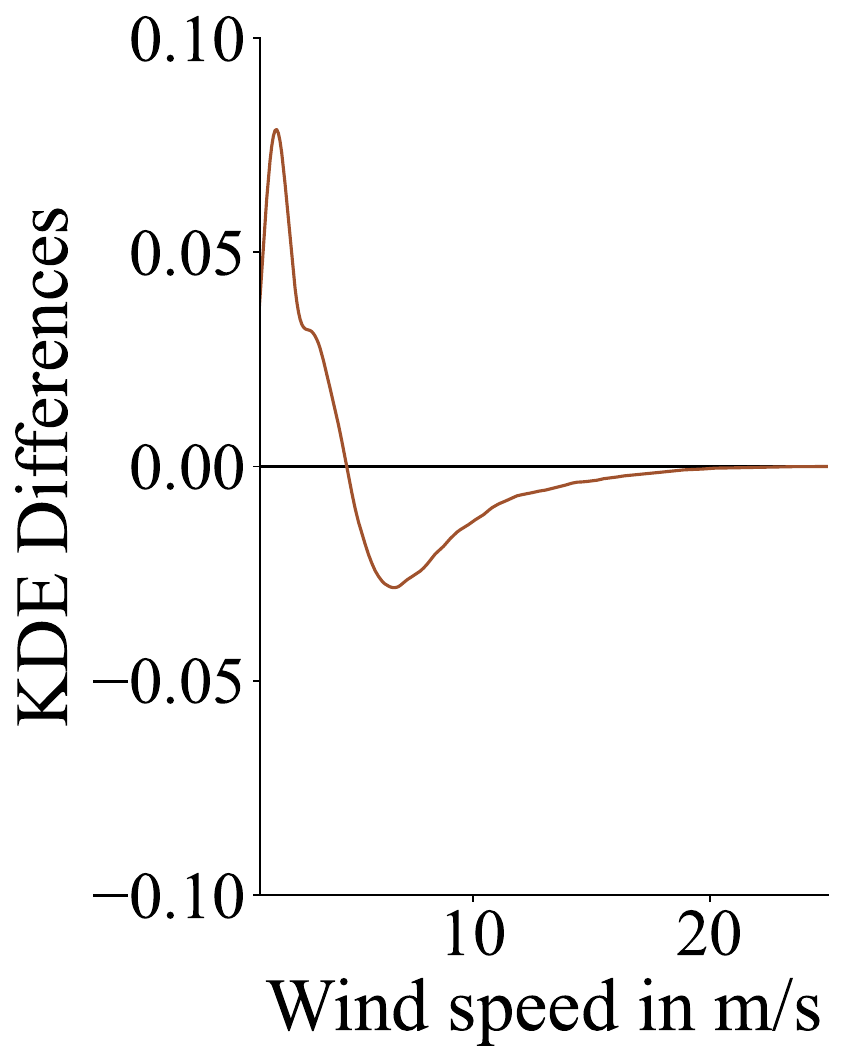}
        \caption{CMCC}
    \end{subfigure}
    \hfill
    \begin{subfigure}[b]{0.19\textwidth}
        \centering
        \includegraphics[width=3cm]{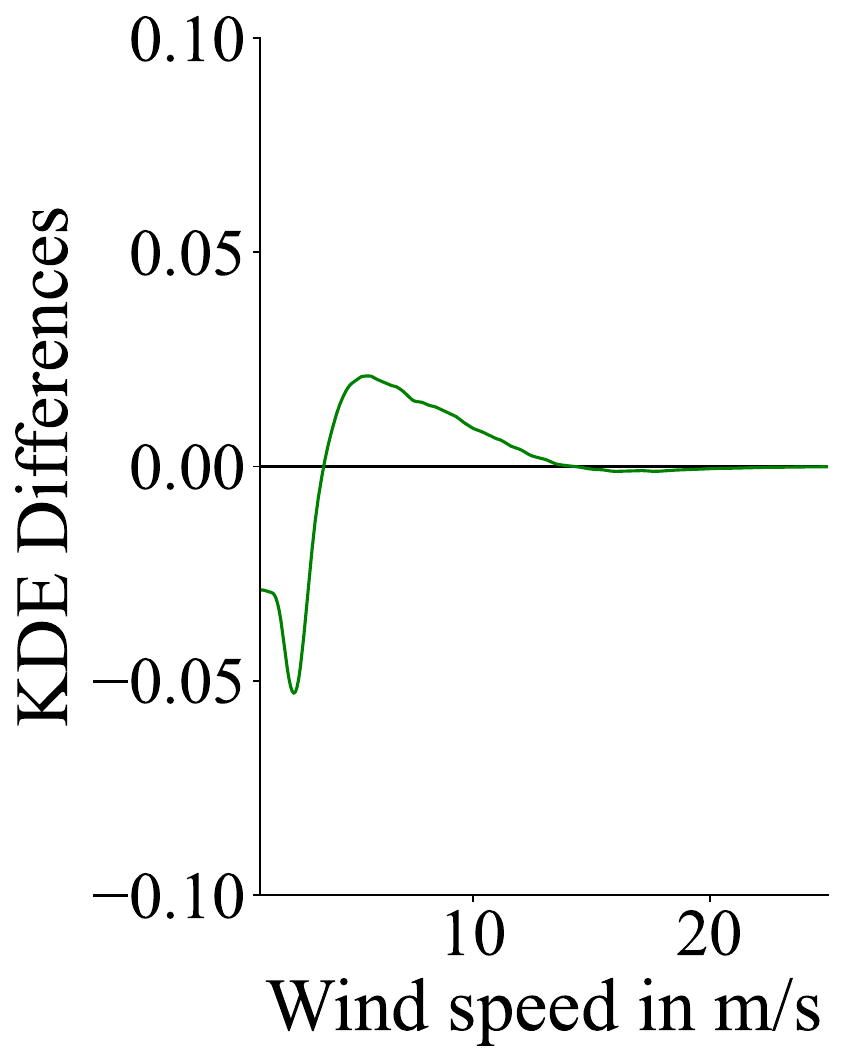}
        \caption{NCC-HR}
    \end{subfigure}
    \hfill
    \begin{subfigure}[b]{0.19\textwidth}
        \centering
        \includegraphics[width=3cm]{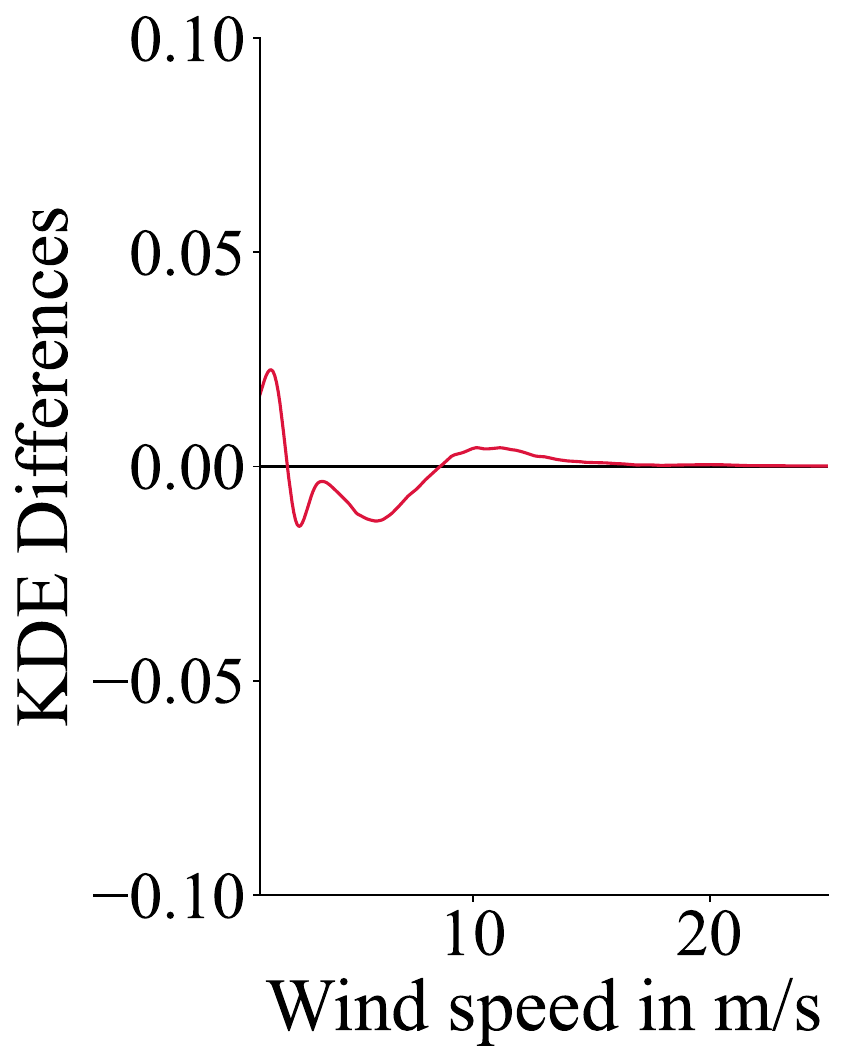}
        \caption{MPI-HR}
    \end{subfigure}
    \hfill
    \begin{subfigure}[b]{0.19\textwidth}
        \centering
        \includegraphics[width=3cm]{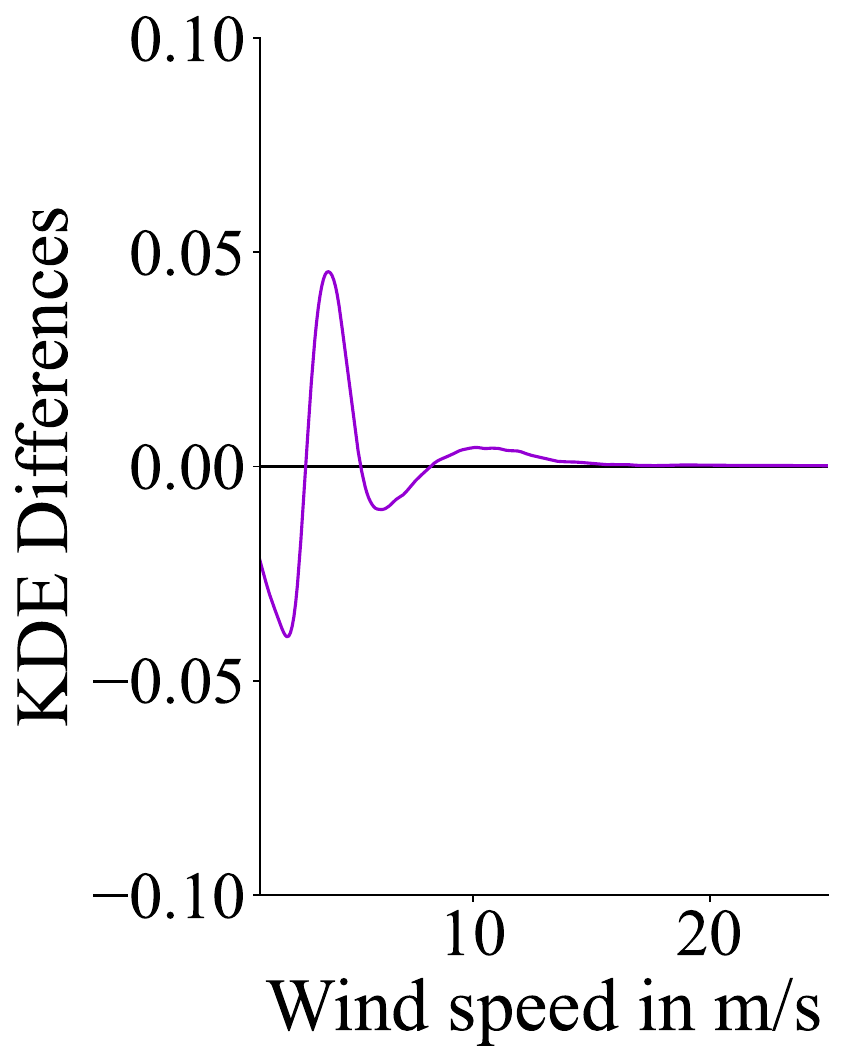}
        \caption{EC-EARTH}
    \end{subfigure}
    \hfill
    \begin{subfigure}[b]{0.19\textwidth}
        \centering
        \includegraphics[width=3cm]{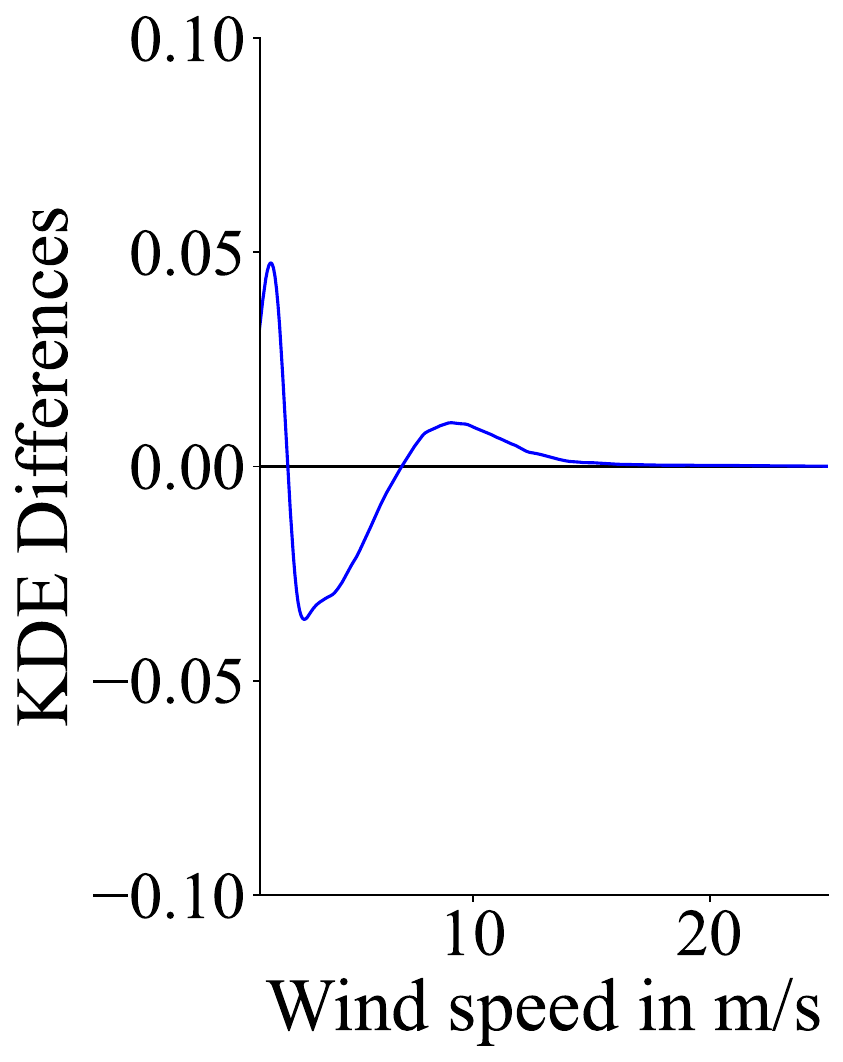}
        \caption{MOHC-HR}
    \end{subfigure}
    \caption{KDE differences between wind speed samples of CMIP6 GCMs and ERA5 in the critical wind speed range from \SI{1}{\meter\per\second} to \SI{25}{\meter\per\second}.}
    \label{fig:kde_differences}
\end{figure}

The critical wind speed range from \SI{1}{\meter\per\second} to \SI{25}{\meter\per\second} is shown in more detail in the KDE difference plots (\Cref{fig:kde_differences}).
IPSL, JAP and CMCC overestimate the small wind speeds and underestimate higher wind speeds in this range, with the underestimation being most pronounced for JAP. 
The NCC data sets and MPI-LR underestimate low wind speeds and overestimate medium wind speeds.
The MOHC data sets underestimate low to medium wind speeds with additional fluctuations and show slight overestimation at higher wind speeds.
EC-EARTH oscillates between positive and negative deviations. 
MPI-HR exhibits minimal differences across the entire wind speed spectrum, indicating little bias towards either underestimating or overestimating wind speeds.

\subsubsection{KDE Ratios and Survival Function in the Upper Tails}

\begin{figure}[htp]
    \centering
    \begin{subfigure}[b]{0.19\textwidth}
        \centering
        \includegraphics[width=3cm]{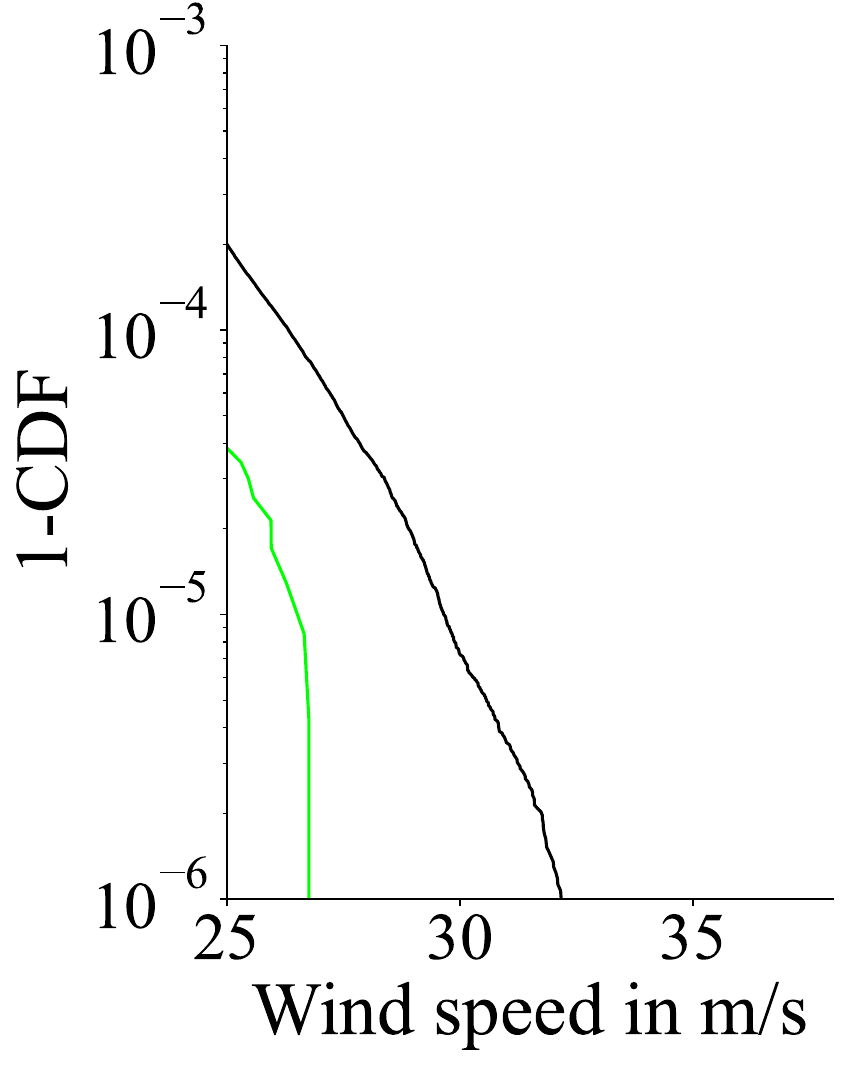}
    \end{subfigure}
    \hfill
    \begin{subfigure}[b]{0.19\textwidth}
        \centering
        \includegraphics[width=3cm]{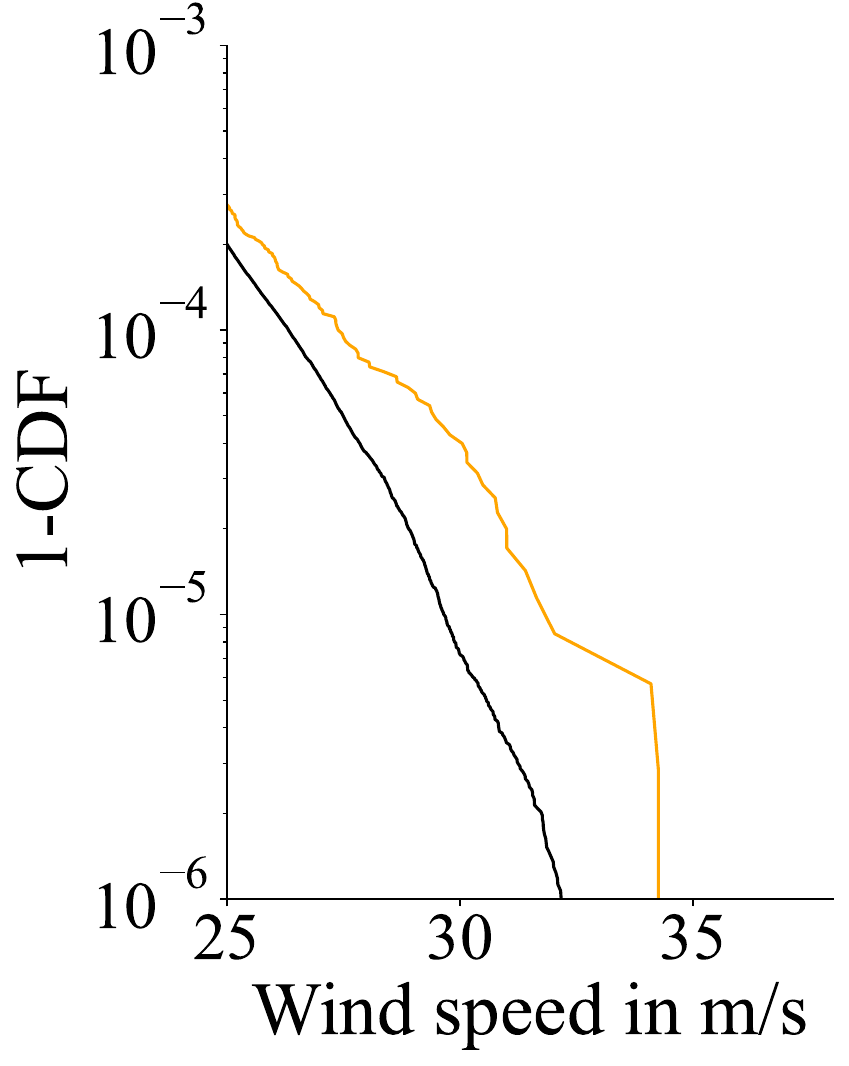}
    \end{subfigure}
    \hfill
    \begin{subfigure}[b]{0.19\textwidth}
        \centering
        \includegraphics[width=3cm]{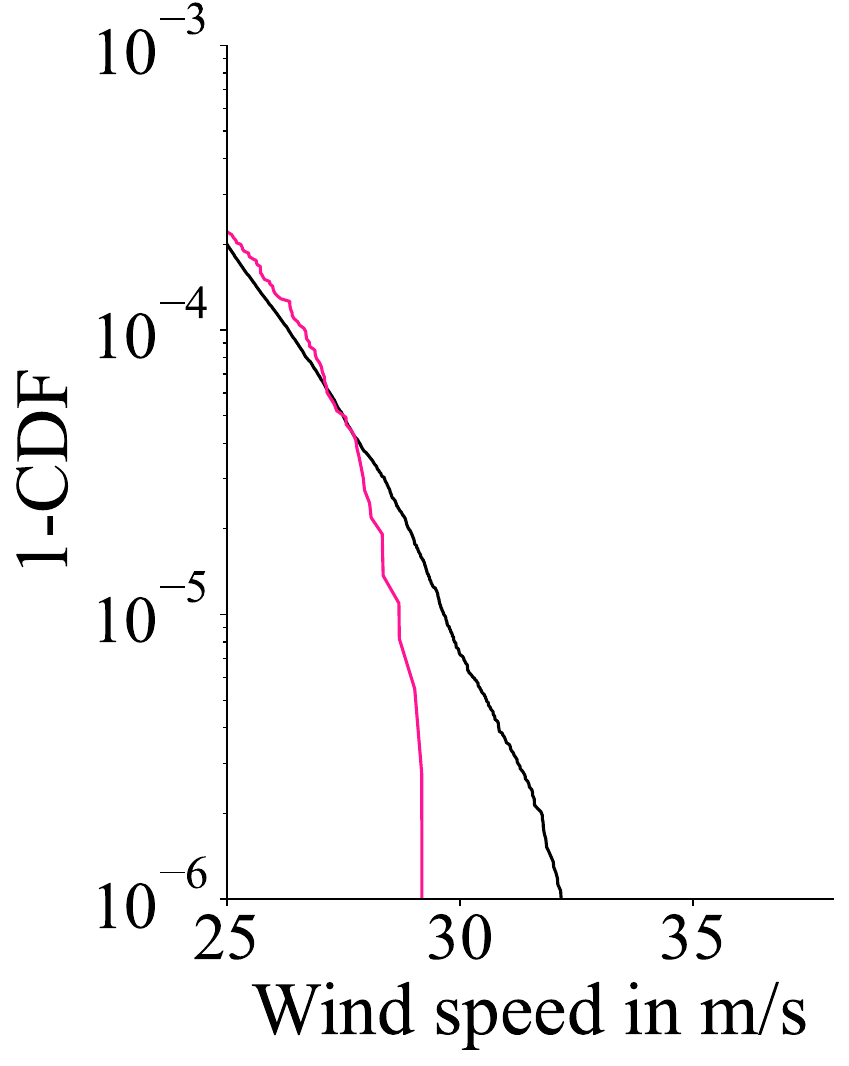}
    \end{subfigure}
    \hfill
    \begin{subfigure}[b]{0.19\textwidth}
        \centering
        \includegraphics[width=3cm]{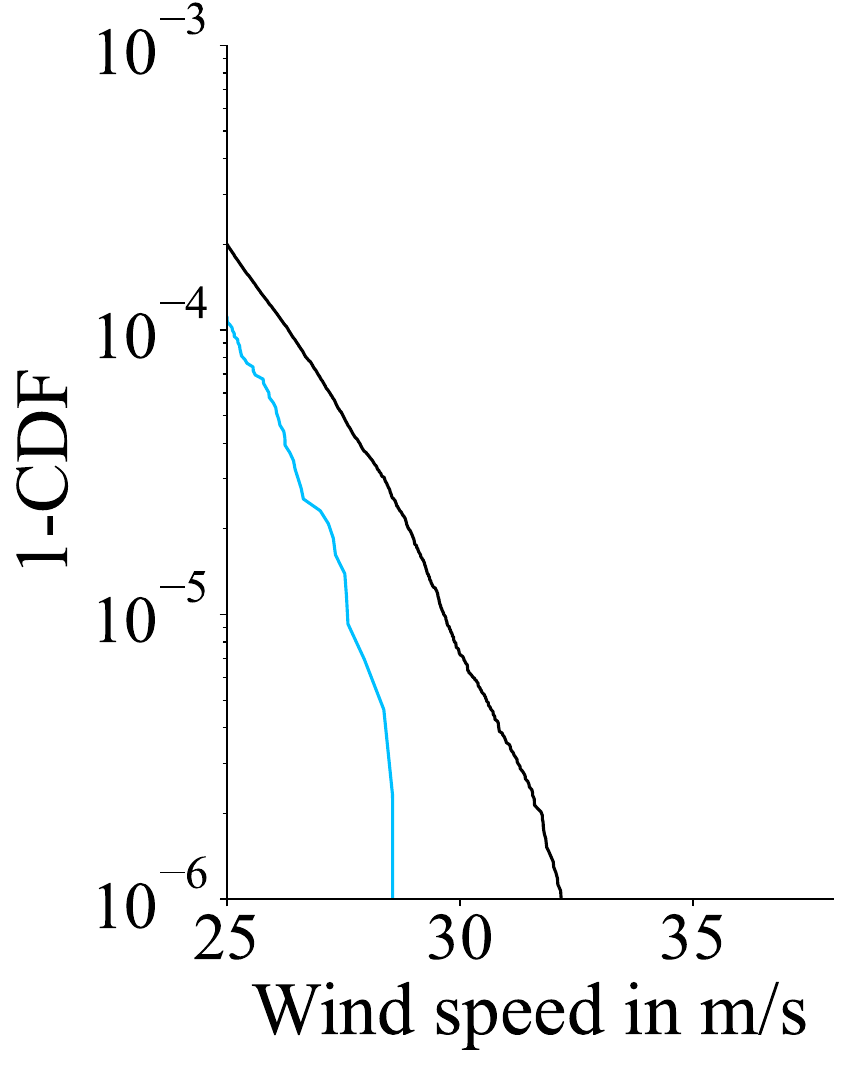}
    \end{subfigure}
    \hfill
    \begin{subfigure}[b]{0.19\textwidth}
        \centering
        \includegraphics[width=3cm]{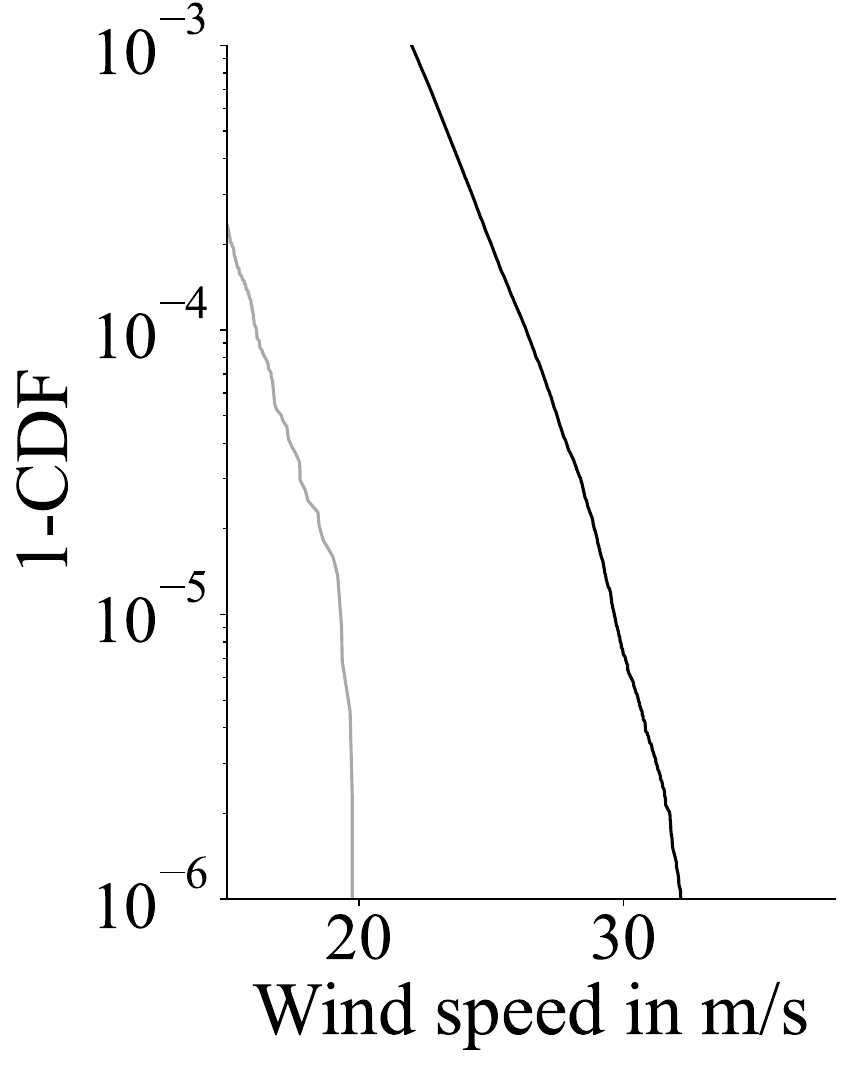}
    \end{subfigure}
    \par\bigskip
    \begin{subfigure}[b]{0.19\textwidth}
        \centering
        \includegraphics[width=3cm]{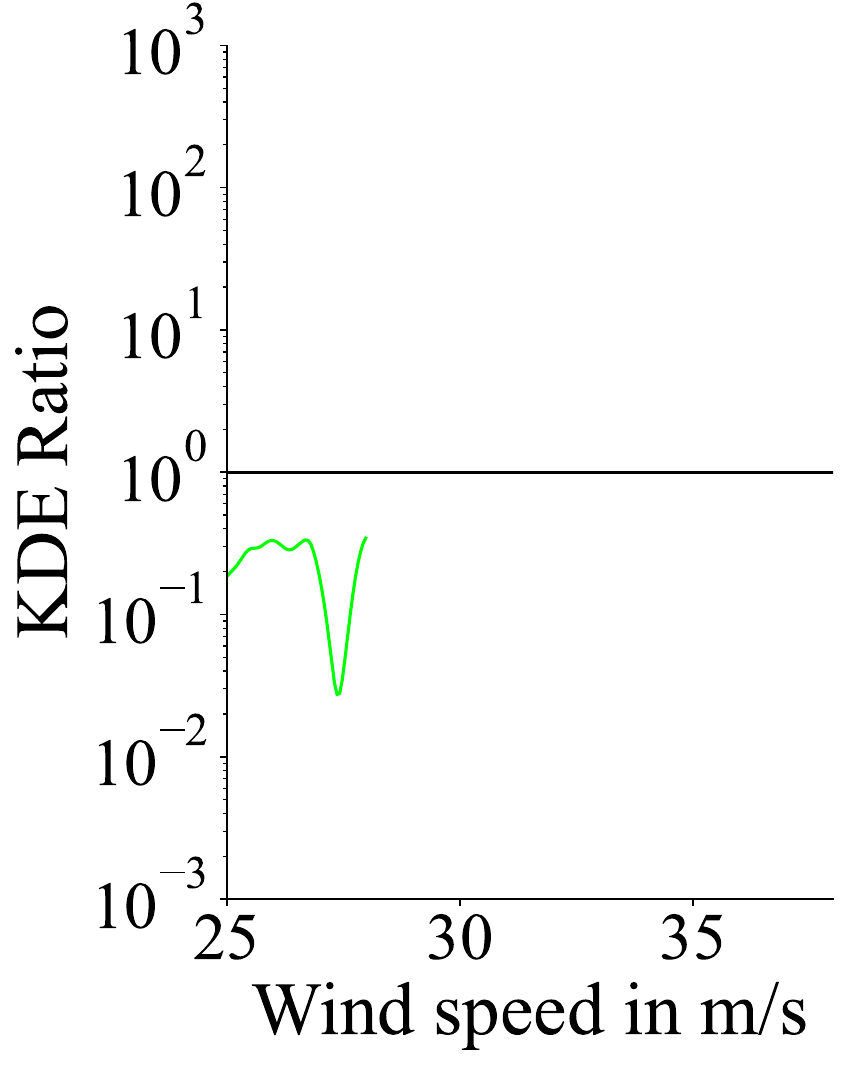}
        \caption{NCC-LR}
    \end{subfigure}
    \hfill
    \begin{subfigure}[b]{0.19\textwidth}
        \centering
        \includegraphics[width=3cm]{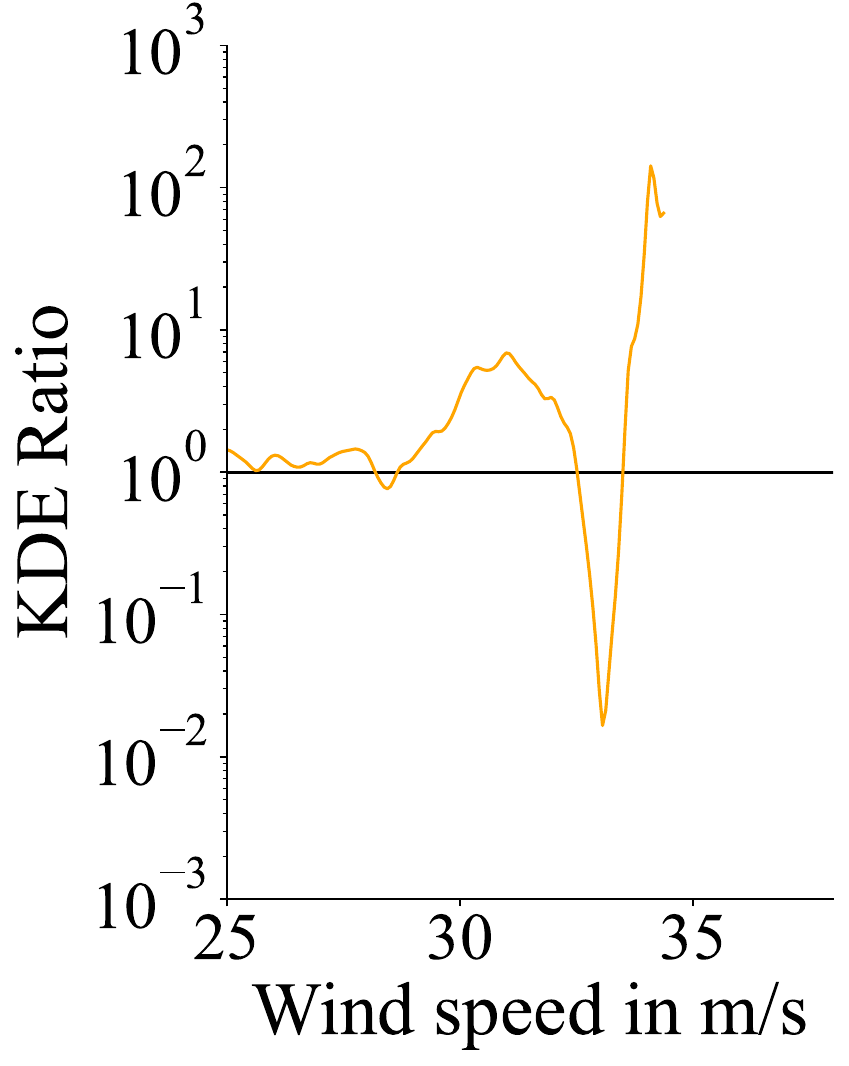}
        \caption{IPSL}
    \end{subfigure}
    \hfill
    \begin{subfigure}[b]{0.19\textwidth}
        \centering
        \includegraphics[width=3cm]{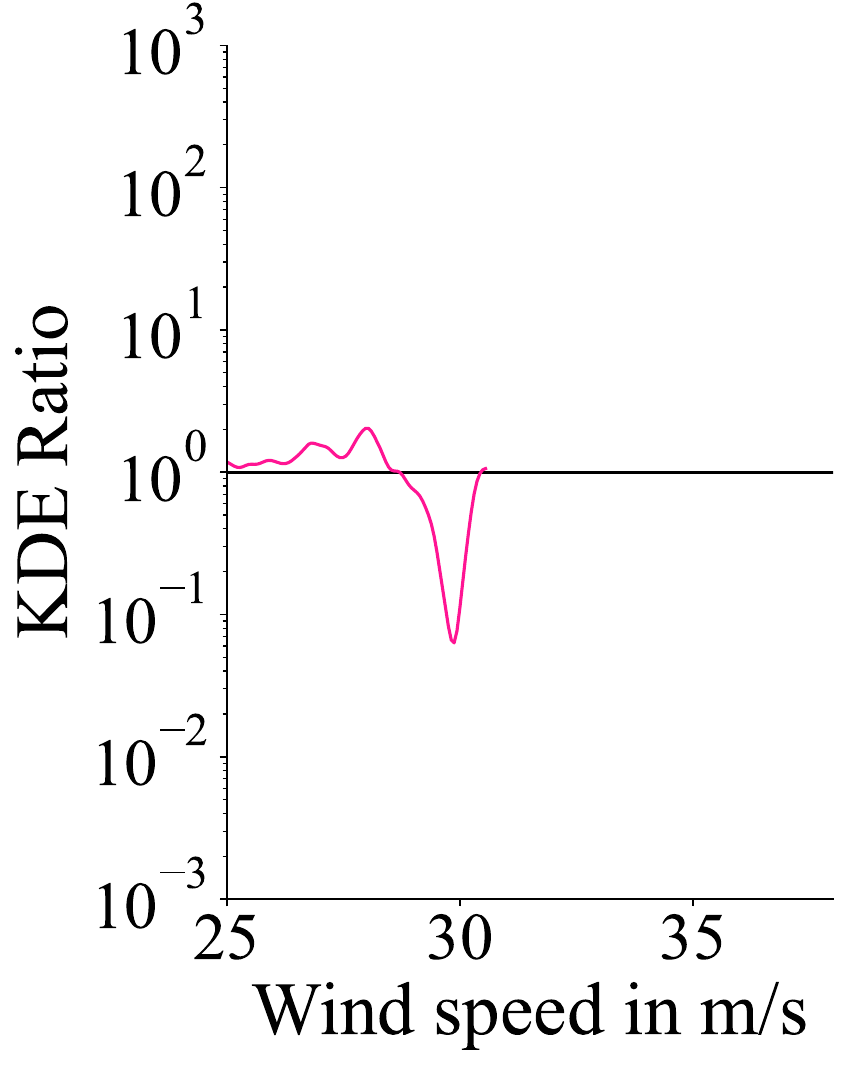}
        \caption{MPI-LR}
    \end{subfigure}
    \hfill
    \begin{subfigure}[b]{0.19\textwidth}
        \centering
        \includegraphics[width=3cm]{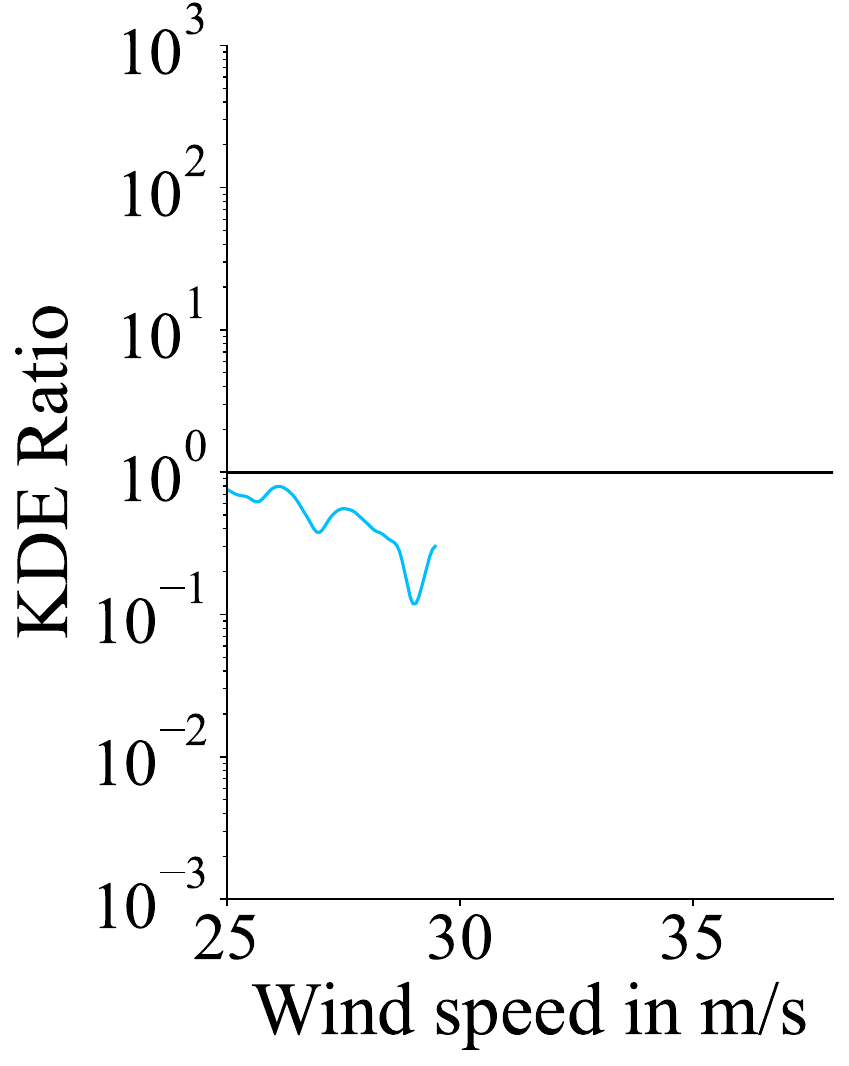}
        \caption{MOHC-LR}
    \end{subfigure}
    \hfill
    \begin{subfigure}[b]{0.19\textwidth}
        \centering
        \includegraphics[width=3cm]{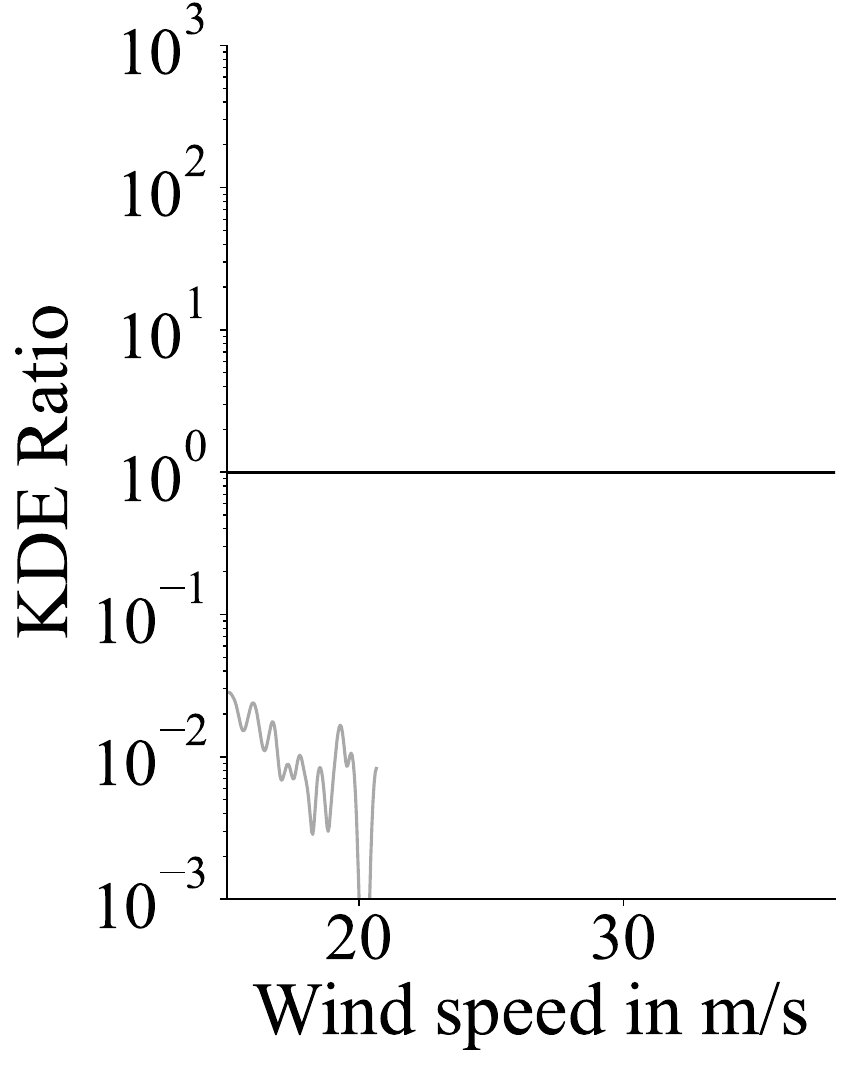}
        \caption{JAP}
    \end{subfigure}
    \par\bigskip
    \begin{subfigure}[b]{0.19\textwidth}
        \centering
        \includegraphics[width=3cm]{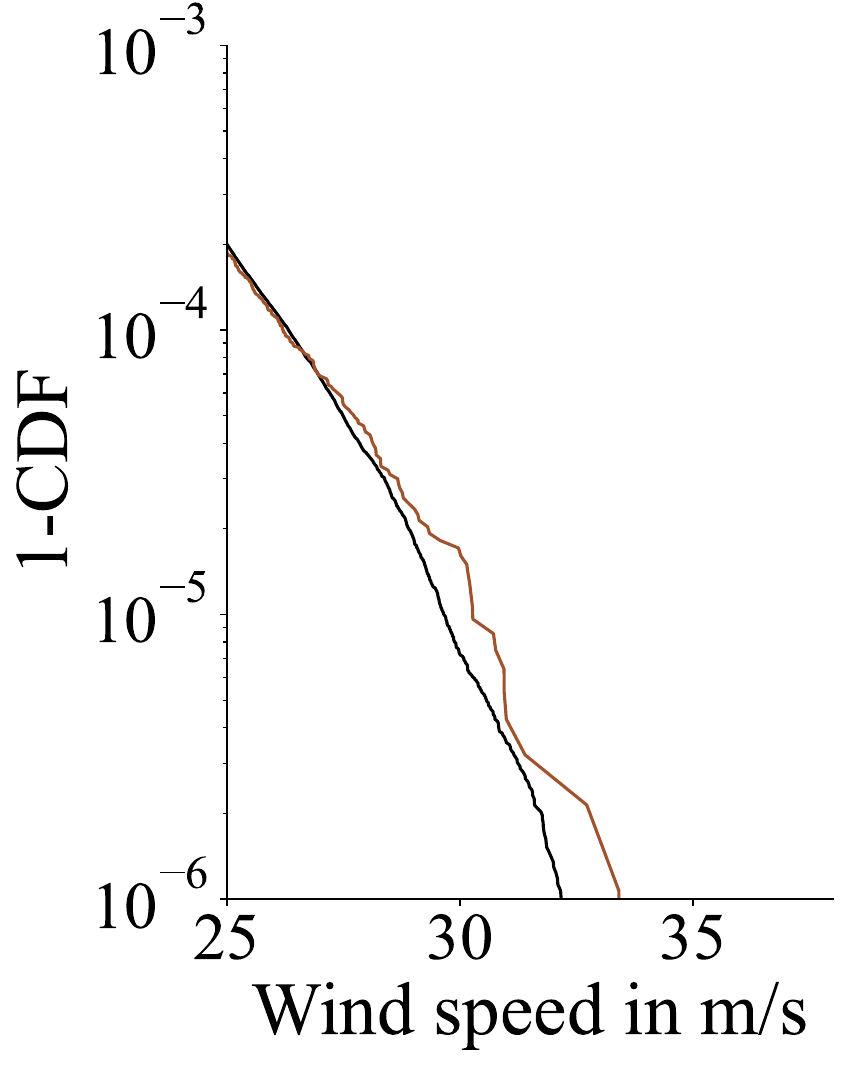}
    \end{subfigure}
    \hfill
    \begin{subfigure}[b]{0.19\textwidth}
        \centering
        \includegraphics[width=3cm]{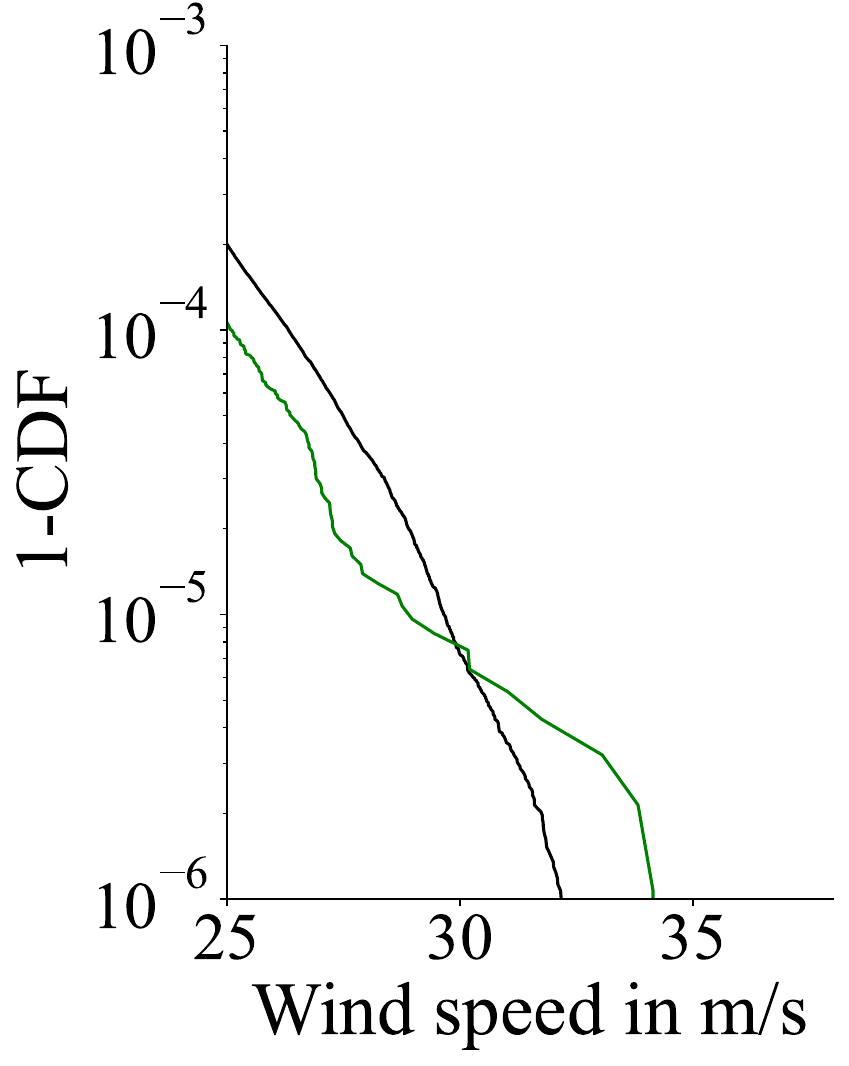}
    \end{subfigure}
    \hfill
    \begin{subfigure}[b]{0.19\textwidth}
        \centering
        \includegraphics[width=3cm]{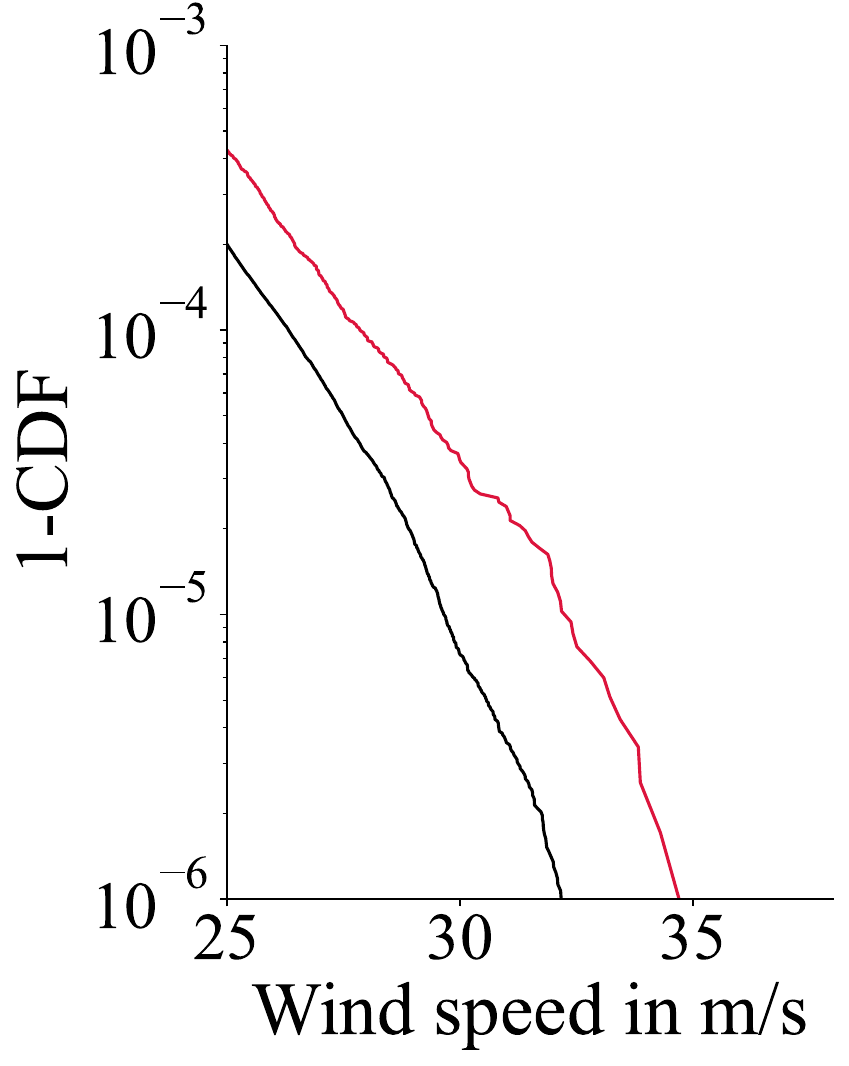}
    \end{subfigure}
    \hfill
    \begin{subfigure}[b]{0.19\textwidth}
        \centering
        \includegraphics[width=3cm]{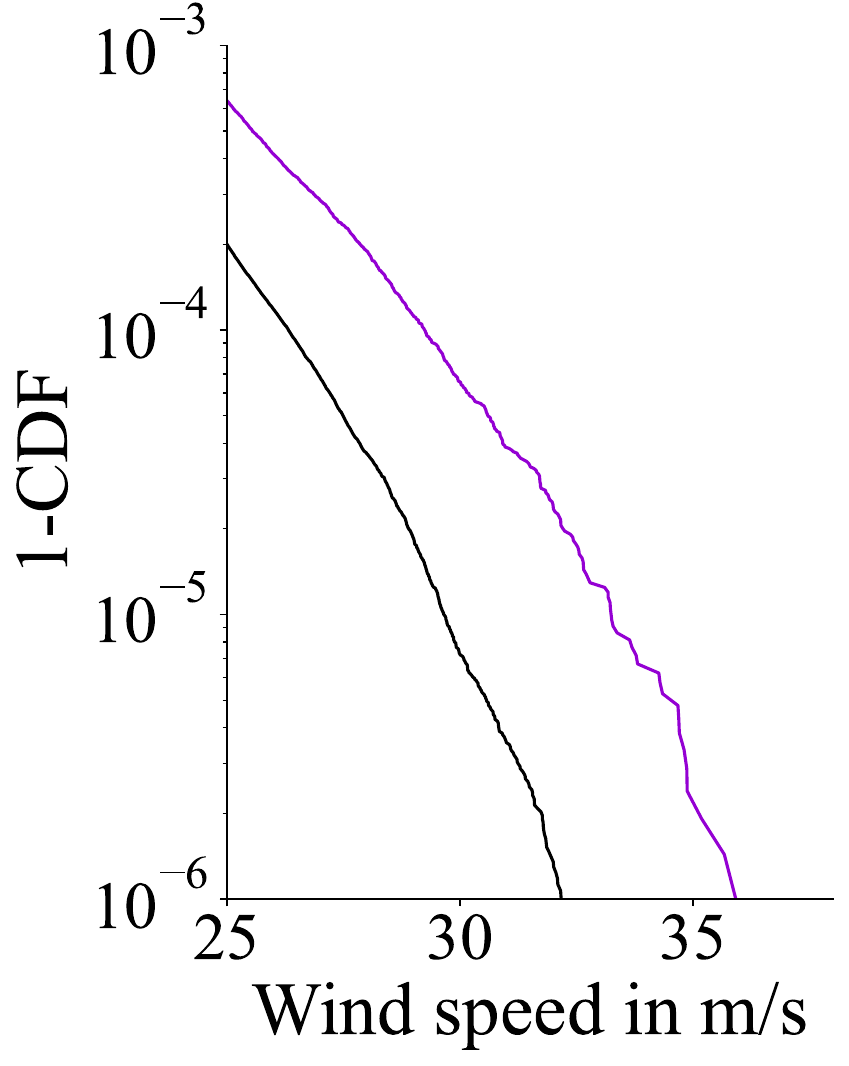}
    \end{subfigure}
    \hfill
    \begin{subfigure}[b]{0.19\textwidth}
        \centering
        \includegraphics[width=3cm]{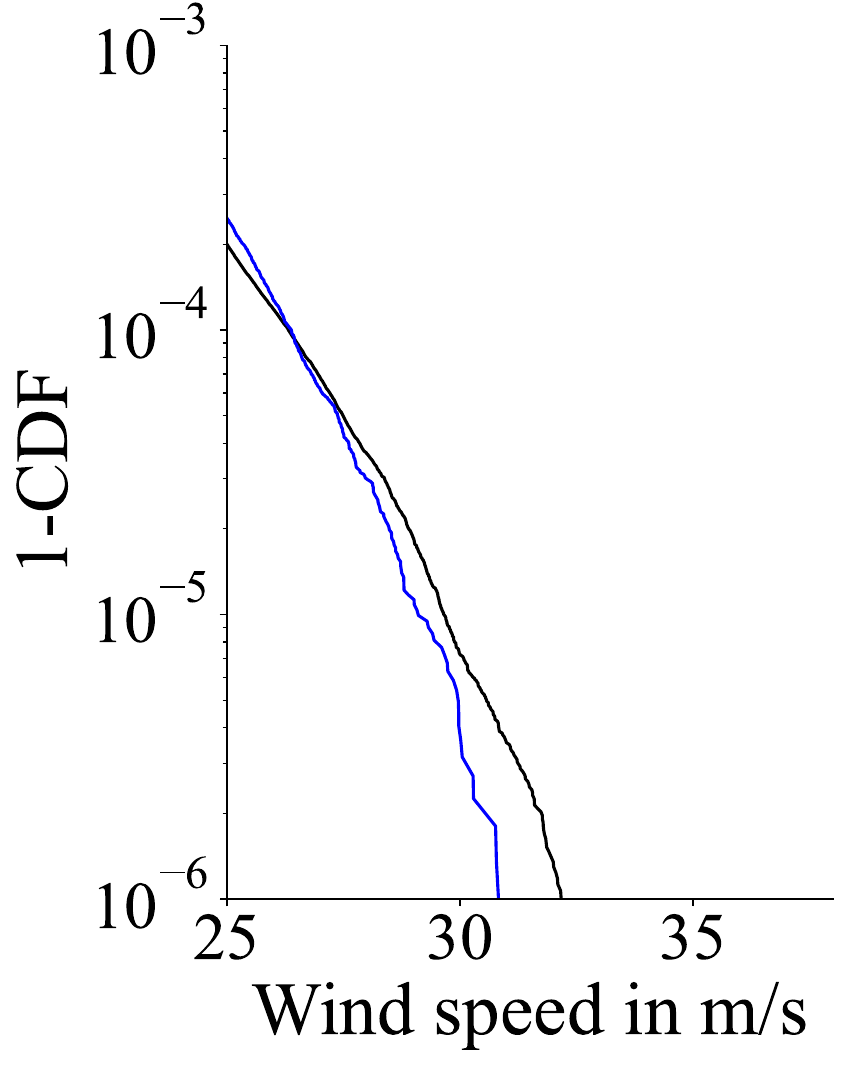}
    \end{subfigure}
    \par\bigskip
    \begin{subfigure}[b]{0.19\textwidth}
        \centering
        \includegraphics[width=3cm]{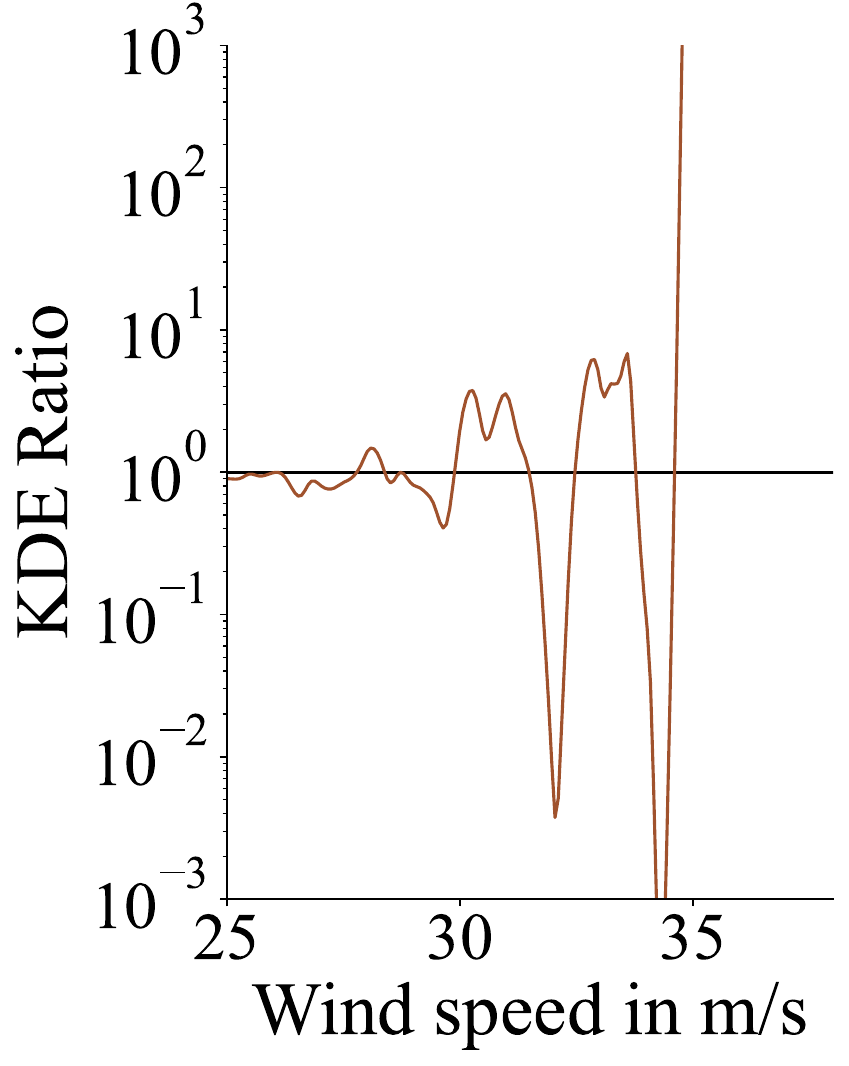}
        \caption{CMCC}
    \end{subfigure}
    \hfill
    \begin{subfigure}[b]{0.19\textwidth}
        \centering
        \includegraphics[width=3cm]{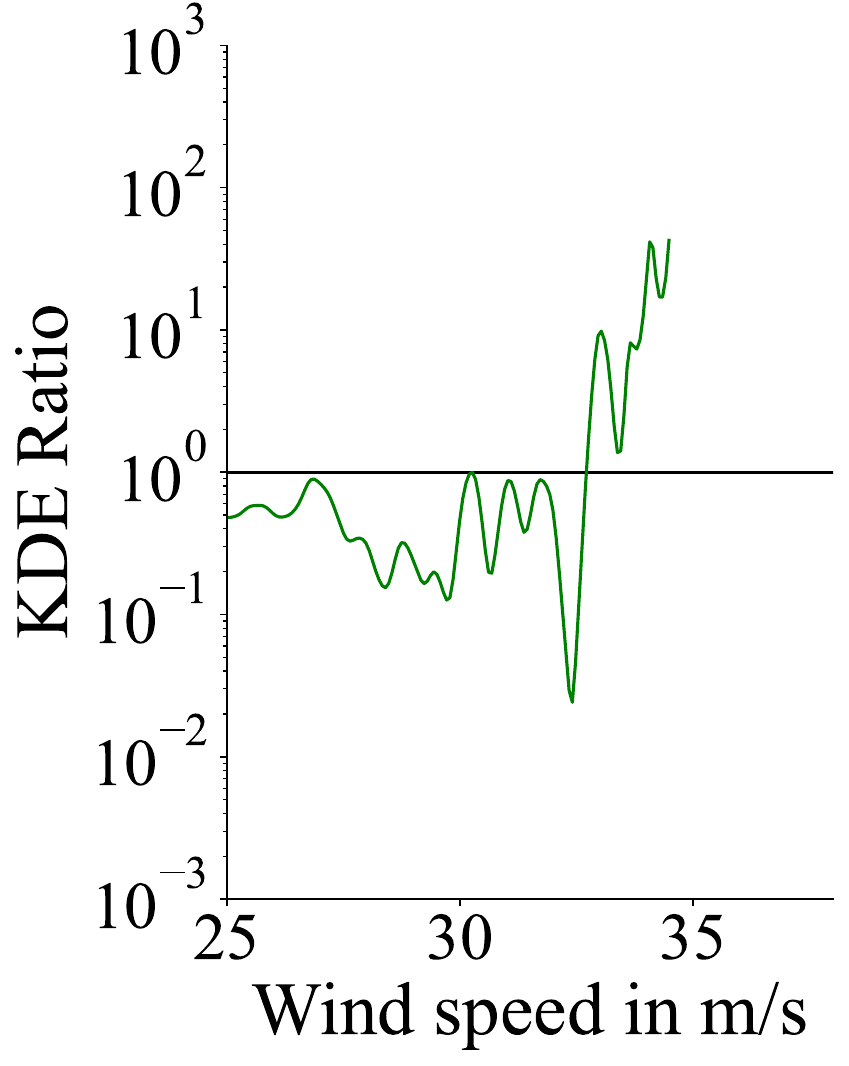}
        \caption{NCC-HR}
    \end{subfigure}
    \hfill
    \begin{subfigure}[b]{0.19\textwidth}
        \centering
        \includegraphics[width=3cm]{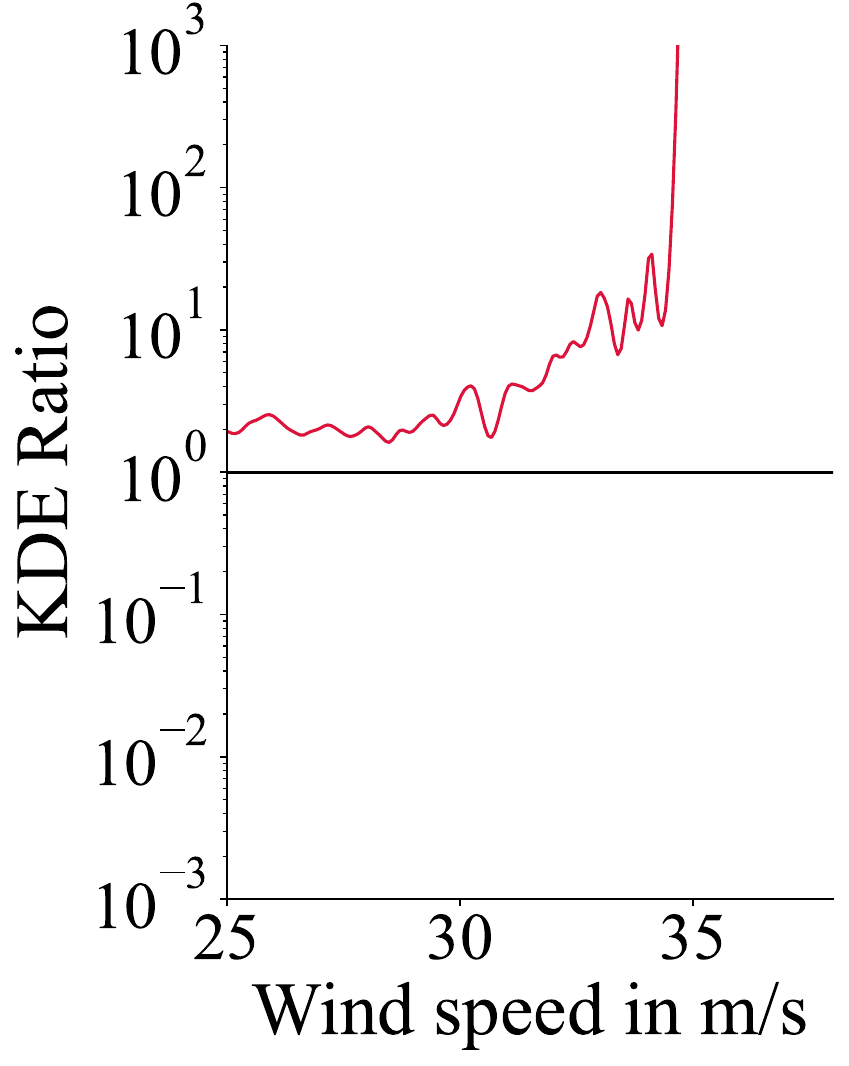}
        \caption{MPI-HR}
    \end{subfigure}
    \hfill
    \begin{subfigure}[b]{0.19\textwidth}
        \centering
        \includegraphics[width=3cm]{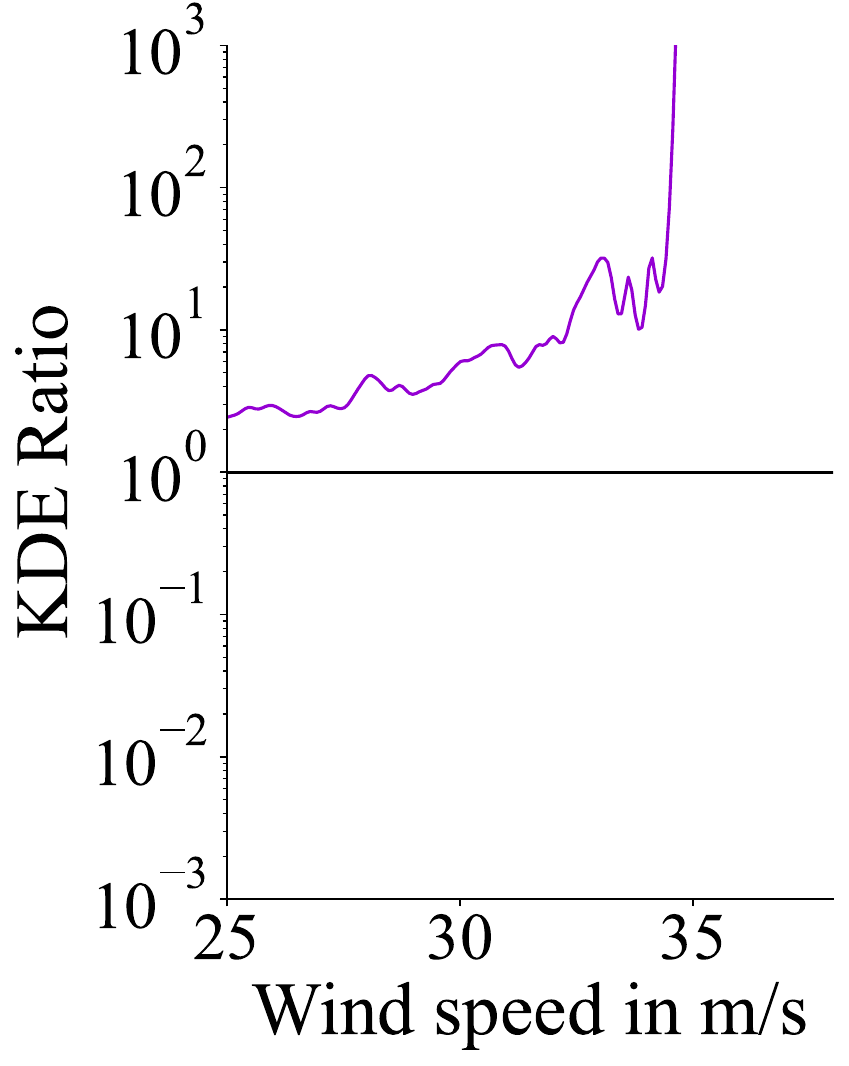}
        \caption{EC-EARTH}
    \end{subfigure}
    \hfill
    \begin{subfigure}[b]{0.19\textwidth}
        \centering
        \includegraphics[width=3cm]{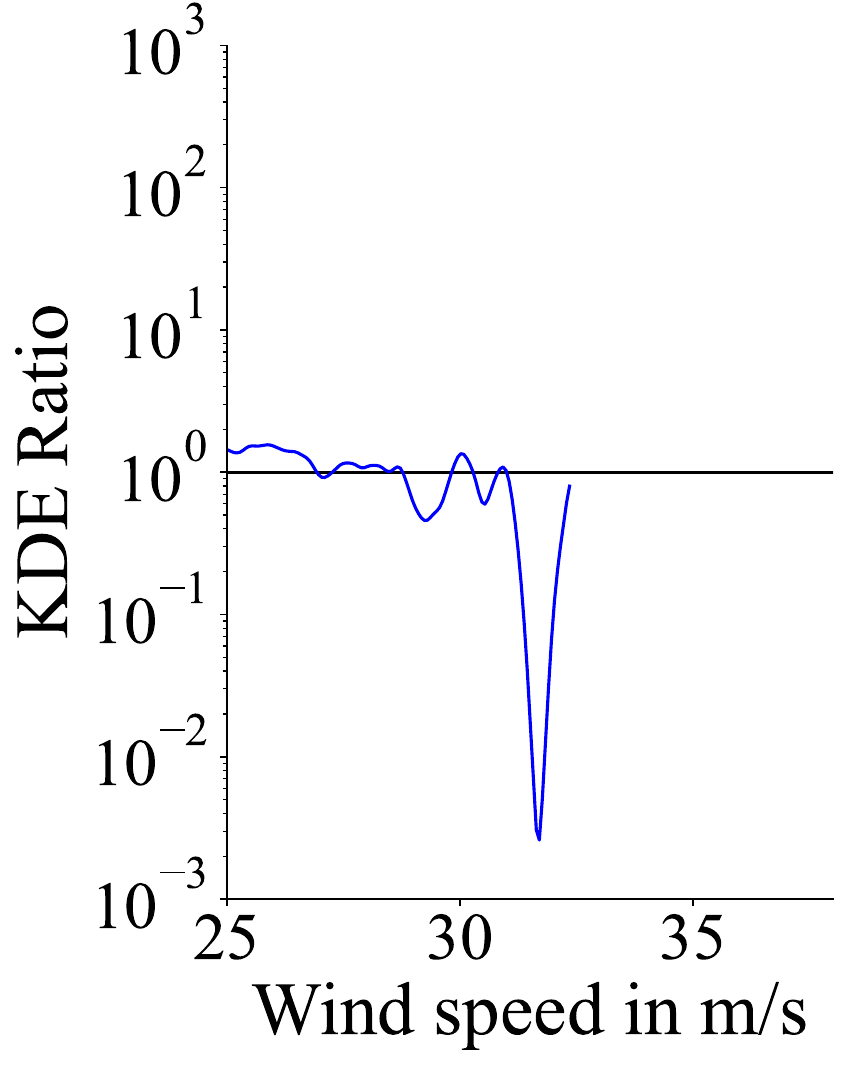}
        \caption{MOHC-HR}
    \end{subfigure}
    \caption{Survival function and the log-KDE-ratio visualizing discrepancies between CMIP6 GCMs and ERA5 in the upper tails of the wind speed distributions. In the first and third rows, the black line represents the survival function of ERA5, while the colored lines are the survival functions of the GCM data sets. In the second and last rows, the colored lines are the log-KDE-ratios, while the black line represents a ratio of $1$ indicating perfect alignment with ERA5. Data sets are sorted by spatial resolution from left to right.}
    \label{fig:tails}
\end{figure}

The tails of the wind speed distributions are compared to ERA5 in \Cref{fig:tails} using the survival function and the log-KDE-ratio. The survival function goes to zero when the estimated cumulative distribution becomes one, i.e.\ at the highest wind speed value of the sample. The log-KDE-ratio amplifies differences whose absolute value would be too small to see. However, it diverges to infinity when the highest wind speed value of ERA5 is reached. The two plots are therefore best interpreted together. 

They show that NCC-LR, MPI-LR, MOHC-LR, and especially JAP underestimate extreme wind speed events. The only low-resolution data set (plots in the first two rows) that captures wind speeds in the range of the highest ERA5 wind speed value is IPSL. The high-resolution models capture more extreme wind speed values.  MPI-HR and EC-EARTH even overestimate them clearly compared to ERA5.

\subsection{Model Performance}
\label{subsec:model_performance}

\begin{figure}[htp]
    \centering
    \begin{subfigure}[b]{0.32\textwidth}
        \centering
        \rotatebox{90}{
    \includegraphics[width=5.6cm]{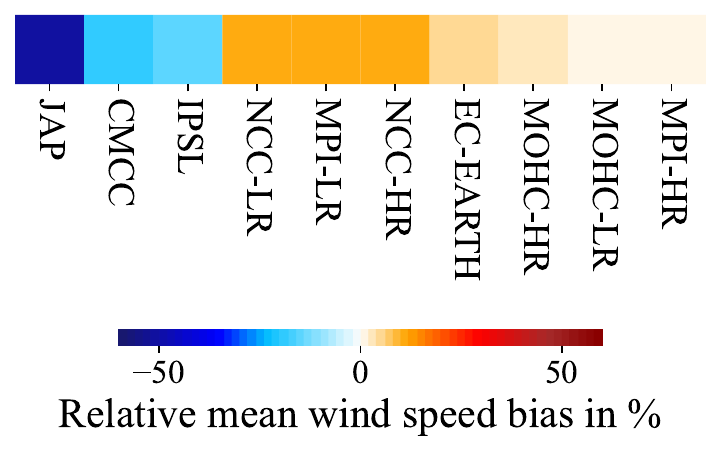}}
        \caption{Relative mean bias}
        \label{subfig:bias_mean_ws}
    \end{subfigure}
    \hfill
    \begin{subfigure}[b]{0.32\textwidth}
        \centering
        \rotatebox{90}{
    \includegraphics[width=5.6cm]{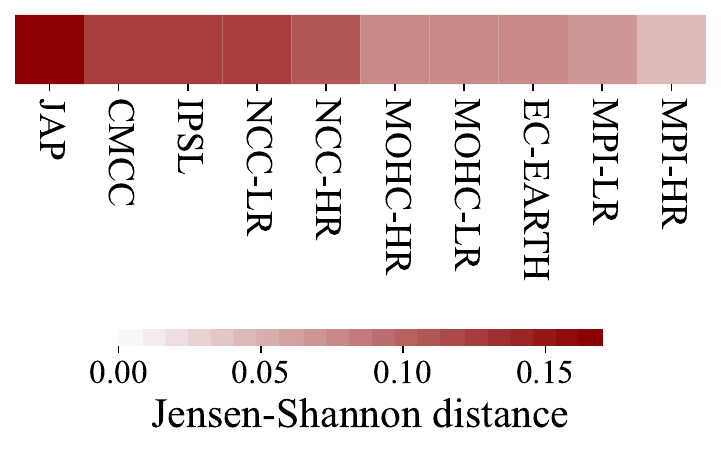}}
        \caption{Jensen-Shannon distance}
        \label{subfig:heatmap_js_div}
    \end{subfigure}
    \hfill
    \begin{subfigure}[b]{0.32\textwidth}
        \centering
        \rotatebox{90}{
    \includegraphics[width=5.6cm]{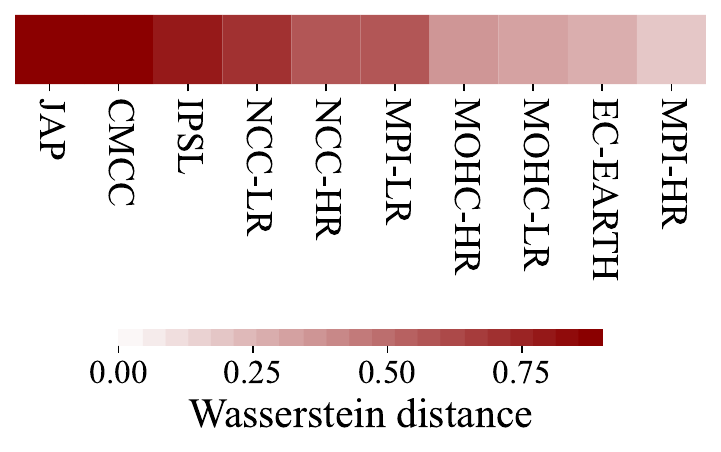}}
        \caption{Wasserstein distance}
        \label{subfig:heatmap_wasserstein}
    \end{subfigure}
    \caption{Model rankings. Relative mean wind speed biases, Jensen-Shannon (JS) and Wasserstein (WS) distances ordered by decreasing similarity to ERA5 (regardless of the direction of the bias).}
    \label{fig:f_distances}
\end{figure}

The mean (\Cref{subfig:bias_mean_ws}) is overestimated by the majority of GCMs compared to ERA5, whereas the models JAP, IPSL, and CMCC-MR exhibit a notable degree of underestimation.  The bias is smallest for MPI-HR ($0.8\%$) and largest for JAP ($-50.3\%$). 

The Jensen-Shannon and the Wasserstein distance estimating the difference between the CMIP6 GCMs and ERA5 are shown in \Cref{fig:f_distances}, for exact values see \Cref{tab:summary_statistics_germany_2005-2015}. Both indicate that the wind speed distribution of the MPI-HR model is most similar to that of ERA5. 

According to the Jensen-Shannon distance, the MPI-LR, EC-EARTH, and both MOHC models follow next, all with similar values, while the IPSL, CMCC, and both NCC models deviate more significantly. The Wasserstein distance suggests a slightly different ranking of these models. According to it, the EC-Earth and both MOHC perform only slightly worse than MPI-HR. MPI-LR, the NCC models, IPSL, and CMCC follow in that order.

Both measures agree that the JAP model exhibits by far the largest distance.

\section{Wind Power Prediction}
\label{sec:results_wpp}

For the wind power prediction, the original wind speed samples were interpolated to the turbine locations in Germany. \Cref{tab:mean_turbine_preprocessing} shows how this pre-processing step affects the mean wind speed values. The interpolation results in an increase in mean wind speed across all models. ERA5 and MPI-HR exhibit the smallest changes, with increases of $1.1\%$. All other data sets show increases exceeding $5\%$, with NCC-LR, CMCC, MOHC-LHR, and MOHC-HR demonstrating increases of more than $10\%$. 

\begin{table}[htp]
\caption{Comparison of wind speed values before and after interpolation to the turbine locations: The first two columns present the mean wind speed at the original grid and at the interpolated sites, respectively. The subsequent two columns illustrate the absolute and relative differences between the two, for each model.}
\label{tab:mean_turbine_preprocessing}
\centering
\begin{tabu}{c|cccr}
\toprule
& Mean original &  Mean turbine & Absolute & Relative   \\
& data &  locations & difference & difference   \\
\midrule
ERA5 & 3.73 & 3.77 & 0.04 & 1.1\% \\
MOHC-HR & 3.84 & 4.25 & 0.41 & 10.7\% \\
EC-EARTH & 3.95 & 4.28 & 0.33 & 8.4\% \\
MPI-HR & 3.77 & 3.81 & 0.04 & 1.1\% \\
NCC-HR & 4.11 & 4.49 & 0.38 & 9.2\% \\
CMCC & 3.01 & 3.39 & 0.38 &  12.6\% \\
MOHC-LR & 3.78 & 4.25 & 0.47 & 12.4\% \\
JAP & 1.86 & 1.96 & 0.10 & 5.4\% \\
MPI-LR & 4.14 & 4.35 & 0.21 & 5.1\% \\
IPSL & 3.17 & 3.39 & 0.22 & 6.9\% \\
NCC-LR & 4.17 & 4.61 & 0.43 & 14.6\% \\
\bottomrule
\end{tabu}
\end{table}

\begin{figure}[htp]
    \centering
    \begin{subfigure}[b]{0.64\textwidth}
        \centering
        \includegraphics[height=6cm]{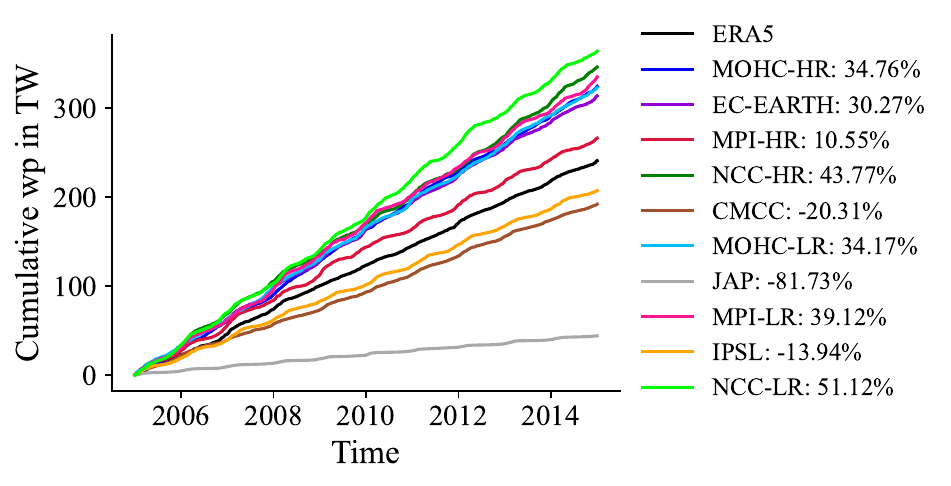}
        \caption{Cumulative wind power}
        \label{subfig:cumulative_wp}
    \end{subfigure}
    \hfill
    \begin{subfigure}[b]{0.34\textwidth}
        \centering
        \rotatebox{90}{
    \includegraphics[width=6cm]{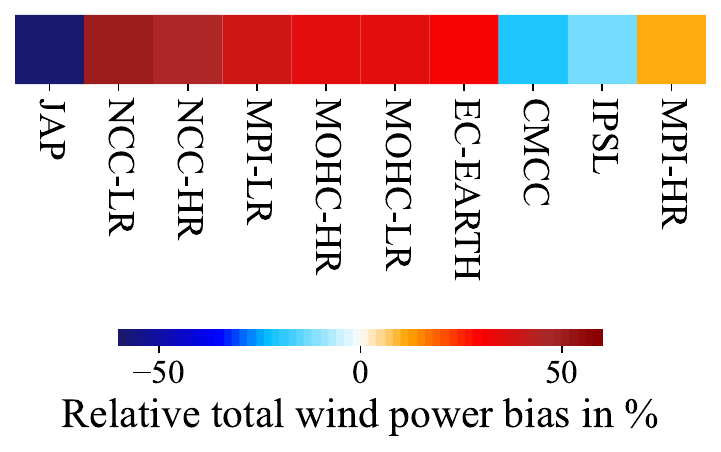}}
        \caption{Relative bias to ERA5}
        \label{subfig:wp_biases}
    \end{subfigure}
    \caption{Wind power comparison. \Cref{subfig:cumulative_wp} shows the predicted cumulative wind power from 2005 to 2015. The black line represents the ERA5 reference data set, against which the GCM predictions are compared. The differences in total wind power at the end of the period are indicated as relative biases to ERA5 in percent. \Cref{subfig:wp_biases} visualizes these biases, with warm colors indicating positive biases and cool colors indicating negative biases. The data sets are sorted by increasing bias, regardless of direction.}
    \label{fig:wp_comparison}
\end{figure}

\Cref{fig:wp_comparison} shows the cumulative curve for 2005 to 2015 and the biases of the total accumulated wind power at the end of this period.
Most models show steeper slopes compared to the ERA5 reference data set, leading to an overestimation of wind power. For instance, NCC-LR, NCC-HR, MOHC-HR, EC-EARTH, MOHC-LR, MPI-LR, and MPI-HR all predict higher total wind power than ERA5, with biases up to $51.12\%$. Conversely, a few models, including IPSL, CMCC, and JAP, display flatter slopes relative to ERA5, resulting in an underestimation of wind power. JAP stands out with the most significant negative bias of $81.73\%$, followed by CMCC with a bias of $-20.31\%$.
MPI-HR and IPSL capture the trend of ERA5 more accurately than other models, resulting in the smallest total wind power biases of $10.55\%$ and $-13.94\%$, respectively.

\chapter{Discussion}\label{chapter:discussion}
The discussion is divided into four parts. First, the methodology is evaluated in \Cref{sec:dis_procedure}. Second, the primary research question regarding the impact of spatial resolution on simulated wind speed distributions and wind power predictions is analyzed in \Cref{sec:dis_spatial_resolution}. Third, additional aspects are examined in \Cref{sec:dis_best_model_performance}, with a focus on identifying the most appropriate metric to measure the model performance and the most accurate wind speed data set for the German region. Finally, open questions for future research are discussed in \Cref{future_work}.

\section{Review on the Methodology}
\label{sec:dis_procedure}

Evaluating which spatial resolution is needed when using global climate model data for multi-decadal wind power predictions presents various difficulties (see \Cref{chapter:research_question}). Part of this work was to develop a procedure that meets these challenges. The chosen approach consisted of the following steps:
\begin{enumerate}
    \item Data selection: 6-hourly near-surface wind speed samples from 2005 to 2015 in a rectangular region covering Germany, derived from historical first ensemble member runs of ten CMIP6 GCMs, extrapolated to a hub height of \SI{100}{\m}
    \item Choice of reference for validation: reanalysis data set ERA5
    \item Comparison of wind speed distributions using various metrics: Kolmogorov-Smirnov tests, kernel density estimates, summary statistics, fits with exponentiated Weibull distribution, Jensen-Shannon and Wasserstein distances
    \item Wind power predictions for bias estimation: interpolation to wind turbine locations in Germany according to \cite{turbineLocations} and transformation of wind speed with power curve
\end{enumerate}
The reasons for the choices made, assumptions, and neglected aspects are discussed in the following.

\subsection{Data Selection}
\label{sec:dis_data}

Wind speed data were derived from CMIP6 GCMs because multi-decadal wind power predictions are typically based on climate projections and CMIP6 currently provides the most recent generation of GCMs. Dynamical downscaling products were not considered because the latest RCMs use the predecessor CMIP5 GCMs as boundary models, potentially introducing outdated assumptions or biases. Statistical downscaling methods were not considered because of the variety of competing techniques and the primary objective of assessing the quality of the original data. This approach allows for an analysis of the inherent characteristics of global model output while the value of downscaling techniques is a separate question not addressed in this thesis.

The analysis in \Cref{sec:run_comparison} showed that the output discrepancy between runs with different initial/boundary conditions of identical models at the same resolution was relatively small. However, averaging over multiple runs resulted in an over-representation of low wind speed values and an under-representation of extremely high wind speed values. Based on these results, a single run per model was chosen. Historical model runs were selected, as these are the only ones that can be validated through comparison with past weather data. 

The use of 6-hourly instantaneous wind speed values aligns with recommendations from \citet{effenberger2023mind} and can be considered a robust choice for temporal resolution. Considering a 10-year period was deemed sufficient to reflect typical wind patterns, particularly given the use of historical data, which is generally more reliable than future projections.

Germany was selected as the area under investigation, primarily due to the availability of turbine location data. The choice of a relatively small region reduces the risk of region-specific structures being averaged out, yet constrains the scope for generalization. Consequently, the results are specific and only reliably applicable to Germany. It is important to note that a model's performance in Germany does not necessarily indicate that it would perform similarly in other regions. However, this geographical focus is not necessarily a limitation. Rather, it acknowledges that different models may have different strengths and focuses.

The specific characteristics of the region under consideration include a temperate climate with pronounced seasonal variations. In addition, Germany has distinctive wind speed patterns with a marked decrease from North to South, influenced by the flat coastal terrain in the North and mountainous inland regions in the South. This trend is particularly evident due to the inclusion of offshore areas. 

To preserve the original spatial resolution of the data, I evaluated the wind speed samples on their native grids. This approach resulted in variations in the extent of the area under consideration between different models, depending on cell size and the centering of each model's coordinate system. Coarser grids, in particular, tended to over- or under-represent certain regions within the area of interest. For example, MIP-LR covered a notably large area, while JAP encompassed a comparatively small area relative to other models.

This issue could potentially be addressed by implementing a more sophisticated method for defining the data range. Instead of simply checking whether a point falls within the region's bounds, a weighting scheme could be employed for grid cells that only partially cover the area under consideration. This approach would provide a more nuanced representation of the region of interest across different model resolutions. However, it would also introduce additional complexity to the analysis.

As an additional pre-processing step, I extrapolated the near-surface wind speeds to a generic hub height of \SI{100}{\meter} with the theoretical wind profile power law at stable conditions. According to \citet{exponent_power_law}, this law provides "realistic but conservative estimates of available wind power, except at extremely rough sites where the estimates can only be conservative". It is therefore appropriate in this context and has the advantage of affecting the wind speed values consistently and straightforwardly: shifting the wind speeds and thus the critical wind speed range by a linear factor. However, this means that choosing a different hub height could alter the results, particularly the model rankings.

\subsection{Choice of Reference}
The challenge of determining the \textit{ground truth} requires careful consideration.  Although the selected reanalysis data set (ERA5) was identified as the best alternative for Germany based on a review of the literature (see \Cref{subsec:era5}), it may still contain biases as some sources have indicated. Thus, basing the analysis on the original ERA5 data represents a strong assumption. The results are, however, supported by the comparison of the near-surface wind speed biases identified in this study between GMCs and ERA5 data with those reported in \citet{NSWSbiases} between GMCs and observations, as detailed in \Cref{appendix:model_biases}.

\subsection{Wind Speed Distribution Comparison}
\label{sec:dis_ws_comaprison}
Given the non-linear relationship between wind speed and wind power, certain wind speed ranges are more relevant for predicting wind power than others. One might conclude that a comparison of the wind power forecasts would thus be most informative. 
However, the prediction process involves numerous assumptions, particularly regarding the power curve's precise shape, which depends on the selected turbine type. This dependency can shift the relevant wind speed ranges. 

Consequently, the most robust wind power predictions are achieved when models accurately capture the entire wind speed distribution. Nonetheless, it is reasonable to pay special attention to some specific ranges, such as those beyond the typical cut-in wind speed and extreme values. To address this, a comprehensive analysis of the wind speed distributions was conducted using various measures to evaluate both the overall agreement in model output and differences in critical ranges.

\subsection{Wind Power Prediction}
\label{sec:dis_wp_prediction}

While the main focus of this work is wind speed, a rudimentary wind power prediction was performed in the final evaluation step. It serves as an illustration of the model bias behavior when the non-linear relationship between wind speed and wind power is taken into account.

The prediction relies on several simplifying assumptions and cannot be considered realistic. I assume a single turbine type and the same hub height at all locations, and thus a power curve that is specific to these choices. Different power curves could alter the wind power prediction and shift the critical wind speed ranges. However, it is beyond the scope and not the goal of this thesis to investigate how this factor affects the results.

Comparing the predicted wind power with actual energy production is not feasible for several reasons. Firstly, assuming a single turbine type at all locations is inadequate given the diversity of wind turbines in use. Secondly, the uniform hub height of \SI{100}{\meter} does not account for the variations in hub heights across different turbines. Finally, the theoretical power curve does not accurately reflect real-world conditions, as discussed in \Cref{subsec:wpp_challenges}. Consequently, I continued to consider ERA5 as the reference.

\section{Impact of the Spatial Resolution}
\label{sec:dis_spatial_resolution}

The results in \Cref{sec:ws_sample_comparison} show that the overall shape of the wind speed distributions varies more between models than between spatial resolutions, indicating the presence of model biases. The only exception is the extreme values, which are better captured by high-resolution models.

Following \citet{effenberger2023mind}, I used  Kolmogorov-Smirnov tests to identify wind speed samples originating from the same distribution. However, no significant agreement was found, indicating that the wind speed samples generally differ greatly. Consequently, the kernel density estimates (KDEs) exhibit considerable variation in terms of the location of the mode, skewness, and width of the distribution.

The two KDEs that look the most alike and whose comparison yielded the comparatively largest $p$-value in the KS tests originate from the same model at different spatial resolutions.  This suggests that spatial resolution plays a minor role in shaping the wind speed distribution compared to the influence of the model itself. This is consistent with the findings of \cite{zhang2024quantify}, who identified the model as the largest contributor to the uncertainty in surface wind speed projections with CMIP6 GCMs.

For most descriptive measures, I did not detect any trends related to spatial resolution. The variance, skewness, and kurtosis are not particularly informative in this context, nor are the fit parameters of the exponentiated Weibull distribution.

An increase in mean wind speed with higher resolution, as reported in \citet{SpatialResoltionRCMs}, was also not observed. On the contrary, the mean wind speed was lower for MPI-HR and NCC-HR and only slightly higher for MOHC-HR compared to the corresponding low-resolution data sets. 
Moreover, these biases are small compared to the ones between data originating from different models. This suggests that model-induced mean biases are prevalent.

Examining the KDE differences between the GCMs and ERA5 in the medium wind speed range confirms the presence of model-specific biases. The patterns vary greatly between data sets from different models, while they are more similar for low- and high-resolution data sets of the same model. This similarity is particularly high for the NCC and MOHC models, whereas it is slightly less pronounced for the MPI model.

A certain degree of consistency of the mean biases is supported by comparing the results with those of \citet{NSWSbiases}, as presented in \Cref{appendix:model_biases}. The comparison shows that the GCM models exhibit similar mean biases for the European region in the period 1978 to 2014 compared to observations, as they do for the German region in the period 2005 to 2015 compared to ERA5.

There was one parameter for which I could identify a trend with the spatial resolution: the maximum wind speed increases with finer resolutions for the considered model outputs. This is consistent with the findings of \citet{SpatialResoltionRCMs} and \citet{ResolutionDependenceofExtremeWindSpeedProjections} who detected a rise in extreme wind speeds with increasing resolution.

Another observation aligns with this trend for extreme wind speeds: the higher-resolution models perform better at representing the upper tail of the wind speed distribution, as shown by the KDE ratio and survival function plots. This is in line with \citet{AddedValueCordex}, who found that the benefits of dynamical downscaling are greater in the upper tail.

In conclusion, the necessity of high-resolution models for accurately representing extremely high wind speeds can be considered a robust finding.

The results presented in \Cref{sec:results_wpp} reveal an additional, indirect influence of spatial resolution on wind power prediction. When wind speeds are interpolated to turbine locations, mean values increase across all models, with the effect being minimal for ERA5 and MPI-HR. This phenomenon is likely attributable to the larger grid cells in GCMs. The interpolation is based on neighboring grid cells. Consequently, low-resolution models with a wider spread of data points incorporate data from larger areas and more distant locations. In the North of Germany, where most turbines are located, the neighboring region is the ocean. Thus, low-resolution models more frequently incorporate high wind speeds from offshore areas in the interpolation process. Additionally, the effect is influenced by the center of the coordinate system, as it determines how many data points considered in the interpolation are placed above the ocean. This observation underscores the importance of spatial resolution and grid choice as a significant factor in wind power forecasting, particularly in coastal regions.

\section{Identifying the Best-Performing Model}
\label{sec:dis_best_model_performance}

Given that the model itself exerted the greatest influence on the GCM wind speed output, and trends related to spatial resolution were only identified for the upper tail of the distribution, this work cannot provide a definitive answer to the question of optimal resolution for multi-decadal wind power forecasting. Consequently, the study shifted focus to an in-depth examination of individual model performance for the region under consideration, namely Germany.

Ranking the models' performance is not straightforward. While the total wind power estimate should ideally be close to that of ERA5, this metric alone is insufficient, as results could change significantly when using a different power curve. Therefore, several metrics were considered to examine the proximity of each model to the ERA5 wind speed distribution. These include a simple summary statistic, the relative mean bias, as well as two probability distances: the Jensen-Shannon distance and the Wasserstein distance.

The relative mean bias, a simple summary statistic, is calculated by dividing the mean of the GCM wind speed sample by the mean of the ERA5 data. This metric effectively summarizes the difference patterns in the medium wind speed range. For instance, models like IPSL, CMCC, and JAP, which underestimate low wind speeds and overestimate high wind speeds compared to ERA5, show negative mean biases. Conversely, models with the opposite pattern exhibit positive biases.

The Jensen-Shannon (JS) distance is a symmetric and bounded metric that focuses on the overlap of the distributions. It is calculated by taking the average Kullback-Leibler distance between each distribution and their mixture distribution, punishing differences in the probability mass within specified bins.
The Wasserstein distance measures the minimum cost of transforming one distribution into another, considering the entire distribution's shape at once. It is robust to small perturbations in the data and particularly effective in capturing differences in the distributions' tails.

All three metrics consistently identify MPI-HR as the top performer and rank NCC-LR, IPSL, CMCC, and JAP as the lowest performers in descending order, with JAP standing out as the worst-performing model. They also all indicate that MPI-HR significantly outperforms MPI-LR, NCC-HR outperforms NCC-LR, and MOHC-LR slightly underperforms compared to MOHC-HR. However, the exact placement of medium-performing models in the ranking varies by metric. For example, mean bias ranks MOHC-LR second, Jensen-Shannon distance favors MPI-LR, and Wasserstein distance places EC-EARTH in the second position. 

When considering the performance according to the total wind power estimate another change in ranking occurs. This can be attributed to the shift towards higher wind speeds in the GCM data when interpolated to the turbine locations. The MPI-HR data set continues to demonstrate the greatest predictive efficacy, with a slightly higher projected increase in cumulative wind power, resulting in an overestimation of $10.55\%$ compared to ERA5 after 10 years. However, the second and third best-performing models in terms of cumulative wind power prediction are IPSL and CMCC, as their underestimation is reduced. In this way, IPSL provided the second smallest total wind speed value of $-13.94\%$.

Disregarding the better performance of underestimating models due to interpolation to turbine locations, the Wasserstein distance ranks the models most similarly to the total wind power bias. Nevertheless, the mean wind speed bias also serves as a surprisingly good indicator of wind power bias, despite its simplicity. 

Synthesizing the results from various metrics and analyses, MPI-HR consistently performs best and JAP worst compared to ERA5 in terms of both wind speed distribution agreement and wind power estimates. This significant performance gap could stem from the models' divergent focus. JAP's developers prioritize improvements in modeling tropical climate systems and midlatitude atmospheric circulation \citep{JAP_paper}. In contrast, the MPI high-resolution model specifically addresses biases in upper-level zonal winds and jet stream positions in the northern extratropics. This focus results in more accurate representations of North Atlantic storm tracks and European weather blocks\footnote{Weather blocks are persistent high-pressure systems that disrupt typical weather patterns, often leading to extended periods of consistent weather conditions.} \citep{MPI-HR_paper}, which are particularly relevant for German weather patterns. Additionally, JAP uses the Japanese 55-year Reanalysis (JRA55) as a reference \citep{Miroc5_JRA55}. This data set differs in wind speed output compared to ERA5 \citep{reanalysis_comparison}, which might contribute to JAP's poorer performance in this comparison.

Other models in the study fall between these two extremes, with IPSL being the second-best model according to the total wind power predictions. However, it should be emphasized that the comparison is based on the original data sets. A bias correction \citep[e.g.][]{bias_correction} could potentially reduce the differences and improve the wind speed accuracy across all models.

\section{Future Work}
\label{future_work}

To assess the generalizability of the findings, extending the analysis to other regions would help determine whether the observed trends and biases are consistent across different geographic locations or whether region-specific factors influence the results. This expanded geographic scope should include areas with different topographic and climatic characteristics to assess the extent to which the results are applicable in different environments. In addition, examining seasonal trends within these regions could reveal important patterns in model performance, as some models may excel at predicting winter wind patterns but struggle with summer conditions, or vice versa. Understanding these seasonal variations could lead to more nuanced model selection and application based on the time of year.

Another crucial aspect to investigate is the impact of different turbine types and hub heights on power output predictions, as variations in turbine specifications can significantly affect the results. This analysis should be coupled with an examination of wind speed distribution shifts over long time scales, considering future climate scenarios in addition to the historical model runs. Such an approach is essential for understanding the potential impacts of climate change on wind patterns and energy production, assessing how well different models capture these long-term shifts, and providing insights into the reliability of wind power under various future climate conditions.

To refine the current approach, exploring downscaling techniques could be valuable. This should include both statistical and dynamical methods, each offering unique advantages in bridging the gap between global model outputs and the finer spatial resolution needed for accurate local predictions. Comparing the efficacy of these downscaling approaches in different contexts could reveal optimal strategies for mitigating the weaknesses of global models and enhancing region-specific predictions. Alongside downscaling, exploring various bias correction methods — both statistical and physical — could further improve model accuracy. Evaluating these techniques across different models, regions, and timescales would help identify the most effective ways to reduce systematic errors and enhance overall prediction quality.

By integrating these additional aspects - geographical expansion, seasonal analysis, long-term distribution shifts including future scenarios, downscaling techniques, and bias correction methods - the study would provide a more comprehensive understanding of model performance and potential improvement strategies.

\chapter{Conclusion}\label{chapter:conclusion}
This study assessed the impact of the spatial resolution of Global Climate Models (GCMs) on the accuracy of simulated wind speeds for multi-decadal wind power predictions. The analysis focused on historical runs of ten CMIP6 GCMs, examining 6-hourly near-surface wind speed outputs from the first ensemble member over the period 2005 to 2015 in Germany extrapolated to a hub height of \SI{100}{\meter}. Validation of the models was conducted using the ERA5 reanalysis data set.

The analysis employed various statistical methods, including Kolmogorov-Smirnov tests, Jensen-Shannon and Wasserstein distances, summary statistics, and fits of the data to the exponentiated Weibull distribution, alongside visualizations like kernel density estimates (KDEs). Wind power was estimated using the Enercon-115 turbine's power curve, after interpolating wind speeds to turbine locations. 

The results show a complex relationship between model resolution and wind power forecast accuracy, with no single resolution emerging as optimal for gridded GCM data. Kolmogorov-Smirnov tests revealed significant differences across all wind speed samples. In particular, deviations in wind speed distributions were more pronounced between models than between resolutions, suggesting that model choice has a greater impact on wind speed output than spatial resolution.

Further analysis using Jensen-Shannon and Wasserstein distances showed a trend of decreasing distance with increasing resolution, though this was not statistically significant. Interestingly, some low-resolution models agreed better with ERA5 than their high-resolution counterparts, underscoring the importance of model-specific characteristics beyond resolution.

Contrary to some previous studies, an increase in mean wind speed with higher resolution was not consistently observed. In some cases, high-resolution models (MPI-HR and NCC-HR) produced lower mean wind speeds than their low-resolution equivalents. Other summary statistics and fit parameters of the exponentiated Weibull distribution also did not exhibit a clear trend with resolution.

However, high-resolution models consistently performed better in capturing extreme wind events. Maximum wind speeds increased with finer resolutions, and high-resolution models demonstrated superior capability in representing the upper tail of the wind speed distribution. This indicates that the benefits of higher resolution are most pronounced for extreme wind events.

When interpolating wind speeds to turbine locations, an additional indirect influence of spatial resolution emerged. Low-resolution models showed larger increases in mean wind speeds post-interpolation, likely due to the incorporation of data from wider areas, including offshore regions with higher wind speeds. 

Different metrics sometimes yielded varying rankings for medium-performing models, highlighting the importance of using multiple evaluation criteria. Notebly, some models that underestimated wind speeds showed improved performance in wind power predictions after interpolation.

MPI-HR consistently emerged as the top-performing model across multiple evaluation metrics, including relative mean bias, Jensen-Shannon divergence, and Wasserstein distance. IPSL demonstrated the second-greatest predictive efficacy for wind power. Conversely, the JAP model consistently ranked as the poorest performer. The performance differences between models, particularly MPI-HR and JAP, can likely be attributed to their divergent focuses in climate system modeling.

These findings underscore the complexity of wind speed modeling for power predictions, highlighting the need to consider both model characteristics and resolution. Increased resolution improves extreme wind speed representation but not all aspects of wind speed simulation. This study emphasizes the careful selection of GCMs for regional climate applications, especially for wind power forecasting, and the importance of appropriate downscaling and bias correction techniques.

Future research should explore other regions and investigate influencing factors such as the hub height and the turbine type to enhance the generalizability of these findings.

\addcontentsline{toc}{chapter}{Acknowledgement}
\chapter*{Acknowledgments}\label{chapter:acknowledgments}
I would like to express my heartfelt gratitude to several individuals who have supported me during my Master's project and beyond:

Nina, for her guidance throughout the project, her availability to discuss questions at any time, and her valuable feedback on how to focus and structure my work. Thank you as well for connecting me with others, for being a great coffee master, and for organizing insightful awareness workshops!

Luca, for her helpful comments and ideas in our weekly discussions, and for always keeping an eye on the details! Also for integrating me into the group, inviting me to Banana walks, and sharing coffees, her experiences, and many useful tips with me!

Chris, for his ideas on visualizing distribution tails and comparing wind speed distributions using f-divergences.

Fabi, for helping me explore the feasibility of constructing a custom divergence for wind power, drawing on his knowledge from Professor Stefan Eckstein's probabilistic distances course. And for being a very fun and chill housemate! 

Nicole, for the possibility of doing the project in her group, for taking the time to discuss ideas, and for guiding me on the research directions.

The whole MLSES group for welcoming me warmly, for insightful talks, fun group events, and interesting conversations at lunch. The same goes for the many other nice people in the building!

The Method Center, Professor Brandt, and Professor Keleva, for establishing the QDS master's program, allowing me to participate and for assisting whenever needed.

My parents, for their unwavering support and encouragement, which paved the way for my academic journey.

Stefano, for always being there when I needed someone to listen, discuss, or motivate me. Grazie, tesoro mio - I don't know what I would have done without you!

\pagebreak

\addcontentsline{toc}{chapter}{Bibliography}
\printbibliography

\newpage
\appendix
\renewcommand{\thesection}{A\arabic{section}}
\chapter*{Appendix}
\addcontentsline{toc}{chapter}{Appendix}
\section{Fluid Dynamics in Numerical Weather Prediction}
\label{sec:fluid_dynamics_nwp}

From a theoretical point of view, the atmosphere can be described as a fluid - a continuous medium characterized by its density $\rho$, velocity $\vec{v}$, and specific internal energy $\epsilon$. Its dynamics is described by a set of coupled partial differential equations. The first version for adiabatic (nonconductive), inviscid, and incompressible ($\rho = 0$) flows was developed by \citet{euler1757principes}. This system of equations comprises the three dimensional balance of momentum based on Newton’s second law of motion
\begin{equation}
    \label{equ:momentum_balance}
    \frac{\partial \vec{v}}{\partial t}+\vec{v} \cdot \nabla \vec{v}+\frac{\nabla p}{\rho}=\vec{g}
\end{equation}
and the conservation of mass in form of the the continiuty equation 
\begin{equation}
    \label{equ:mass_consv_incompr}
    \frac{\partial \rho}{\partial t}+\vec{v} \cdot \nabla \rho+\rho \nabla \cdot \vec{v}=0
\end{equation}
where the last term is zero due the condition for incompressible fluids
\begin{equation}
    \label{equ:incomp_constraint}
     \nabla \cdot \vec{v}=0.
\end{equation}
Thereby, $p$ is the pressure and $\vec{g}$ is the gravitational acceleration. The Nabla operator represents the partial spatial derivatives and is defined as
\begin{equation}
    \left(
    \frac{\partial}{\partial x},
    \frac{\partial}{\partial y},
    \frac{\partial}{\partial z}
    \right)^T
\end{equation}
in cartesian coordinates. 
The case of compressible fluids can be established by omitting the constraint of the \Cref{equ:incomp_constraint} and introducing the thermodynamic energy equation
\begin{equation}
\frac{\partial \epsilon}{\partial t}+\vec{v} \cdot \nabla \epsilon+\frac{p}{\rho} \nabla \cdot \vec{v}=0.
\end{equation}
The system of equations is completed by the material-dependent equation of state, which relates pressure $p$, temperature $T$, and density $\rho$. If the ideal gas law
\begin{equation}
    p = \rho RT
\end{equation}
with the ideal gas constant $R$ is used, the internal energy can be substituted for the temperature in all above equations, since they are related proportionally by the molar heat capacity $c_V$ and the molar amount of substance $N$:
\begin{equation}
    \epsilon = c_VNT
\end{equation}
The extension for viscous fluids is achieved by changing the momentum balance (Equation \ref{equ:momentum_balance}) to a form known as the Navier-Stokes equations \citep{navier_stokes_euquations}. Furthermore, diabatic effects and friction terms can be included. When considering the atmosphere, the spherical shape of the Earth and its rotation must be taken into account. Moreover, a comprehensive description encompasses additional continuity equations for water vapor, cloud water, cloud ice, and different types of precipitation, incorporating terms for the exchange of water through phase changes. Altogether, these so-called governing equations constitute the dynamic core of atmospheric models. \citep{book_Warner_2010} 

Physical processes that occur on scales smaller than the spatial resolution are incorporated into the model through parameterizations \citep{book_parametrization}. Examples are cloud processes (formation, growth, and precipitation), atmospheric boundary layer processes (turbulence, mixing, and surface-atmosphere interactions), and radiative transfer processes (absorption, emission, and transfer of solar and terrestrial radiation). They are added as additional time tendencies $\partial X/\partial t$ to the governing equations, where $X$ is one of the model variables.

To solve differential equations numerically, they must be discretized into so-called difference equations so that computers operating in binary systems can process them. Two fundamentally different approaches can be distinguished \citep{timeIntegration}.
The first approach, known as Eulerian, treats spatial and temporal integration independently. In the first step, the equations are spatially discretized with finite difference and finite volume schemes \citep{durran2010numerical}. The temporal component is then solved using numerical integrators for ordinary differential equations (ODEs), such as the Runge-Kutta method \citep{rungeKutta}. 
Due to the varying propagation speeds of waves in the atmosphere, split explicit schemes are typically employed \citep{splitExplicit}. Among these, the Horizontally-Explicit Vertically-Implicit (HEVI) method is often used, where horizontal terms are treated explicitly and vertical terms implicitly to balance computational efficiency and stability.

The second approach is the Lagrangian, which addresses the entire time integral simultaneously. Instead of considering fluxes within a fixed grid, this method focuses on the trajectories of individual particles. However, in weather and climate models, a semi-Lagrangian approach is often preferred \citep{SemiLagrangianIntegrationSchemes}. This approach involves repeatedly interpolating the particle positions to their spatially fixed departure points. The selection of appropriate grids for the discretization of the atmosphere is addressed in \Cref{sec:numerical_grids}.

\section{Dynamics of Global Climate System}
\label{subsec:climate_system}

The global climate system is divided into five subsystems: the atmosphere (layer of gas surrounding the Earth), the hydrosphere (liquid water), the cryosphere (frozen water), the lithosphere (solid land surfaces), and the biosphere (living organisms). The atmosphere and the ocean (most of the hydrosphere) form large-scale circulation systems of thermal energy that interact with each other and the other spheres through various physical and biochemical processes.

The source of the energy is electromagnetic radiation from the Sun which reaches the Earth mostly in the visible and infrared spectrum. On average, there is an incoming energy flux of \SI{343}{\watt\per\meter^2}, referred to as the energy budget.
When the climate system is in equilibrium, the same amount of energy is eventually radiated back into space. Approximately $30\%$ of the incoming solar radiation is immediately reflected, with $23\%$ reflected by clouds and the atmosphere and $7\%$ reflected by the Earth's surface. The remaining $70\%$ is initially absorbed, with $47\%$ being absorbed by the Earth's surface and $23\%$ by the atmosphere. \citep{EarthsGlobalEnergyBudget}

The energy absorbed by the Earth’s surface can be stored for some time, but eventually reaches the atmosphere through three distinct pathways: the direct convective transfer of heat from the surface to the air (sensible heat flow), the convective transfer through a phase change into water via evapotranspiration (latent heat flow), or emission of infrared radiation. Some short-wave infrared radiation can escape directly into space through atmospheric windows, i.e. wavelength ranges in which the atmosphere is permeable to radiation. The majority, however, is absorbed by gases and clouds and re-emitted as infrared radiation in all directions. This phenomenon creates what is known as the greenhouse effect: heat is trapped by the Earth's atmosphere raising the average global temperature. \citep{earthClimateSystem}

Spatial temperature variations leading to the circulation of air and water arise due to the Earth's spherical shape. Regions near the equator receive the most direct insolation, leading to higher average temperatures in these areas than close to the poles. Additionally, the radiation only reaches the side of the Earth facing the Sun. In conjunction with Earth's rotation around its axis, this causes the night-day cycle. Furthermore, the inclination of the Earth's axis results in the occurrence of different seasons depending on the Earth's position on its elliptical orbit around the Sun.  

The atmosphere and ocean circulation systems are directly linked, with the Bjerknes feedback loop serving as a prominent example. This feedback loop leads to the El Niño Southern Oscillation, which has a profound impact on the weather in the Pacific \citep{ENSO}. The other spheres also play an important role in the circulation dynamics. The topography of the lithosphere influences the atmospheric jet and rain belts due to the barrier effect mountain ranges exert on the airflow \citep{mountains}. 
The biosphere modifies the composition of the atmosphere through various processes, including plant photosynthesis (replacing carbon dioxide with oxygen) and animal digestion (e.g. cattle emitting methane). These exchanges of substances like carbon and methane are known as biogeochemical cycles \citep{book_BiochemicalCycles}. Another important such cycle is the hydrological/water cycle \citep{waterCycle}, which encompasses the following processes: Liquid water from the hydrosphere and plants in the biosphere is converted into vapor through evaporation and transpiration. This vapor rises into the atmosphere and is carried over long distances by atmospheric rivers, acting as the primary atmospheric absorber of both short-wave and long-wave radiation.  When the vapor condenses in the air, a large amount of latent heat is released, and clouds or fog form. Through precipitation, the water returns to the ground in liquid or frozen form.  
Furthermore, the freezing and melting of snow and ice from the cryosphere contribute to the ocean’s water circulation.

In addition, various external factors can alter the dynamics of the global climate system by imposing an imbalance in the energy budget. These are called climate forcings \citep{book_radiativeForcing}. Natural forcings include changes in the Earth's orbit or solar activity \citep{book_radiativeForcing} and volcanic eruptions that release sulfur gases into the atmosphere \citep{volcanicErruptions}. These forcings have resulted in significant climate changes, such as the Medieval Climate Anomaly (ca. 950-1250) and the Little Ice Age (ca. 1450-1850) \citep{climateLastMillenium}. Over the past two centuries, however, the Earth has experienced a sharp rise in global surface temperature attributed predominantly to anthropogenic forcing \citep{IPCC2023}. The primary contributor to this change is energy generation through fossil fuel combustion \citep{fossilFuelCombustion}. This process releases greenhouse gases, such as CO2 and methane, intensifying the greenhouse effect and thus increasing the global average temperature. In addition, human land use, including deforestation, a decline in biodiversity, mass livestock farming, and land sealing, contributes to climate change \citep{landUse}. 

\section{Numerical Grid Construction}
\label{sec:numerical_grids}

Grids are a set of cells consisting of a finite number of vertices connected by edges covering the area under consideration.  In the case of atmospheric science, this area is the Earth, approximated as a sphere. 

One potential method for defining the numerical grid is to map the sphere to a rectangular plane. The most natural approach is to align the axes with the geographic coordinates. 
The geographic coordinate system defines the locations on the surface based on the Earth's axis of rotation. The axis' intersection with the Earth's surface determines the geographic north and south poles. The intersection with the plane perpendicular to it which runs through the Earth's core, defines the Equator. Locations are specified in angles called latitudes and longitudes, where latitude is the distance to the Equator in the north-south direction and longitude is the distance to the prime axis passing through Greenwich in the west-east direction. The axes are referred to as circles of latitude and meridians.

In a regular geographic grid, the grid points are distributed uniformly along the latitudes and longitudes, resulting in equal cell sizes on the plane. However, on the sphere, this arrangement results in a higher resolution near the poles where the meridians converge. This is not only computationally inefficient but also creates singularities that are difficult to handle \citep{steppeler_mathematics_2022}. To mitigate this issue, one can employ a rectilinear grid in which the values for latitude and/or longitude vary in distance. 

In the case where only a restricted region is to be considered, the use of a curvilinear grid represents an alternative option. Unlike linear grids, which are defined by points situated on straight lines, curvilinear grids are defined by points on curved lines. Regional models frequently employ curvilinear grids with rotated poles that are situated outside the region of interest.
Tripolar grids \citep{tripolarGridMadec, tripolarGridMurray}, are often used in ocean modeling, as they allow for the exclusion of pole singularities by positioning them over land.

An alternative approach is to divide the sphere into patches \citep{steppeler_mathematics_2022}. This is achieved by projecting the edges of a regular polyhedron (e.g., tetrahedron, cube, octahedron, dodecahedron, and icosahedron) onto the sphere and further subdividing the resulting areas into finer grids. One approach that has been employed is the use of cubed-sphere grids \citep{cubeSphereGrid}, which project the sphere onto the six faces of a cube, exchanging the polar singularities for grid distortion near the cube's vertices. Another approach is the triangular discretization of an icosahedral projection, which is gaining popularity \citep{gridsForEarthSystemModels}. 

A completely different possibility is the formulation of the governing equations as a truncated series of spherical harmonics, using a Fourier transformation with Legendre polynomials $P_n^m$ as basis functions \citep{durran2010numerical}. Different truncation schemes can be applied to limit the series expansion by restricting the range of $m$ from $(-\infty, \infty)$ to $(-M, M)$ and the range of $n$ from $(m, \infty)$ to $(m, N(M))$. Triangular truncation, where $N(m) = M$, ensures consistent spatial resolution across the entire spherical surface. The corresponding polar coordinates form a regular Gaussian grid with uniformly spaced longitudes and latitudes defined by Gaussian quadrature \citep{gridsForEarthSystemModels}. This ensures that the circles of latitude become less dense at the poles. In a reduced (or thinned) Gaussian grid, the number of grid points additionally diminishes when approaching the poles. This adjustment ensures that the spacing between grid points is nearly uniform across the entire spherical surface.

\newpage
\section{CMIP6 Institutions}

The institutions providing the data sets investigated in this study are listed in \Cref{tab:institutions}.

\begin{table}[htp]
    \centering
    \caption{Institutions providing climate model data for CMIP6 including the abbreviations used in the data set labels (first column), the full names of the institutions (second column), and their locations (third column).}
    \label{tab:institutions}
    \begin{tabu} to \textwidth { l | X  l }
    \toprule
        Abbreviation & Full Name & Seat \\
        \midrule
        CMCC & Euro-Mediterranean Center on Climate Change  &  Lecce, Italy  \\
        & (Centro Euro-Mediterraneo sui Cambiamenti Climatici) & \\
        EC-Earth & EC-Earth Consortium (30 research institutes from 12 European countries) &   \\
        IPSL & Institut Pierre Simon Laplace   & Paris, France \\
        & (Pierre Simon Laplace Institute) & \\
        JAP &  Japanese modeling community for MIROC (Model for Interdisciplinary Research on Climate): Japan Agency for Marine-Earth Science and Technology, Atmosphere and Ocean Research Institute, University of Tokyo, National Institute for Environmental Studies, RIKEN Center for Computational Science & Yokohama, Japan \\
        MPI & Max Planck Institute for Meteorology  & Hamburg, Germany \\
        & (Max-Planck-Institut für Meteorologie)& \\
        MOHC & Met Office Hadley Centre for Climate Science and Services & Exeter, UK \\
        NCC & NorESM (Norwegian Earth System Model) Climate modeling Consortium: Center for International Climate and Environmental Research, Norwegian Meteorological Institute, Nansen Environmental and Remote Sensing, Norwegian Institute for Air Research, University of Bergen, University of Oslo, Uni Research &  Oslo, Norway \\
        \bottomrule
    \end{tabu}
\end{table}

\section{Weibull Fit Curves}
\label{sec:weibull_fits}

\begin{figure}[htp]
    \centering
    \begin{subfigure}[b]{0.24\textwidth}
        \centering
        \includegraphics[width=3cm]{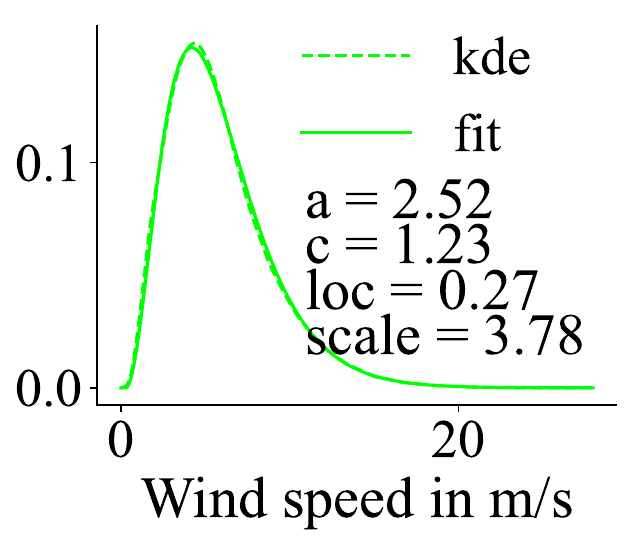}
    \end{subfigure}
    \hfill
    \begin{subfigure}[b]{0.24\textwidth}
        \centering
        \includegraphics[width=3cm]{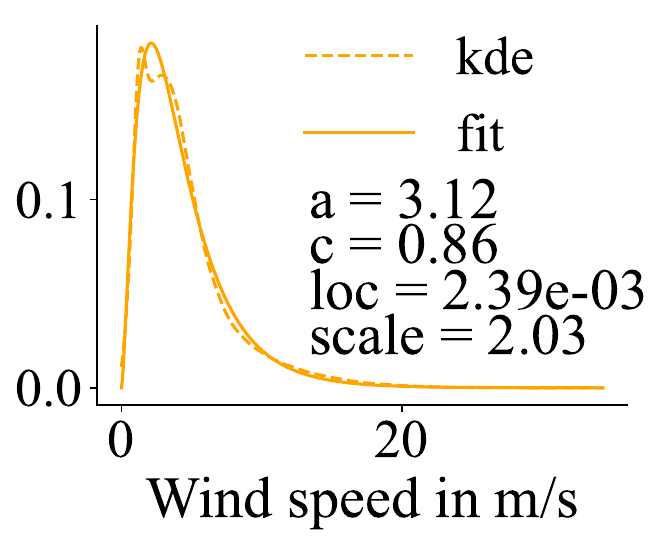}
    \end{subfigure}
    \hfill
    \begin{subfigure}[b]{0.24\textwidth}
        \centering
        \includegraphics[width=3cm]{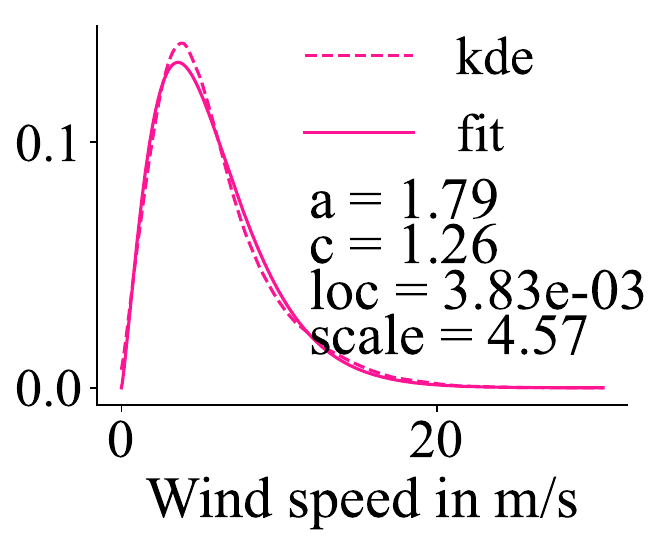}
    \end{subfigure}
    \hfill
    \begin{subfigure}[b]{0.24\textwidth}
        \centering
        \includegraphics[width=3cm]{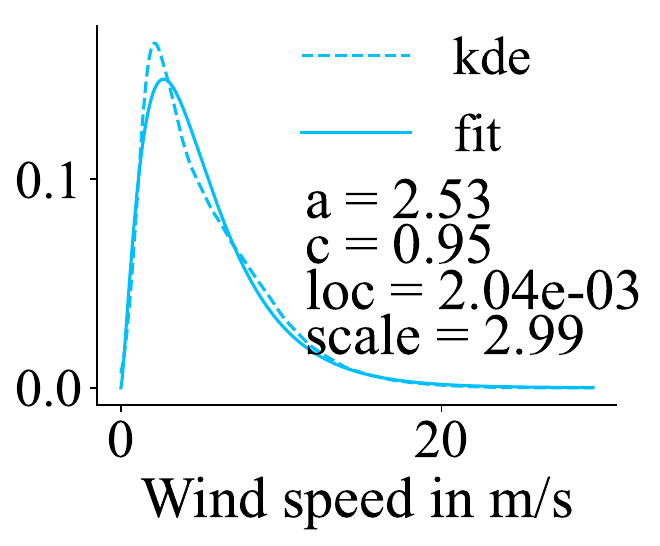}
    \end{subfigure}
    \begin{subfigure}[b]{0.24\textwidth}
        \centering
        \includegraphics[width=3cm]{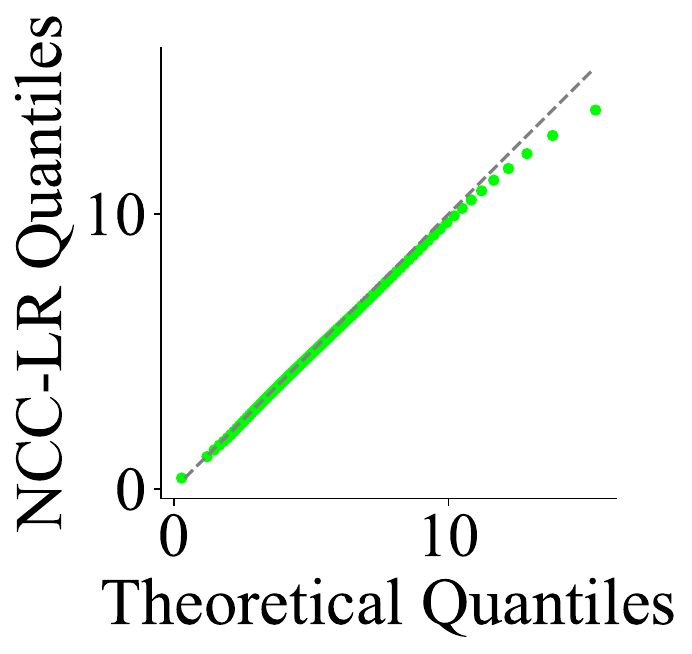}
        \caption{NCC-LR}
    \end{subfigure}
    \hfill
    \begin{subfigure}[b]{0.24\textwidth}
        \centering
        \includegraphics[width=3cm]{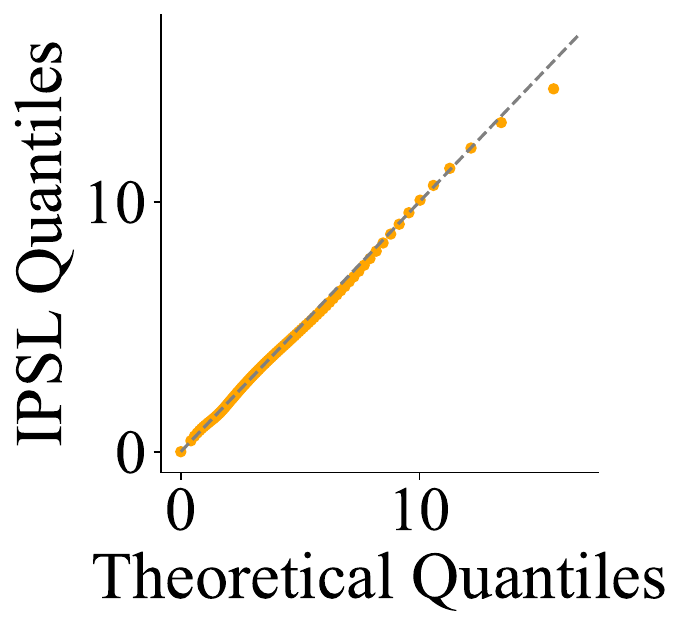}
        \caption{IPSL}
    \end{subfigure}
    \hfill
    \begin{subfigure}[b]{0.24\textwidth}
        \centering
        \includegraphics[width=3cm]{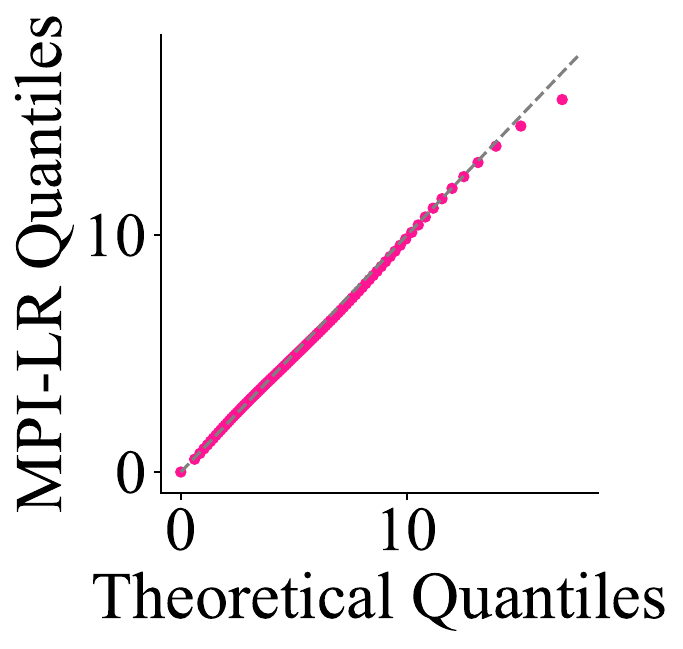}
        \caption{MPI-LR}
    \end{subfigure}
    \hfill
    \begin{subfigure}[b]{0.24\textwidth}
        \centering
        \includegraphics[width=3cm]{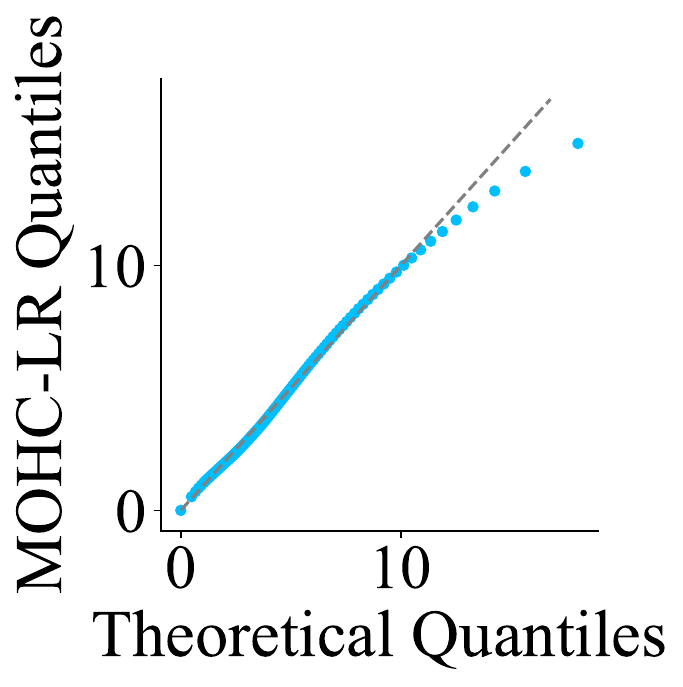}
        \caption{MOHC-LR}
    \end{subfigure}
    \par\bigskip
    \begin{subfigure}[b]{0.24\textwidth}
        \centering
        \includegraphics[width=3cm]{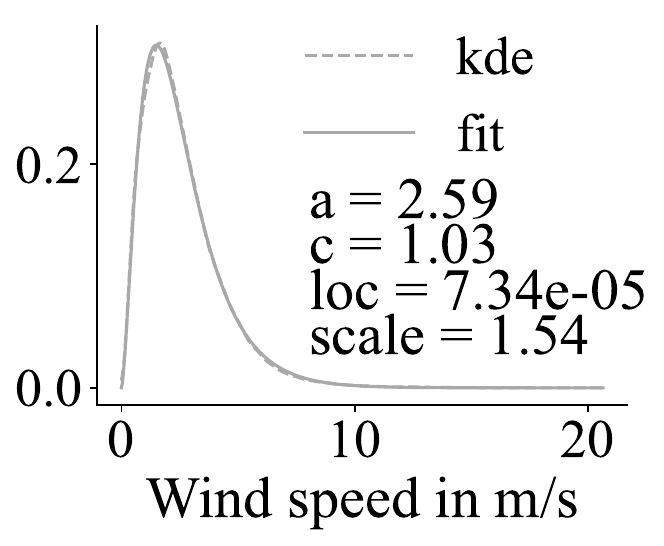}
    \end{subfigure}
    \hfill
    \begin{subfigure}[b]{0.24\textwidth}
        \centering
        \includegraphics[width=3cm]{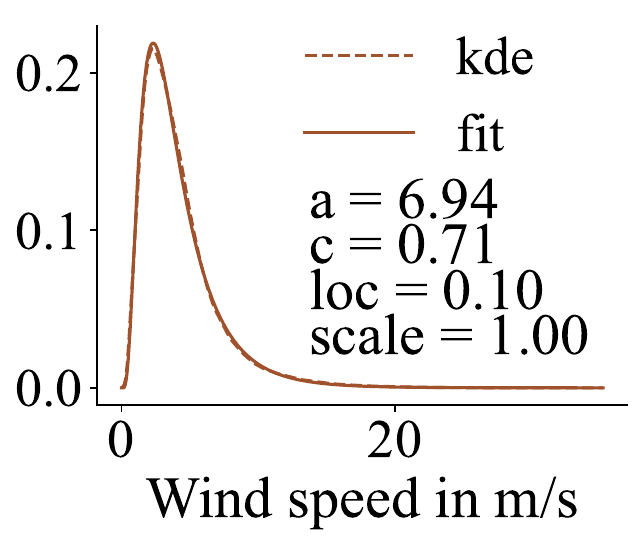}
    \end{subfigure}
    \hfill
    \begin{subfigure}[b]{0.24\textwidth}
        \centering
        \includegraphics[width=3cm]{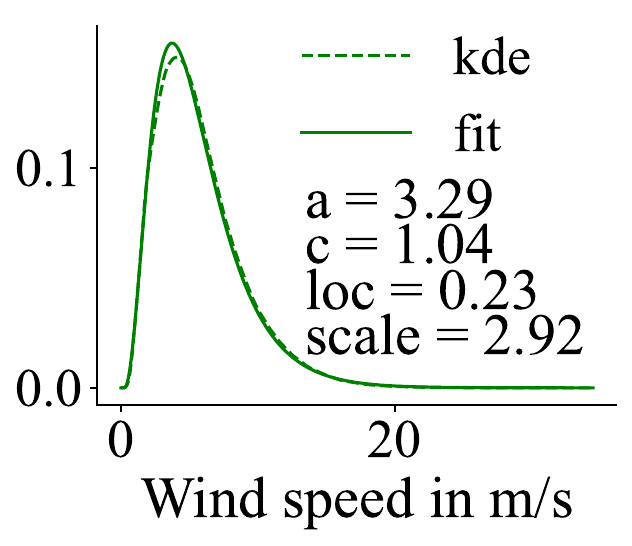}
    \end{subfigure}
    \hfill
    \begin{subfigure}[b]{0.24\textwidth}
        \centering
        \includegraphics[width=3cm]{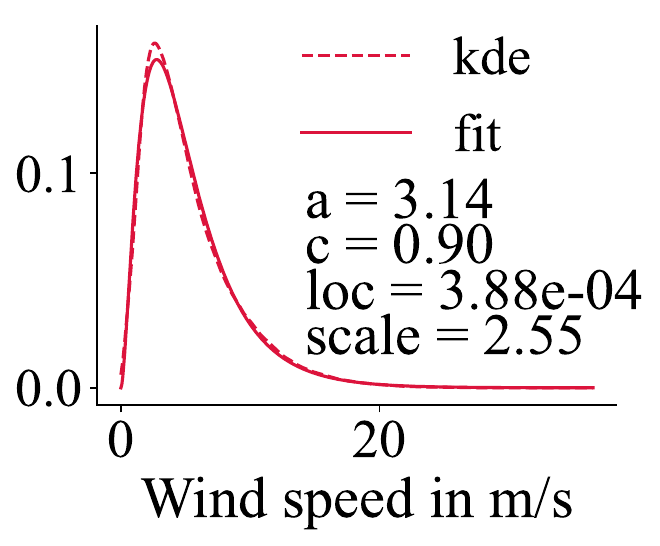}
    \end{subfigure}
    \begin{subfigure}[b]{0.24\textwidth}
        \centering
        \includegraphics[width=3cm]{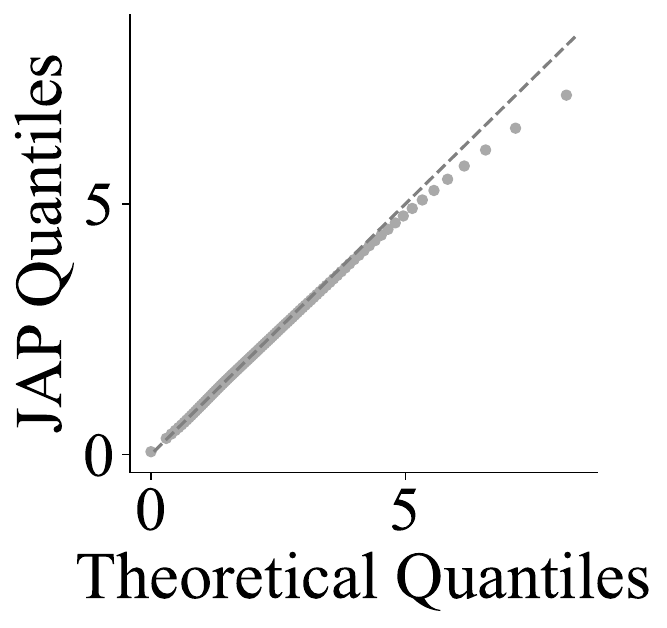}
        \caption{JAP}
    \end{subfigure}
    \hfill
    \begin{subfigure}[b]{0.24\textwidth}
        \centering
        \includegraphics[width=3cm]{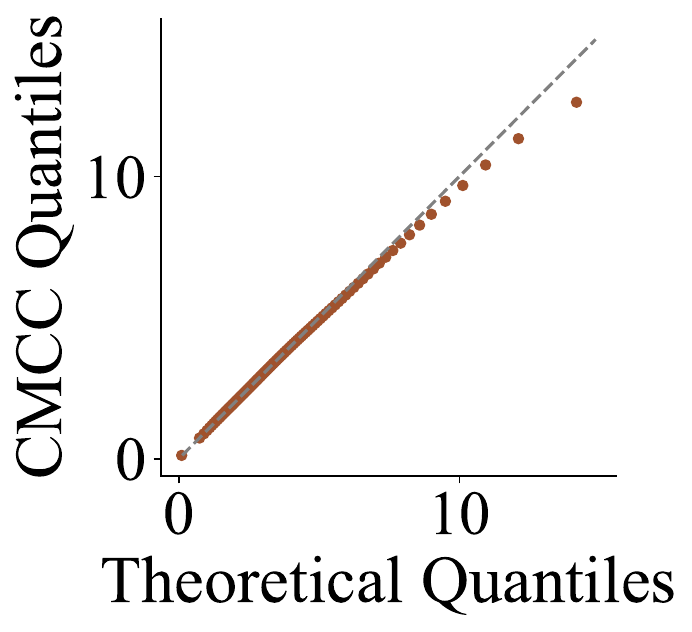}
        \caption{CMCC}
    \end{subfigure}
    \hfill
    \begin{subfigure}[b]{0.24\textwidth}
        \centering
        \includegraphics[width=3cm]{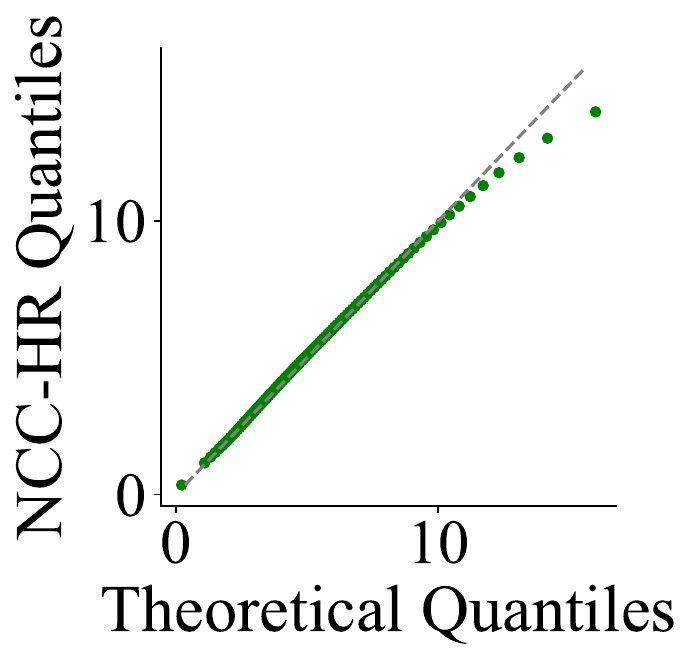}
        \caption{NCC-HR}
    \end{subfigure}
    \hfill
    \begin{subfigure}[b]{0.24\textwidth}
        \centering
        \includegraphics[width=3cm]{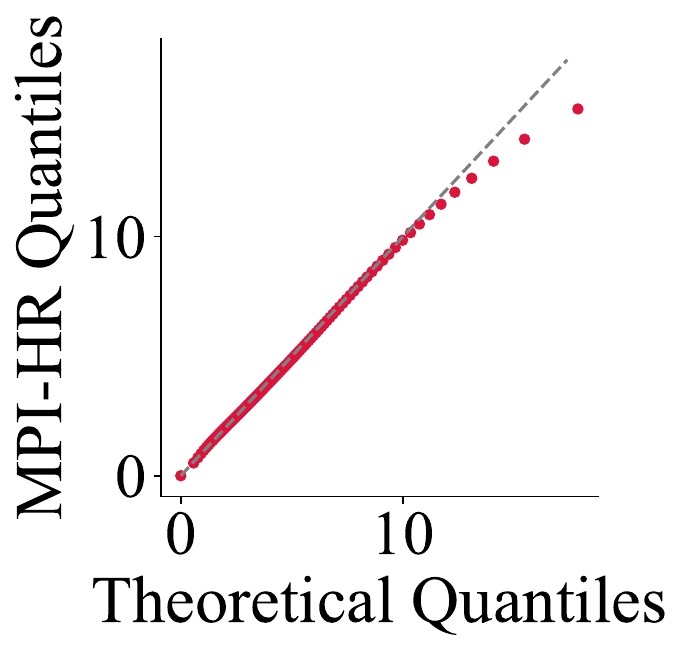}
        \caption{MPI-HR}
    \end{subfigure}
    \par\bigskip
    \begin{subfigure}[b]{0.32\textwidth}
        \centering
        \includegraphics[width=3cm]{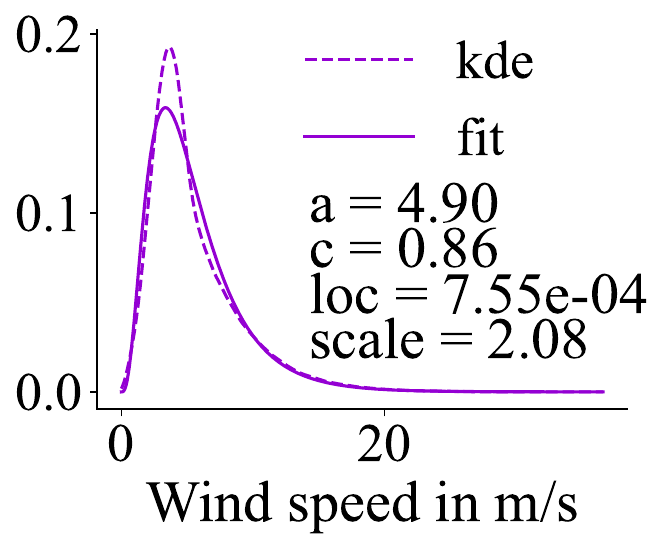}
    \end{subfigure}
    \hfill
    \begin{subfigure}[b]{0.23\textwidth}
        \centering
        \includegraphics[width=3cm]{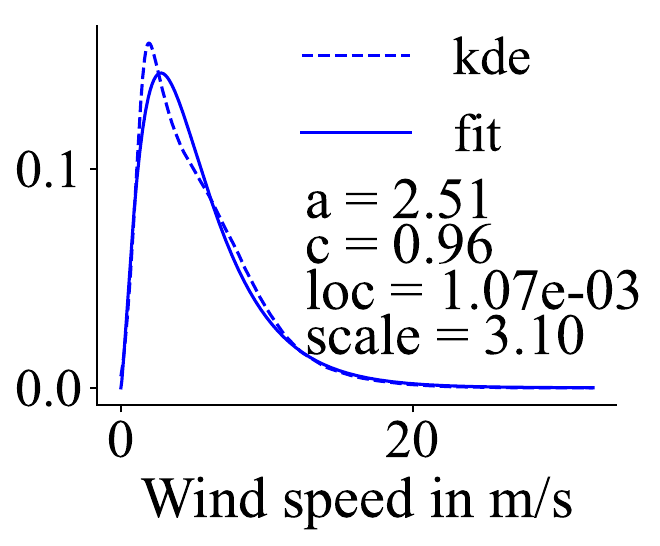}
    \end{subfigure}
    \hfill
    \begin{subfigure}[b]{0.32\textwidth}
        \centering
        \includegraphics[width=3cm]{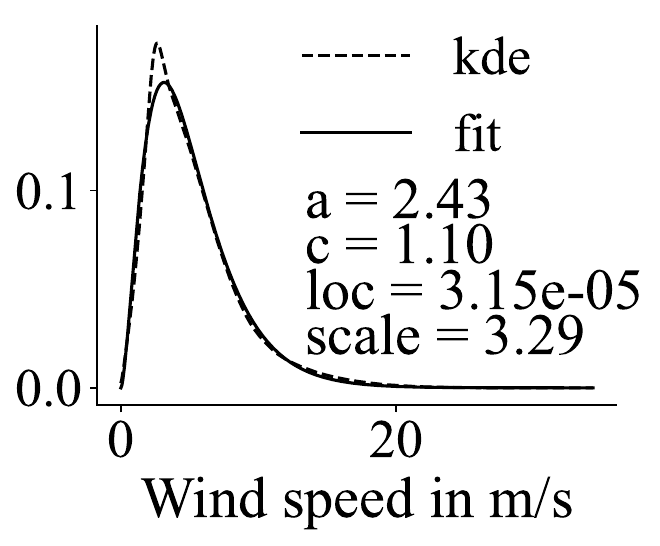}
    \end{subfigure}
    \hfill
    \begin{subfigure}[b]{0.32\textwidth}
        \centering
        \includegraphics[width=3cm]{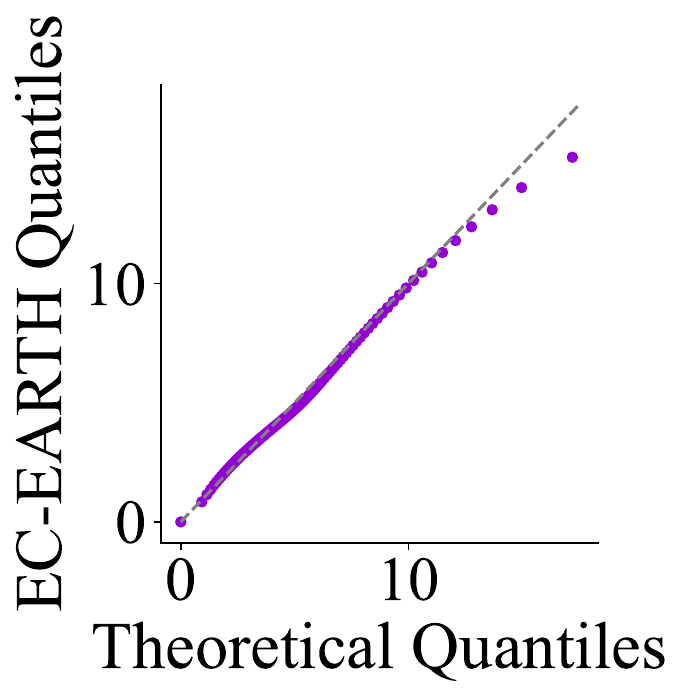}
        \caption{EC-EARTH}
    \end{subfigure}
    \hfill
    \begin{subfigure}[b]{0.32\textwidth}
        \centering
        \includegraphics[width=3cm]{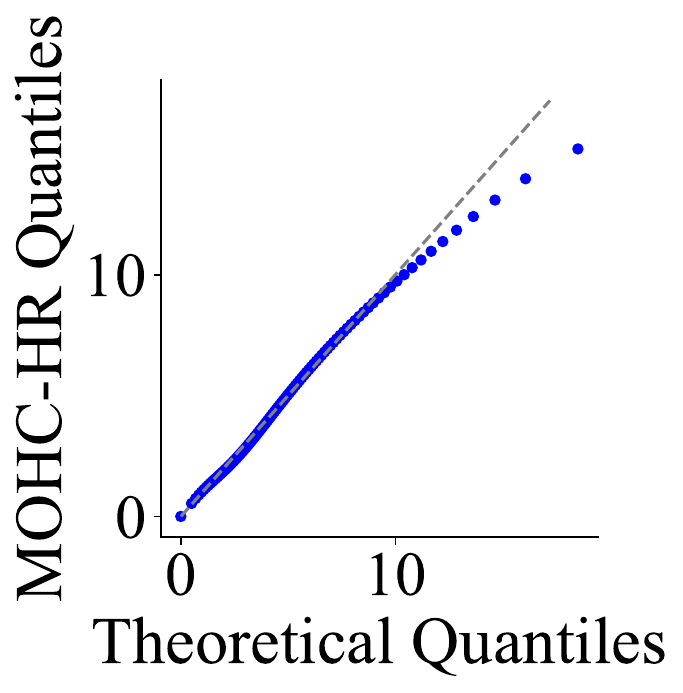}
        \caption{MOHC-HR}
    \end{subfigure}
    \hfill
    \begin{subfigure}[b]{0.33\textwidth}
        \centering
        \includegraphics[width=3cm]{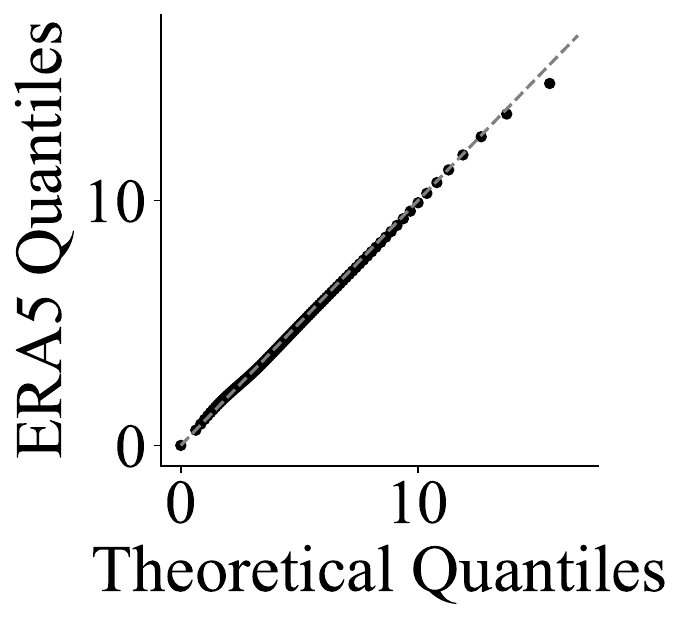}
        \caption{ERA5}
    \end{subfigure}
    \caption{Weibull fit curves with specified parameter values along with kernel density estimates of the wind speed samples and QQ plots comparing the fit to the original data.}
    \label{fig:weibull_fits}
\end{figure}

In \Cref{fig:weibull_fits}, the Weibull fits of the wind speed distributions are plotted along with the kernel density estimates, demonstrating good agreement. The QQ plots show that the fits generally capture the data well, but slightly underestimate the upper quantiles.

\section{Ensemble Run Comparison}
\label{sec:run_comparison}

\begin{figure}[htp]
    \centering
    \begin{subfigure}[b]{0.19\textwidth}
        \centering
        \includegraphics[height=3.5cm]{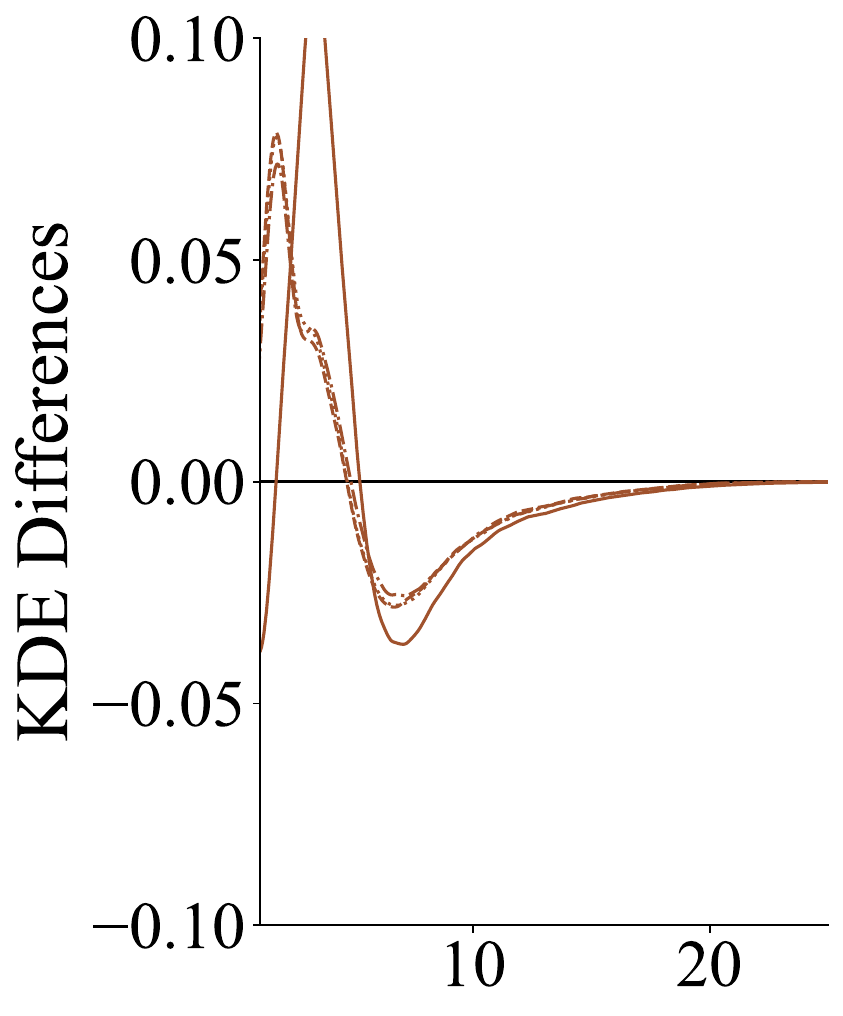}
    \end{subfigure}
    \hfill
    \begin{subfigure}[b]{0.19\textwidth}
        \centering
        \includegraphics[height=3.5cm]{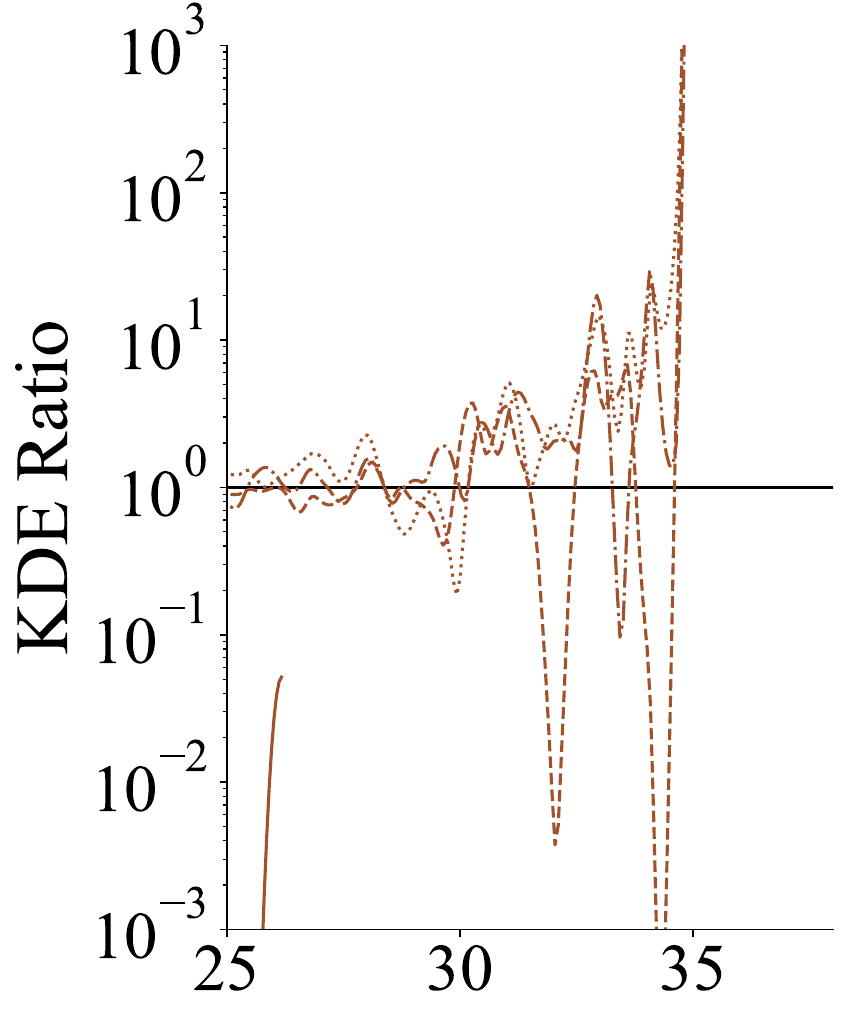}
    \end{subfigure}
    \hfill
    \begin{subfigure}[b]{0.19\textwidth}
        \centering
        \includegraphics[height=3.5cm]{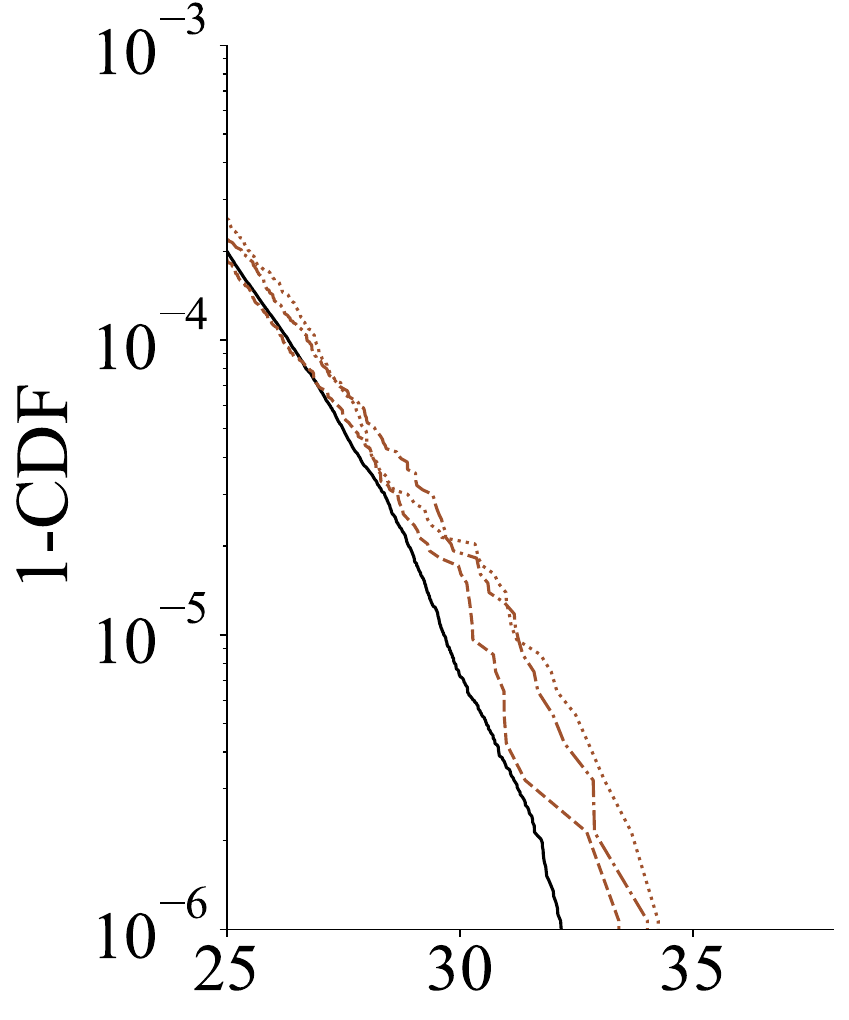}
    \end{subfigure}
    \begin{subfigure}[b]{0.39\textwidth}
        \centering
        \includegraphics[height=3.5cm]{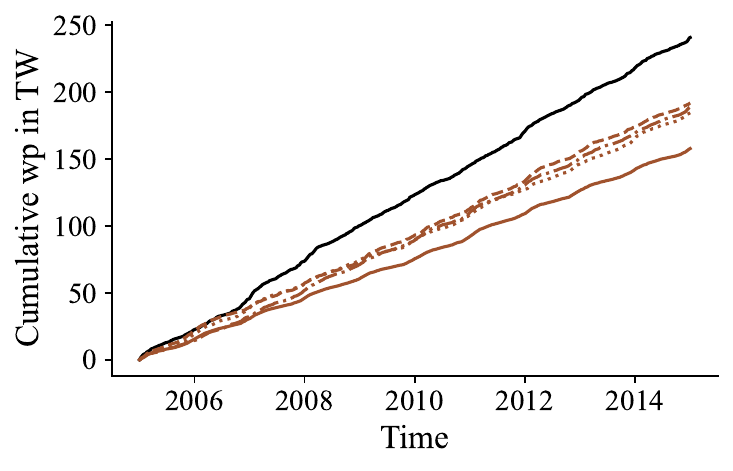}
    \end{subfigure}
    \begin{subfigure}[b]{\textwidth}
        \centering
        \includegraphics[height=1.2cm]{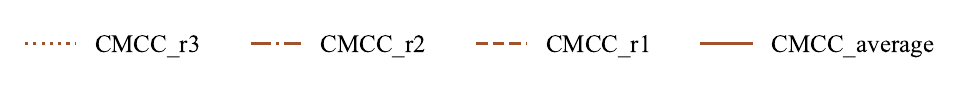}
    \end{subfigure}
    \par\bigskip
    \begin{subfigure}[b]{0.19\textwidth}
        \centering
        \includegraphics[height=3.5cm]{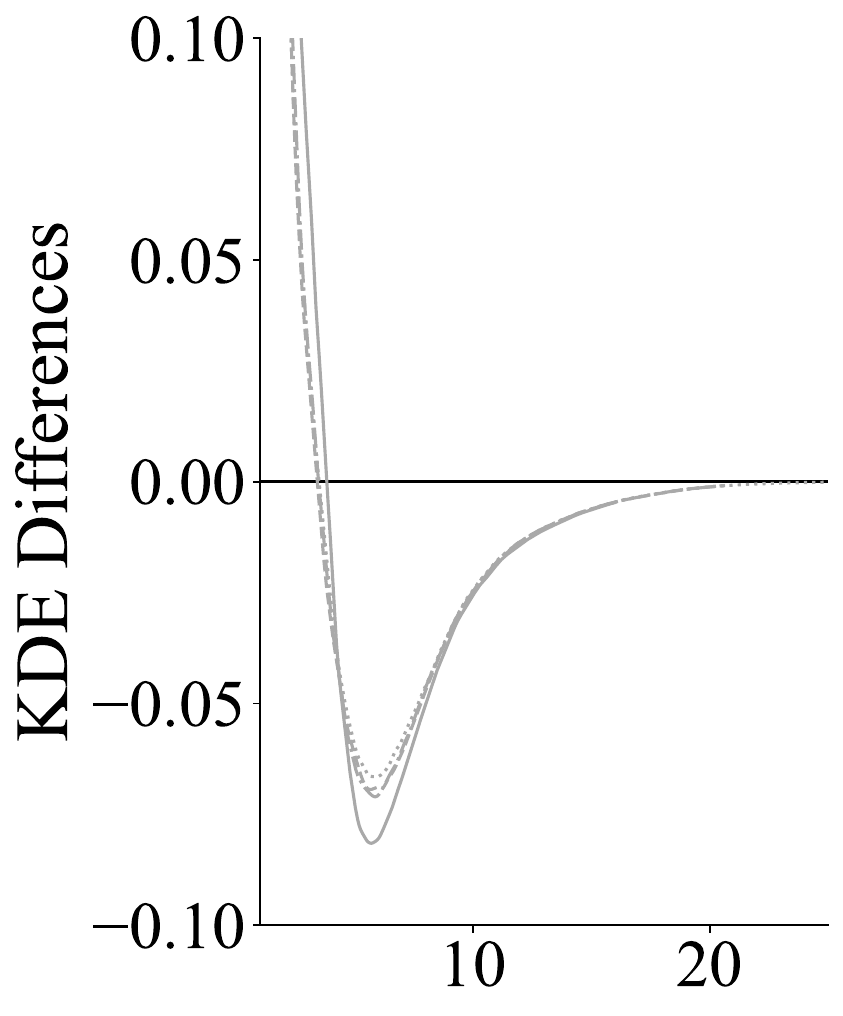}
    \end{subfigure}
    \hfill
    \begin{subfigure}[b]{0.19\textwidth}
        \centering
        \includegraphics[height=3.5cm]{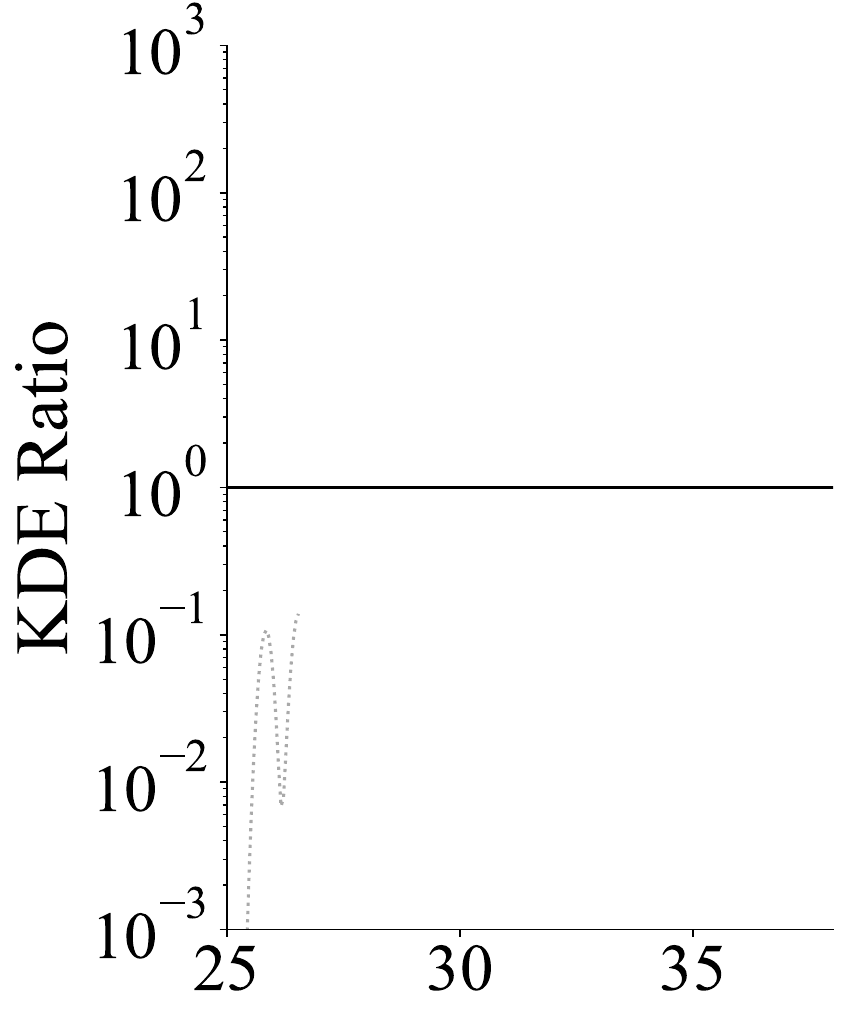}
    \end{subfigure}
    \hfill
    \begin{subfigure}[b]{0.19\textwidth}
        \centering
        \includegraphics[height=3.5cm]{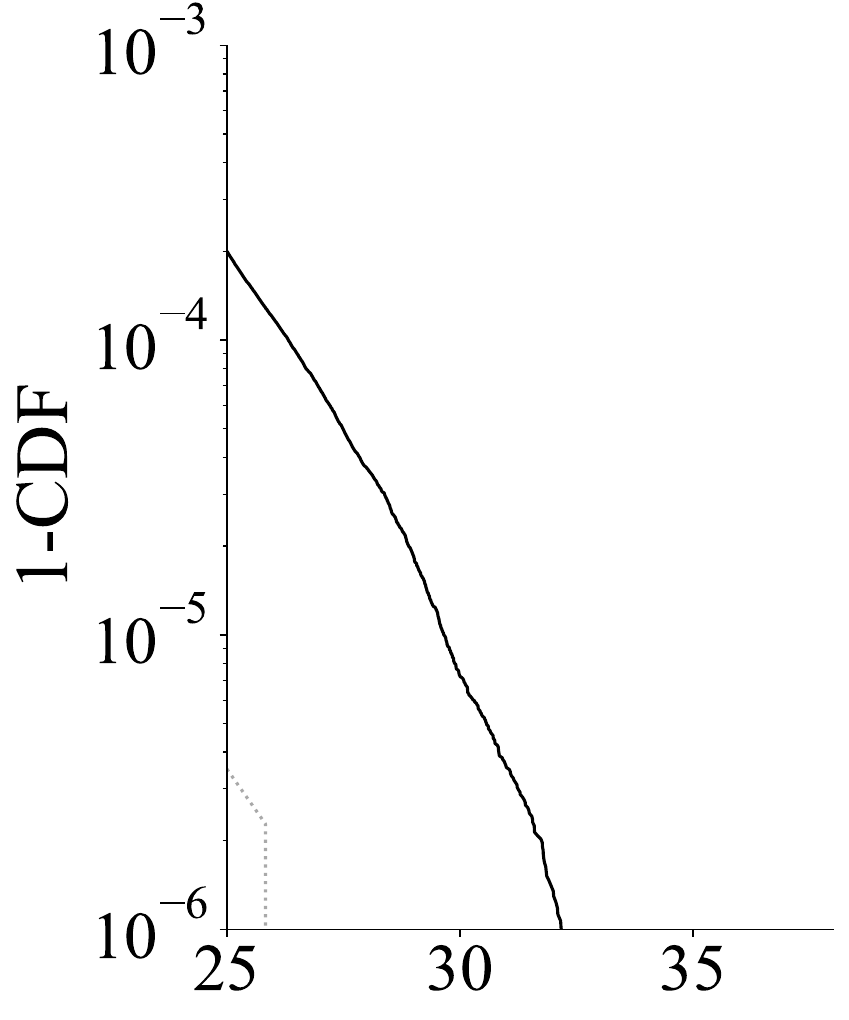}
    \end{subfigure}
    \begin{subfigure}[b]{0.39\textwidth}
        \centering
        \includegraphics[height=3.5cm]{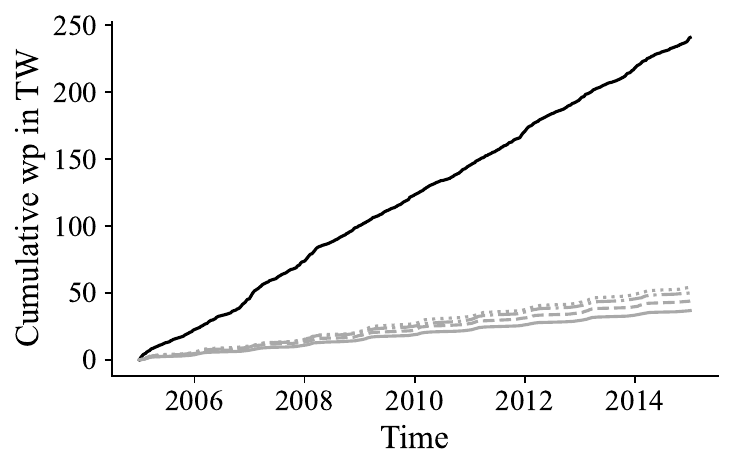}
    \end{subfigure}
    \begin{subfigure}[b]{\textwidth}
        \centering
        \includegraphics[height=1.2cm]{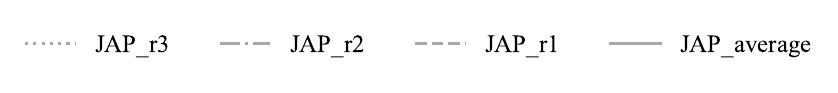}
    \end{subfigure}
    \caption{Performance comparison between model runs. Columns from left to right show: KDE difference to ERA5 in medium wind speed range, log-KDE-ratio with ERA5, survival function at the tail of the wind speed distribution, and cumulative wind power. The first row represents data from the CMCC and the second from the JAP model, displaying three runs with varying initial/boundary conditions and their ensemble average.}
    \label{fig:ensemble_runs_1}
\end{figure}

\begin{figure}[htp]
    \centering
    \begin{subfigure}[b]{0.19\textwidth}
        \centering
        \includegraphics[height=3.5cm]{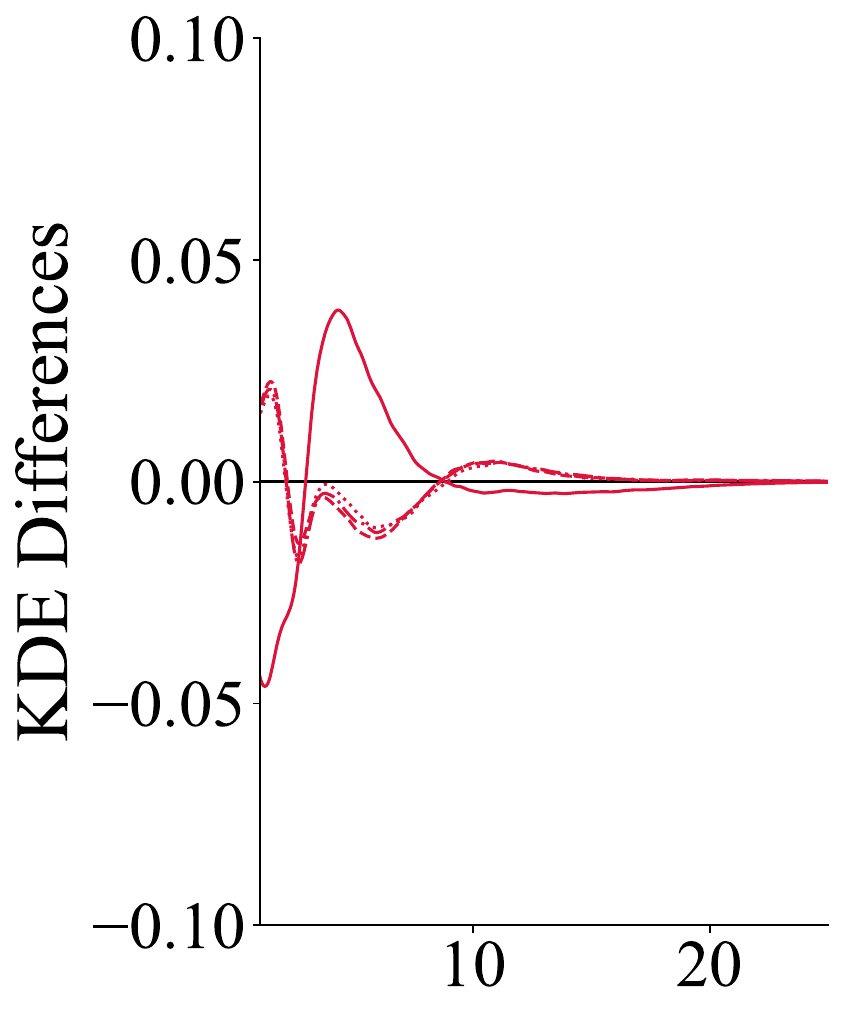}
    \end{subfigure}
    \hfill
    \begin{subfigure}[b]{0.19\textwidth}
        \centering
        \includegraphics[height=3.5cm]{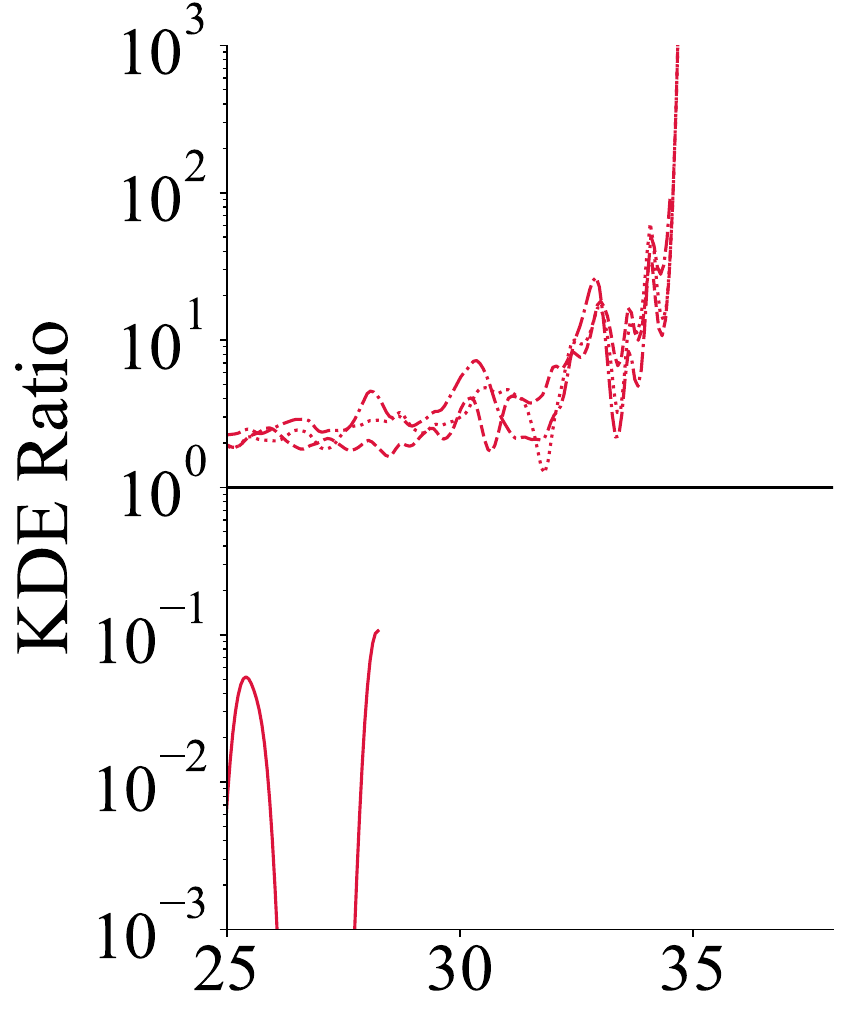}
    \end{subfigure}
    \hfill
    \begin{subfigure}[b]{0.19\textwidth}
        \centering
        \includegraphics[height=3.5cm]{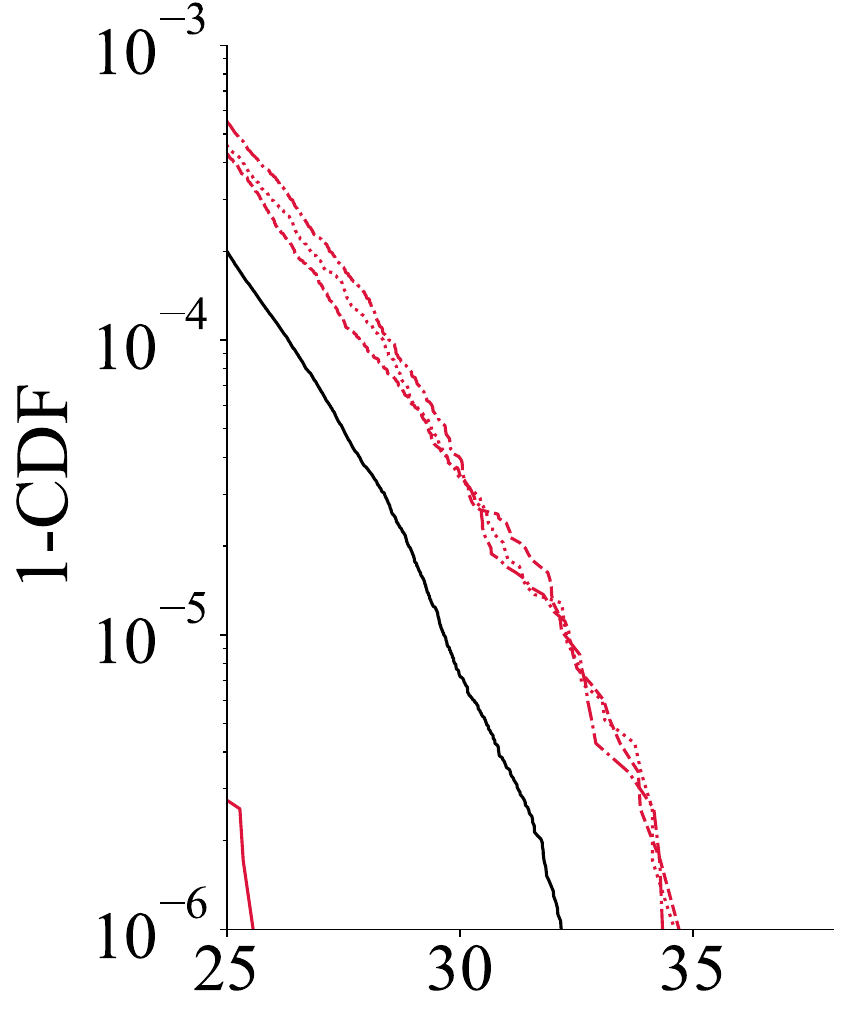}
    \end{subfigure}
    \begin{subfigure}[b]{0.39\textwidth}
        \centering
        \includegraphics[height=3.5cm]{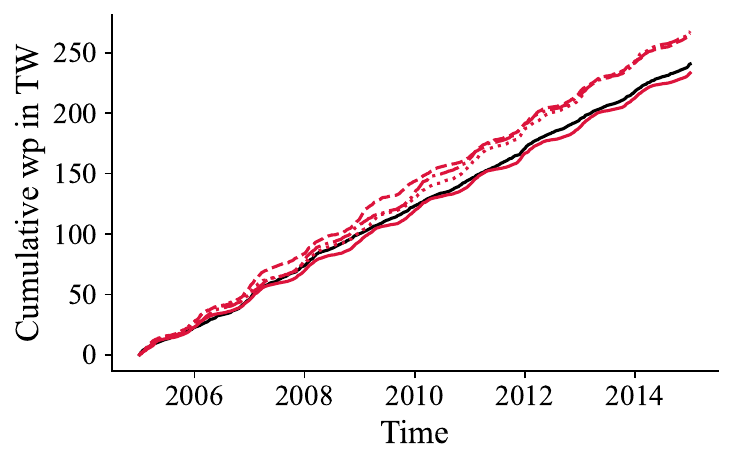}
    \end{subfigure}
    \begin{subfigure}[b]{\textwidth}
        \centering
        \includegraphics[height=1.2cm]{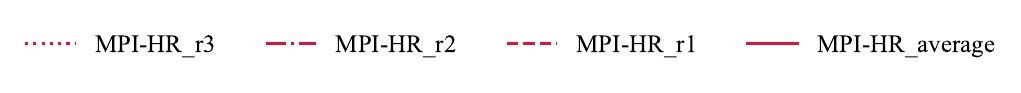}
    \end{subfigure}
    \par\bigskip
    \begin{subfigure}[b]{0.19\textwidth}
        \centering
        \includegraphics[height=3.5cm]{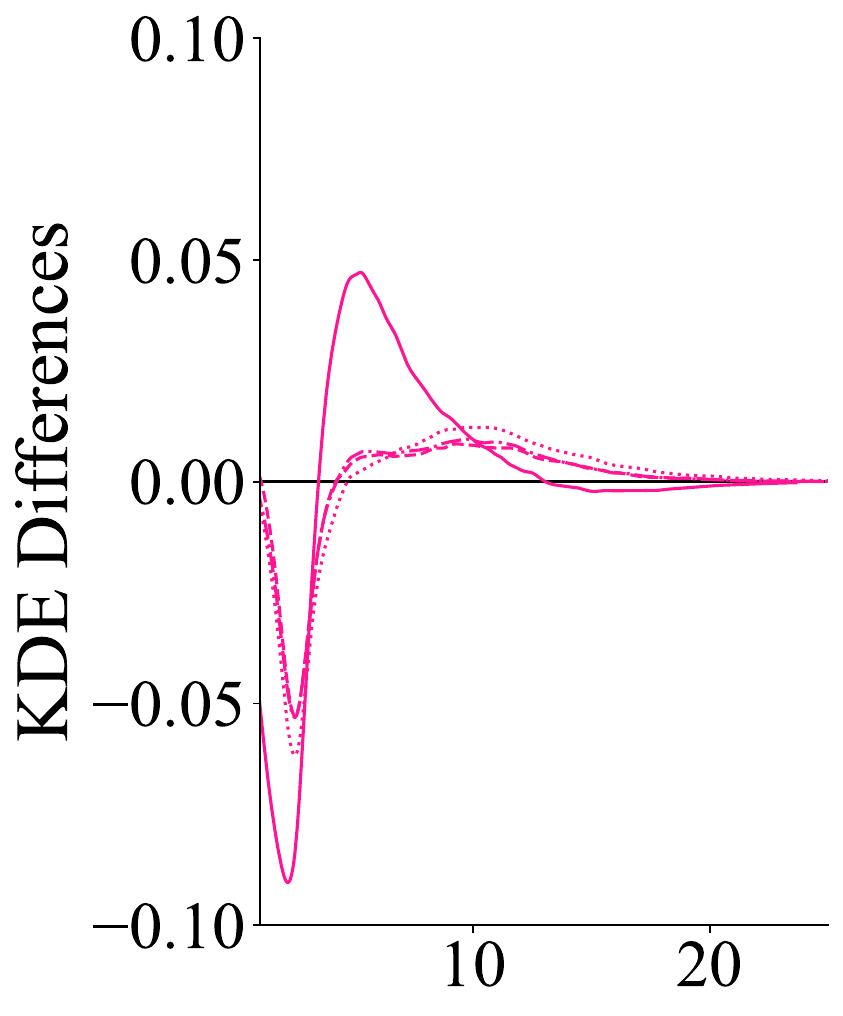}
    \end{subfigure}
    \hfill
    \begin{subfigure}[b]{0.19\textwidth}
        \centering
        \includegraphics[height=3.5cm]{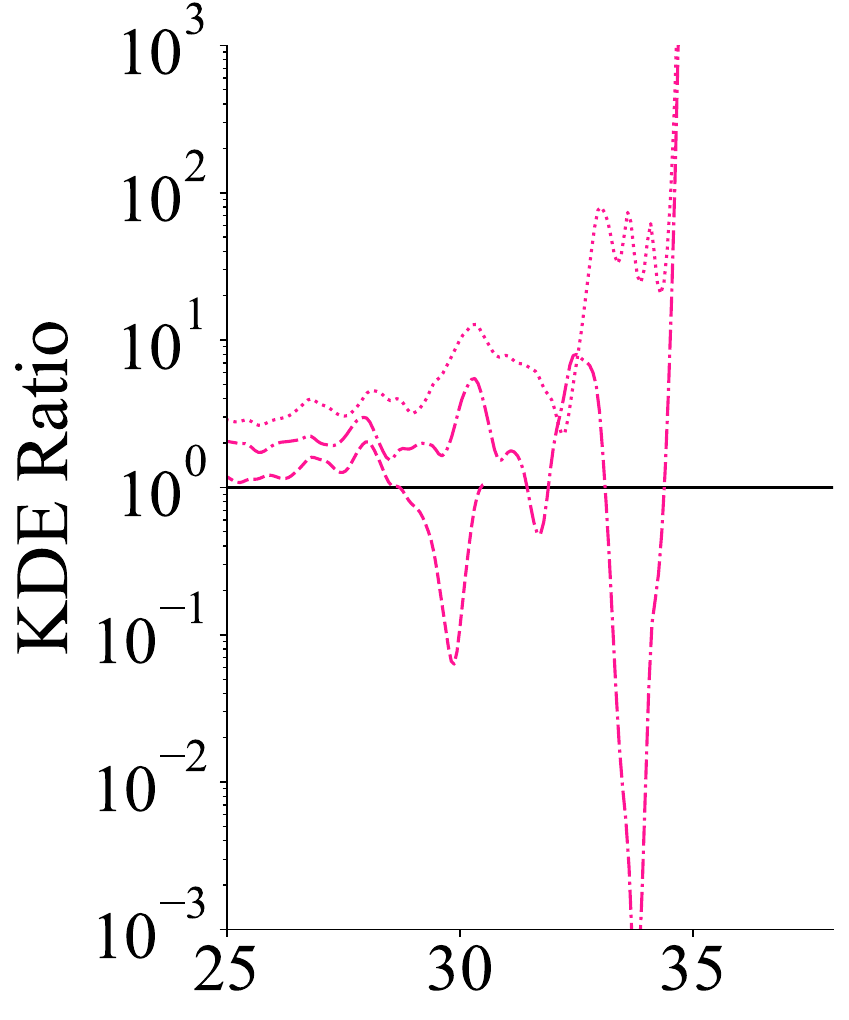}
    \end{subfigure}
    \hfill
    \begin{subfigure}[b]{0.19\textwidth}
        \centering
        \includegraphics[height=3.5cm]{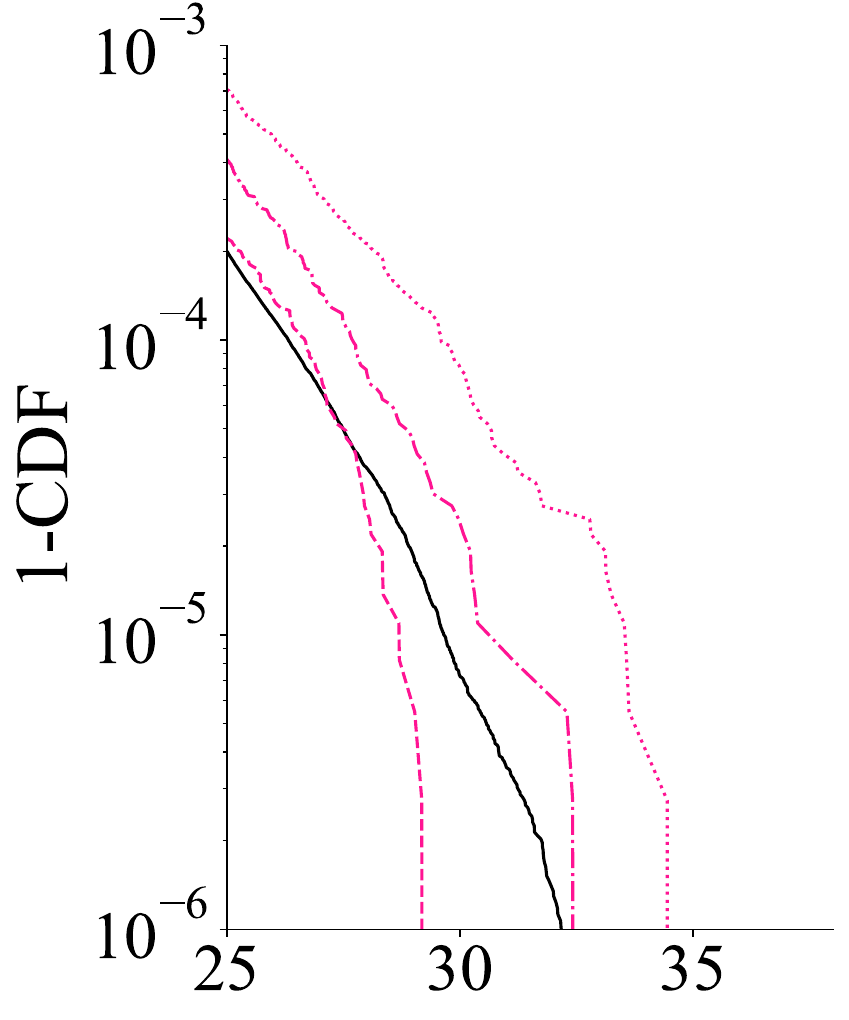}
    \end{subfigure}
    \begin{subfigure}[b]{0.39\textwidth}
        \centering
        \includegraphics[height=3.5cm]{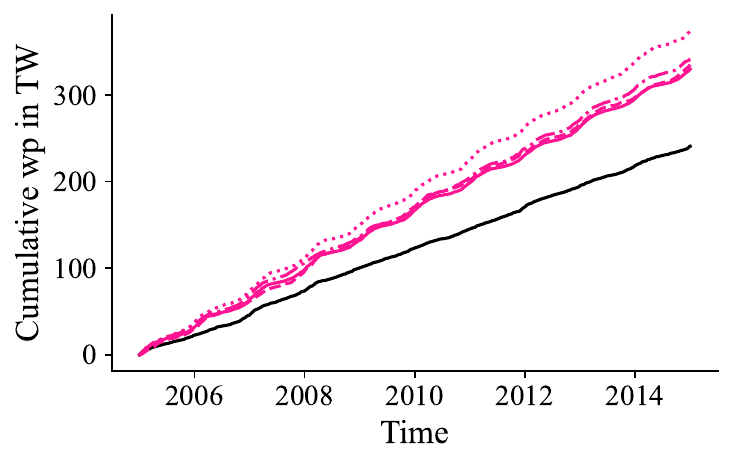}
    \end{subfigure}
    \begin{subfigure}[b]{\textwidth}
        \centering
        \includegraphics[height=1.2cm]{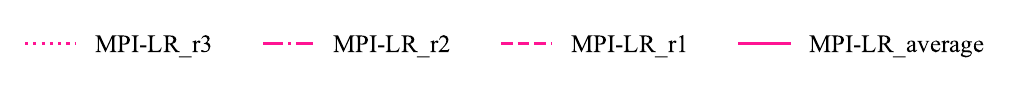}
    \end{subfigure}
    \par\bigskip
    \begin{subfigure}[b]{0.19\textwidth}
        \centering
        \includegraphics[height=3.5cm]{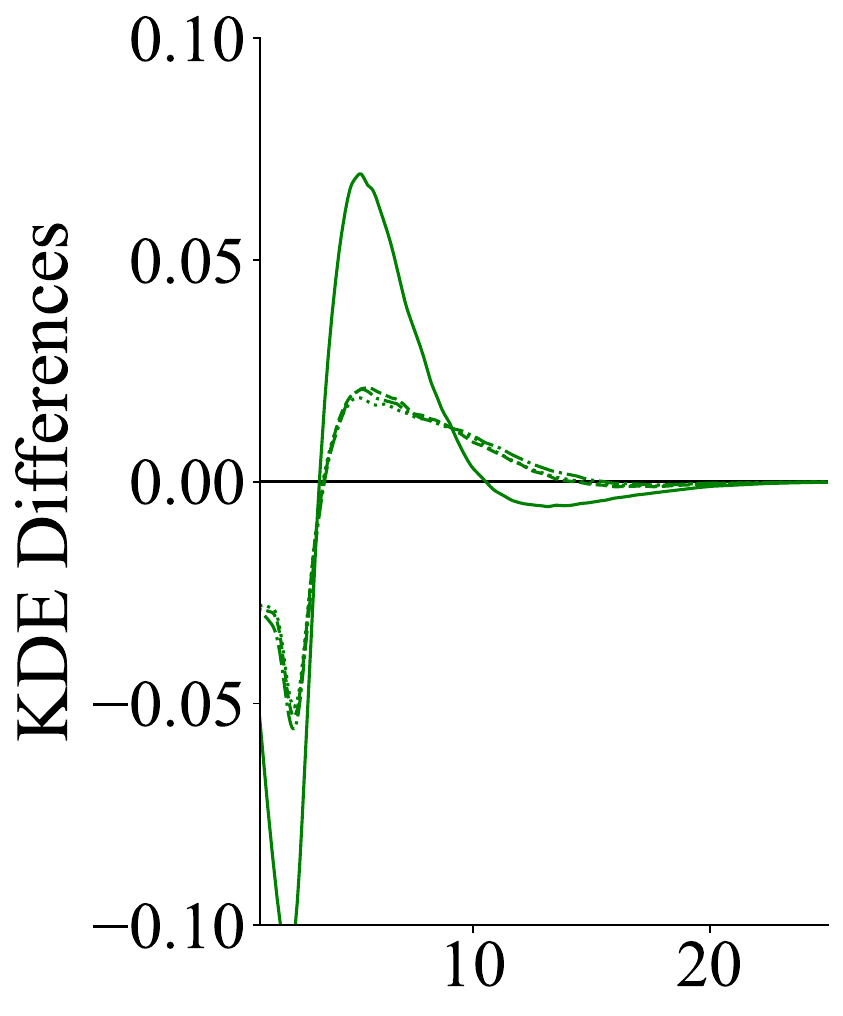}
    \end{subfigure}
    \hfill
    \begin{subfigure}[b]{0.19\textwidth}
        \centering
        \includegraphics[height=3.5cm]{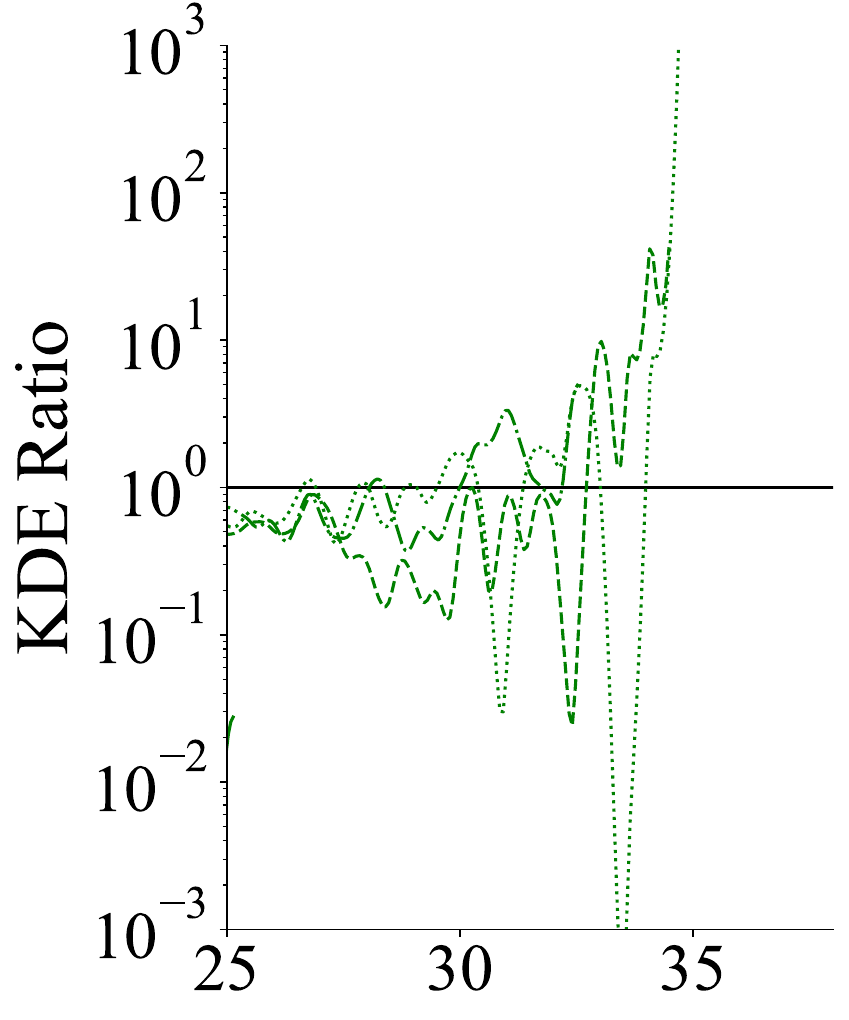}
    \end{subfigure}
    \hfill
    \begin{subfigure}[b]{0.19\textwidth}
        \centering
        \includegraphics[height=3.5cm]{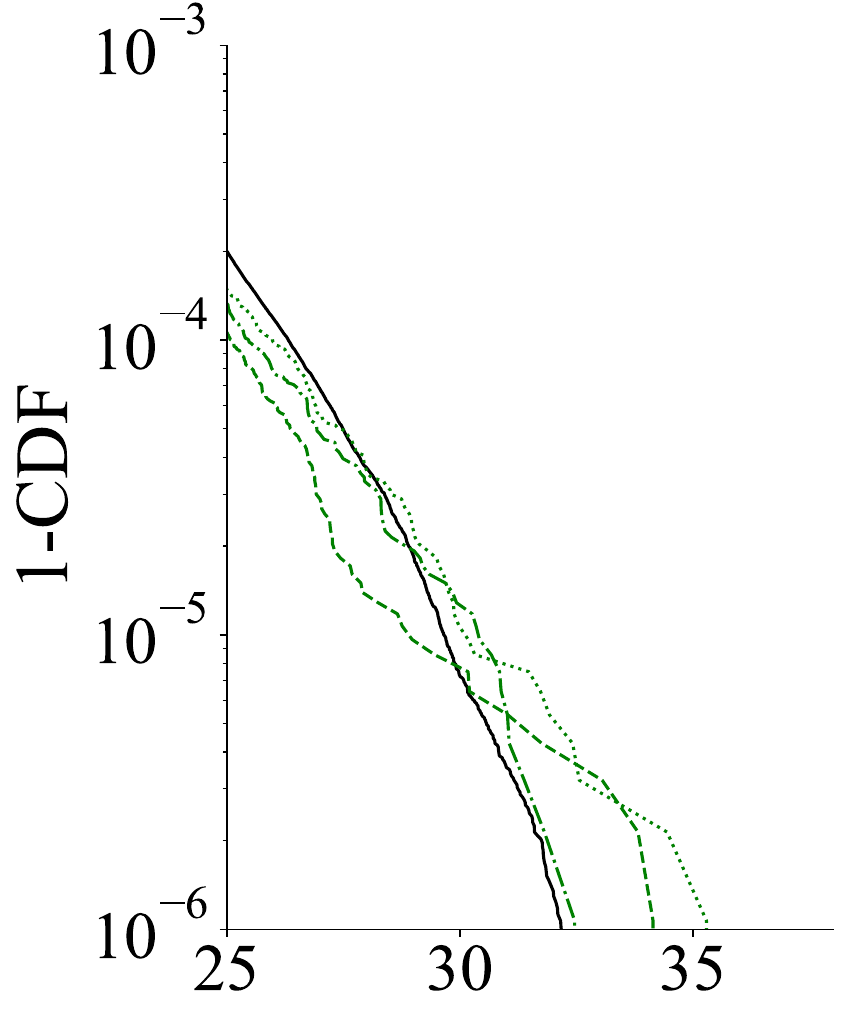}
    \end{subfigure}
    \begin{subfigure}[b]{0.39\textwidth}
        \centering
        \includegraphics[height=3.5cm]{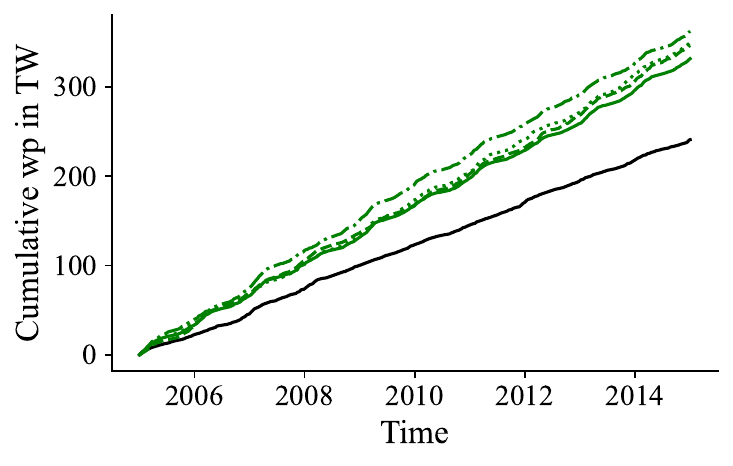}
    \end{subfigure}
    \begin{subfigure}[b]{\textwidth}
        \centering
        \includegraphics[height=1.2cm]{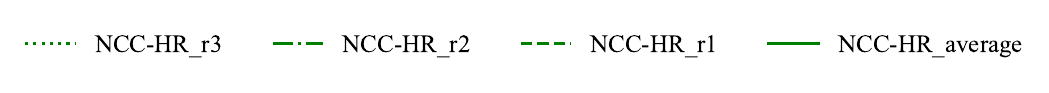}
    \end{subfigure}
    \par\bigskip
    \begin{subfigure}[b]{0.19\textwidth}
        \centering
        \includegraphics[height=3.5cm]{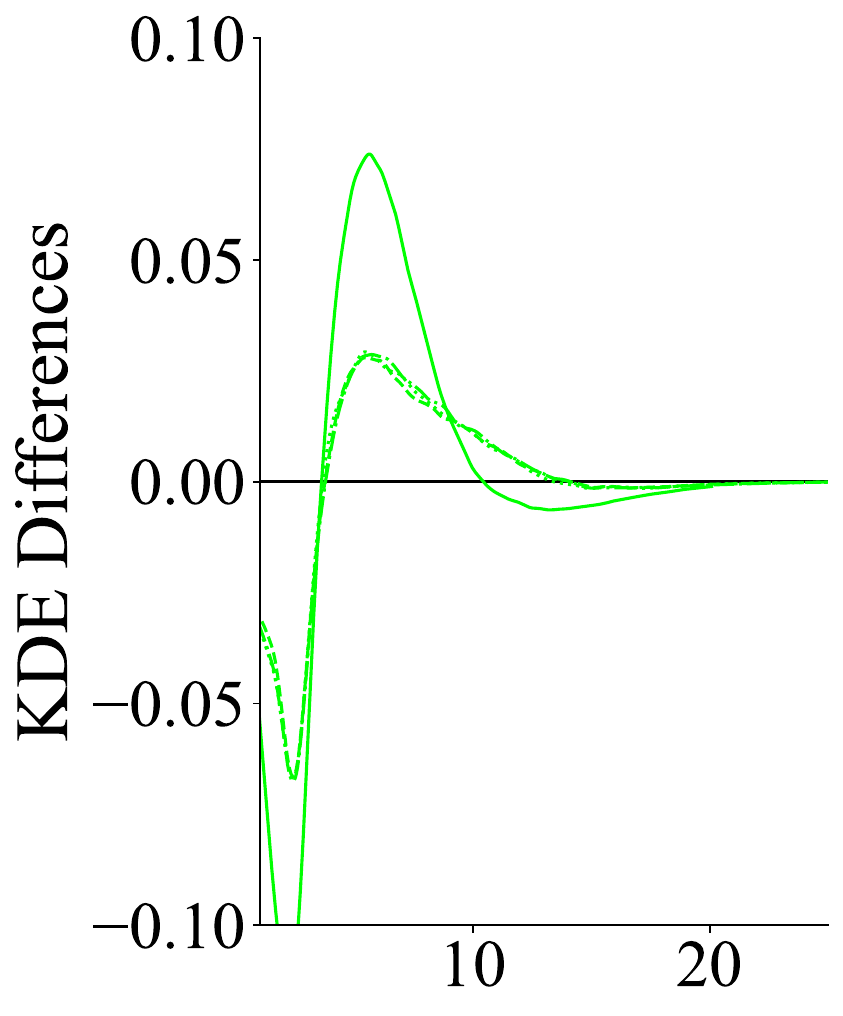}
    \end{subfigure}
    \hfill
    \begin{subfigure}[b]{0.19\textwidth}
        \centering
        \includegraphics[height=3.5cm]{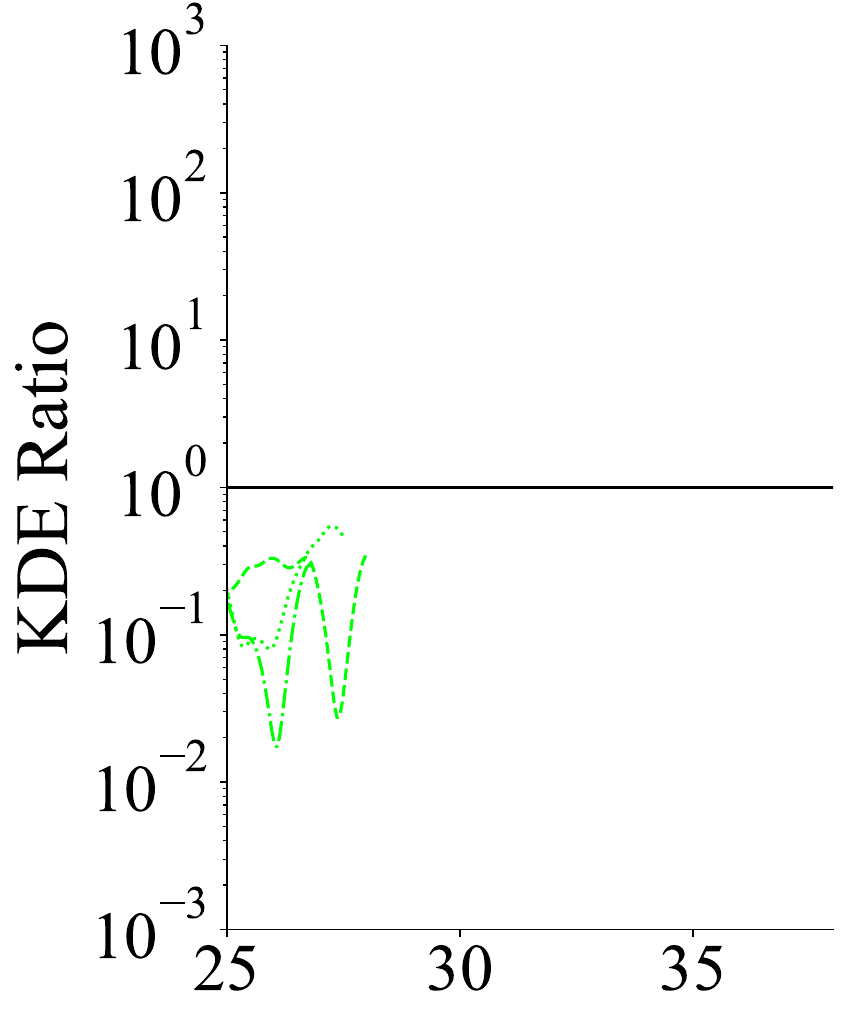}
    \end{subfigure}
    \hfill
    \begin{subfigure}[b]{0.19\textwidth}
        \centering
        \includegraphics[height=3.5cm]{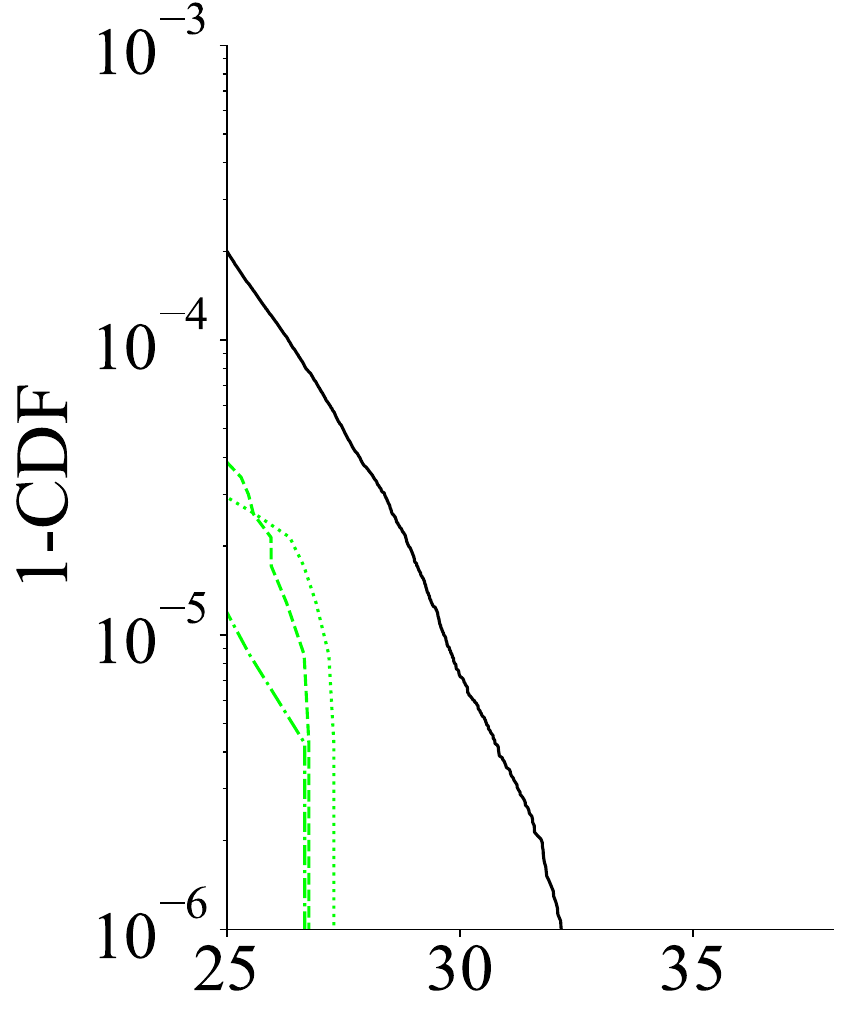}
    \end{subfigure}
    \begin{subfigure}[b]{0.39\textwidth}
        \centering
        \includegraphics[height=3.5cm]{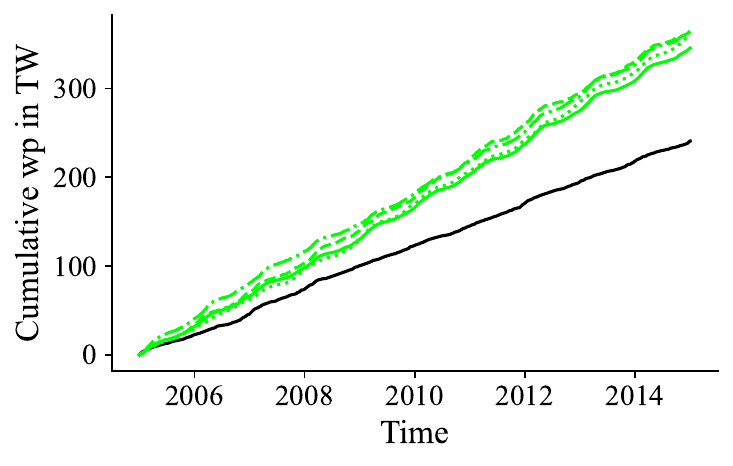}
    \end{subfigure}
    \begin{subfigure}[b]{\textwidth}
        \centering
        \includegraphics[height=1.2cm]{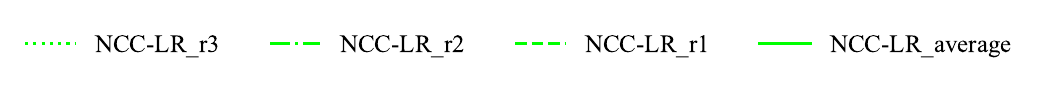}
    \end{subfigure}
    \caption{Performance comparison between model runs as in \Cref{fig:ensemble_runs_1}. Each row represents data from a single GCM (MPI or NCC) at a specific resolution (HR or LR).}
    \label{fig:ensemble_runs_2}
\end{figure}

\Crefrange{fig:ensemble_runs_1}{fig:ensemble_runs_2} illustrate how the wind speed distribution differences to ERA5 and the cumulative wind power changes across runs with varying initial/boundary conditions using the same GCM models at a fixed spatial resolution. 

The first column displays the KDE differences in the medium wind speed range. While the curves of the individual runs are barely distinguishable, averaging over them alters the difference pattern. The averages overestimate low wind speeds more than the individual runs. In addition, averaging reduces the occurrence of higher wind speeds in the medium range for the MPI and the NCC models.

The plots of the distribution tail in the second and third columns demonstrate that the averages underestimate extreme wind speeds, even when individual model runs represent them well, as in the cases of MPI-HR, NCC-HR, and CMCC. Differences between runs become more apparent in these plots but remain small, such that one run can be considered representative of the others.

The cumulative wind power plots in the last column reveal that the averages consistently provide lower wind power estimates compared to the individual runs. They represent a lower boundary rather than a median estimate of wind power.

\section{Model Biases in Previous Study}
\label{appendix:model_biases}
\begin{figure}[htp]
    \centering
    \includegraphics[width=0.7\textwidth]{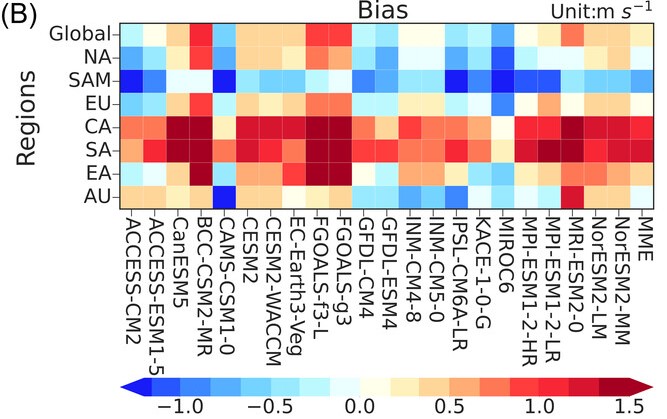}
    \caption{Bias between the mean of simulated and observed near-surface wind speeds from 1978 to 2014 as found in \citet{NSWSbiases}}
    \label{fig:mean_biases}
\end{figure}

\citet{NSWSbiases} estimated the near-surface wind speed biases of historical CMIP6 simulation data from 1978 to 2014 compared to observations listed in the Global Surface Summary of the Day database (derived from the United States Air Force DATSAV3 surface data set and the Federal Climate Complex Integrated Surface Hourly data set). Their results are shown in \Cref{fig:mean_biases}.

For Europe, \citet{NSWSbiases} found a moderate positive bias in near-surface wind speeds for the NCC and EC-Earth models, a slightly larger positive bias for the MPI-LR model, and a small positive bias for the MPI-HR model. For the IPSL and JAP models they found negative biases, with a particularly large negative bias for the JAP model. These results are in line with my results for the historical GCM output from 2005 to 2015 in the region of Germany compared to ERA5 data in \Cref{subfig:bias_mean_ws}.

\renewcommand{\thechapter}{\arabic{chapter}}

\end{document}